\documentclass[twocolumn]{aastex63}

\newcommand{\mrm}[1]{\mathrm{#1}}
\newcommand{\nuc}[2]{$\mrm{^{#2}#1}$}

\newcommand{\gr}{$\gamma$-ray\,}
\newcommand{\grs}{$\gamma$-rays\,}

\usepackage{amsmath}
\usepackage{lineno}

\usepackage{mathrsfs}
\usepackage{xurl}
\PassOptionsToPackage{hyphens}{url}\usepackage{hyperref}
\usepackage{lipsum}

\received{April 11, 2020}
\revised{May 20, 2020}
\accepted{May 21, 2020}
\submitjournal{ApJ}

\shorttitle{511\,keV imaging with COSI}
\shortauthors{Siegert et al.}

\begin{document}

\title{Imaging the 511\,keV positron annihilation sky with COSI}

\correspondingauthor{Thomas Siegert}
\email{tsiegert@ucsd.edu}

\author{Thomas Siegert}
\affiliation{Center for Astrophysics and Space Sciences, University of California, San Diego, 9500 Gilman Dr, La Jolla, CA 92093-0424, USA}

\author{Steven E. Boggs}
\affiliation{Center for Astrophysics and Space Sciences, University of California, San Diego, 9500 Gilman Dr, La Jolla, CA 92093-0424, USA}
\affiliation{Space Sciences Laboratory, University of California, Berkeley, 7 Gauss Way, Berkeley, CA 94720-7450, USA}

\author{John A. Tomsick}
\affiliation{Space Sciences Laboratory, University of California, Berkeley, 7 Gauss Way, Berkeley, CA 94720-7450, USA}

\author{Andreas C. Zoglauer}
\affiliation{Space Sciences Laboratory, University of California, Berkeley, 7 Gauss Way, Berkeley, CA 94720-7450, USA}
\affiliation{Berkeley Institute for Data Science, University of California, Berkeley, CA 94720-7450, USA}

\author{Carolyn A. Kierans}
\affiliation{NASA Goddard Space Flight Center, Greenbelt, MD 20771, USA}

\author{Clio C. Sleator}
\affiliation{Space Sciences Laboratory, University of California, Berkeley, 7 Gauss Way, Berkeley, CA 94720-7450, USA}

\author{Jacqueline Beechert}
\affiliation{Space Sciences Laboratory, University of California, Berkeley, 7 Gauss Way, Berkeley, CA 94720-7450, USA}

\author{Theresa J. Brandt}
\affiliation{NASA Goddard Space Flight Center, Greenbelt, MD 20771, USA}

\author{Pierre Jean}
\affiliation{IRAP, 9 Av colonel Roche, BP44346, 31028 Toulouse Cedex 4, France}

\author{Hadar Lazar}
\affiliation{Space Sciences Laboratory, University of California, Berkeley, 7 Gauss Way, Berkeley, CA 94720-7450, USA}

\author{Alex W. Lowell}
\affiliation{Center for Astrophysics and Space Sciences, University of California, San Diego, 9500 Gilman Dr, La Jolla, CA 92093-0424, USA}

\author{Jarred M. Roberts}
\affiliation{Center for Astrophysics and Space Sciences, University of California, San Diego, 9500 Gilman Dr, La Jolla, CA 92093-0424, USA}

\author{Peter von Ballmoos}
\affiliation{IRAP, 9 Av colonel Roche, BP44346, 31028 Toulouse Cedex 4, France}

\begin{abstract}

The balloon-borne Compton Spectrometer and Imager (COSI) had a successful 46-day flight in 2016.
The instrument is sensitive to photons in the energy range $0.2$--$5$\,MeV.
Compton telescopes have the advantage of a unique imaging response and provide the possibility of strong background suppression.
With its high-purity germanium detectors, COSI can precisely map \gr line emission.
The strongest persistent and diffuse \gr line signal is the 511\,keV emission line from the annihilation of electrons with positrons from the direction of the Galactic centre.
While many sources have been proposed to explain the amount of positrons, $\dot{N}_{\mrm{e^+}} \sim 10^{50}\,\mrm{e^+\,yr^{-1}}$, the true contributions remain unsolved.
In this study, we aim at imaging the 511\,keV sky with COSI and pursue a full-forward modelling approach, using a simulated and binned imaging response.
For the strong instrumental background, we describe an empirical approach to take the balloon environment into account.
We perform two alternative methods to describe the signal: Richardson-Lucy deconvolution, an iterative method towards the maximum likelihood solution, and model fitting with pre-defined emission templates.
Consistently with both methods, we find a 511\,keV bulge signal with a flux between $0.9$ and $3.1 \times 10^{-3}\,\mrm{ph\,cm^{-2}\,s^{-1}}$, confirming earlier measurements, and also indications of more extended emission.
The upper limit we find for the 511\,keV disk, $< 4.3 \times 10^{-3}\,\mrm{ph\,cm^{-2}\,s^{-1}}$, is consistent with previous detections.
For large-scale emission with weak gradients, coded aperture mask instruments suffer from their inability to distinguish isotropic emission from instrumental background, while Compton-telescopes provide a clear imaging response, independent of the true emission.

\end{abstract}


\section{Introduction}\label{sec:intro}

The `511\,keV positron puzzle' is one of the long-standing unresolved problems in current astrophysics \citep[see, e.g.,][for the latest review]{Prantzos2011_511}.
In the centre of the Galaxy, the strongest, persistent, diffuse \gr line signal originates from the annihilation of electrons with positrons \citep{Johnson1973_511,Leventhal1978_511}.
The true origin of these positrons, however, is unknown and difficult to determine.
While the emission itself is bright, on the order of $10^{-3}\,\mrm{ph\,cm^{-2}\,s^{-1}}$ \citep[e.g.][]{Purcell1997_511,Knoedlseder2005_511,Churazov2005_511,Jean2006_511,Weidenspointner2008_511b,Bouchet2010_511,Churazov2011_511,Skinner2014_511,Siegert2016_511,Siegert2019_lv511}, the annihilation morphology alone is believed to show only the annihilation sites and not the positron sources.
The propagation of positrons away from candidate sources possibly leads to a smearing effect, which in turn might result in the diffuse 511\,keV emission associated with the warm and partially ionised interstellar medium \citep[e.g.][]{Guessoum2006_MQ511,Prantzos2006_511,Higdon2009_511,Jean2009_511ISM,Alexis2014_511ISM,Panther2018_pos_transport}.
Nevertheless, it is still reasonable to assume that not all positrons escape their production sites and annihilate in situ \citep[e.g.][]{Milne1997_511}, which could lead to a quasi-diffuse emission built from many point-like sources, such as flaring stars \citep{Bisnovatyi-Kogan2017_511} or low-energy pair-plasma production in X-ray binaries \citep{Bouchet1991_mq511,Sunyaev1992_xrb511,Guessoum2006_MQ511,Weidenspointner2008_511b,Siegert2016_V404}.

\begin{figure}[ht!]
	\centering
	\includegraphics[trim=1.0in 1.0in 0.8in 1.0in, clip=True, width=1.0\columnwidth]{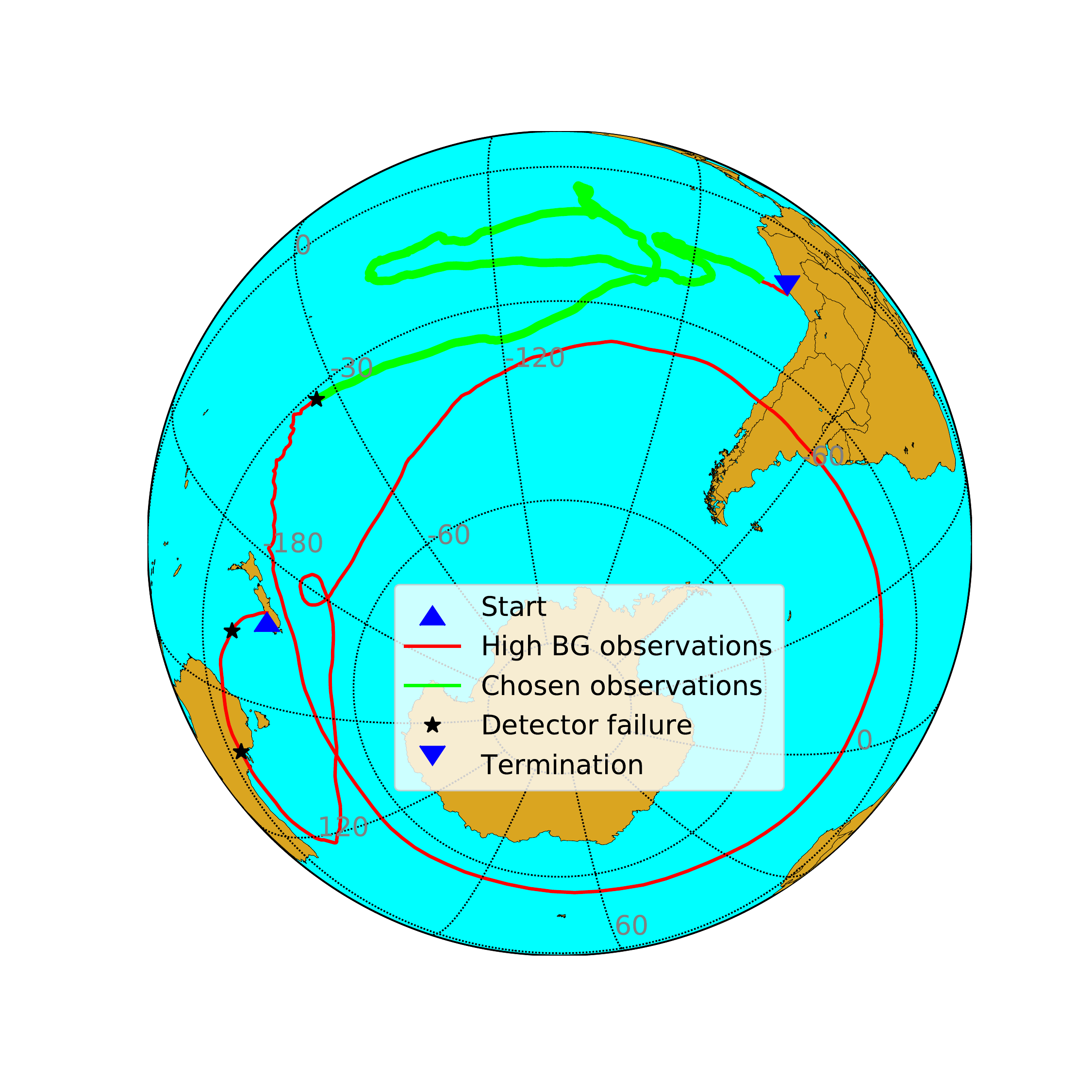}
	\caption{COSI flight path around Earth from its launch in Wanaka, New Zealand $(45^{\circ}\,\mrm{S}, 169^{\circ}\,\mrm{E}$, UTC 2016-05-16 23:35$)$, until termination in Peru $(16^{\circ}\,\mrm{S}, 72^{\circ}\,\mrm{W}$, UTC 2016-07-02 19:54$)$. The green line shows the chosen and analysed data set. High background observations (red, see also Fig.\,\ref{fig:countrate_tracers}) are excluded from the analysis. The failures of three main detectors are marked by black star symbols.}
	\label{fig:COSI2016_flightpath}
\end{figure}

The distinction between true diffuse emission and the cumulative effect of a population of point-like sources is difficult to measure in \grs because the sensitivity of today's instruments suffers from strong instrumental background, and the apertures can only provide a spatial resolution of the order of degrees.
The pioneering instruments OSSE aboard CGRO \citep{Johnson1993_OSSE} and SPI aboard INTEGRAL \citep{Winkler2003_INTEGRAL,Vedrenne2003_SPI} provided valuable insights into the true morphology of the positron annihilation emission.
OSSE, with its four scintillation collimators (spatial resolution $3.8^{\circ} \times 11.4^{\circ}$, spectral resolution $\approx 7\,\%$ at 511\,keV), provided a first image reconstruction of the Galactic 511\,keV line, showing a bright bulge and a possibly truncated disk \citep{Purcell1993_511,Purcell1997_511}.
After initial observations from balloon experiments found the Galactic emission to be apparently variable with time, results from OSSE finally resolved the signal to truly be extended and steady \citep{Lingenfelter1989_511,Purcell1997_511}.
The possible mono-polar emission towards the Galactic North pole that was reported by OSSE, however, has not been verified by other instruments.
SPI has been operating in space for 18 years, and with its high-purity germanium (Ge) detectors, the 511\,keV line and other positron annihilation emission features have been finely resolved \citep[0.4\,\% spectral resolution; e.g.][]{Jean2006_511,Churazov2005_511,Churazov2011_511,Weidenspointner2008_511b,Siegert2016_511,Siegert2019_lv511}.
SPI's $2.7^{\circ}$ resolution is achieved by a coded aperture mask, and it could possibly identify individual 511\,keV point sources.
Such `smoking-gun' evidence is still missing.
Instead, after several years of observation, SPI found the long-sought Galactic disk in positron emission \citep{Bouchet2010_511,Skinner2014_511,Siegert2016_511}, which was expected from the proposed origins of positrons related to star formation.
A study of possible `granularity' in the emission has been restricted to the bright bulge region \citep[see discussion in][]{Knoedlseder2005_511} but a clear characterisation is still missing.
Neither spiral arms nor individual positron production sites have been consistently detected.
Nevertheless, different Galactic sources, such as massive stars \citep[e.g.][]{Oberlack1996_26Al,Diehl2006_26Al,Kretschmer2013_26Al,Pleintinger2019_26Al}, core-collapse supernovae \citep[e.g.][]{Iyudin1997_CasA,Vink2001_CasA,Grebenev2012_SN1987A,Grefenstette2014_CasA,Grefenstette2017_CasA,Boggs2015_SN1987A,Siegert2015_CasA,Tsygankov2016_44Ti}, and thermonuclear supernovae  \citep[e.g.][]{Morris1995_SN1991TCOMPTEL,Churazov2014_SN2014J,Churazov2015_2014JCo,Diehl2014_SN2014J_Ni,Diehl2015_SN2014J_Co,Isern2016_SN2014J} have been shown to produce $\beta^+$-unstable nuclei, and microquasars have been claimed to produce pair-plasma \citep{Bouchet1991_mq511,Sunyaev1992_xrb511,Siegert2016_V404}.

A development towards a better understanding of this puzzle is provided by the usage of modern Compton telescopes in combination with high resolution detectors.
The Compton Spectrometer and Imager \citep[COSI, ][]{Tomsick2019_COSI} is designed as a compact Compton telescope, which utilises multiple Compton scatters in cross-strip Ge detectors to identify the direction of incoming photons.
COSI mounts 12 detectors, each measuring $8\,\mrm{cm} \times 8\,\mrm{cm} \times 1.5\,\mrm{cm}$, in a $2 [x] \times 2 [y] \times 3 [z]$ configuration, leading to a total active volume of $972\,\mrm{cm^3}$.
Five sides of the detector array are surrounded by a CsI anti-coincidence shield, leading to a field of view of $\approx \pi\,\mrm{sr}$.
COSI is a non-pointing, i.e. free-floating, survey instrument, operating as a payload of a super-pressure balloon.
After shorter previous flights \citep[see, e.g.,][for an overview]{Bandstra2011_NCT}, COSI observed the southern sky for 46 days between May and July, 2016 \citep{Kierans2016_COSI}.
The current COSI design leads to a spatial resolution of $\approx 5^{\circ}$, with a spectral resolution of $\approx 0.7\,\%$ ($\approx 3.5\,\mrm{keV}$ FWHM) at 511\,keV.
With an upgraded future version in space, COSI would be a leading next-generation $\gamma$-ray telescope with superior background rejection, and thus increased sensitivity.
This is further supported by having more detectors and therefore a larger active volume, resulting in better event reconstruction \citep[e.g.][]{vonBallmoos1989_ComptonTelescope,Boggs2000_EventReconstruction} and better spatial resolution\footnote{Note that the angular resolution of Compton telescopes is ultimately restricted to $\approx 1^{\circ}$ due to the intrinsic motion of electrons in the Ge lattice, leading to an inevitable Doppler-broadening \citep{Zoglauer2003_DopplerCompton}. Beyond this resolution, either narrow collimators or Laue lenses would be required.}.

\begin{figure*}[ht!]
	\centering
	\includegraphics[trim=1.0in 0.5in 1.0in 1.0in, clip=True, width=1\textwidth]{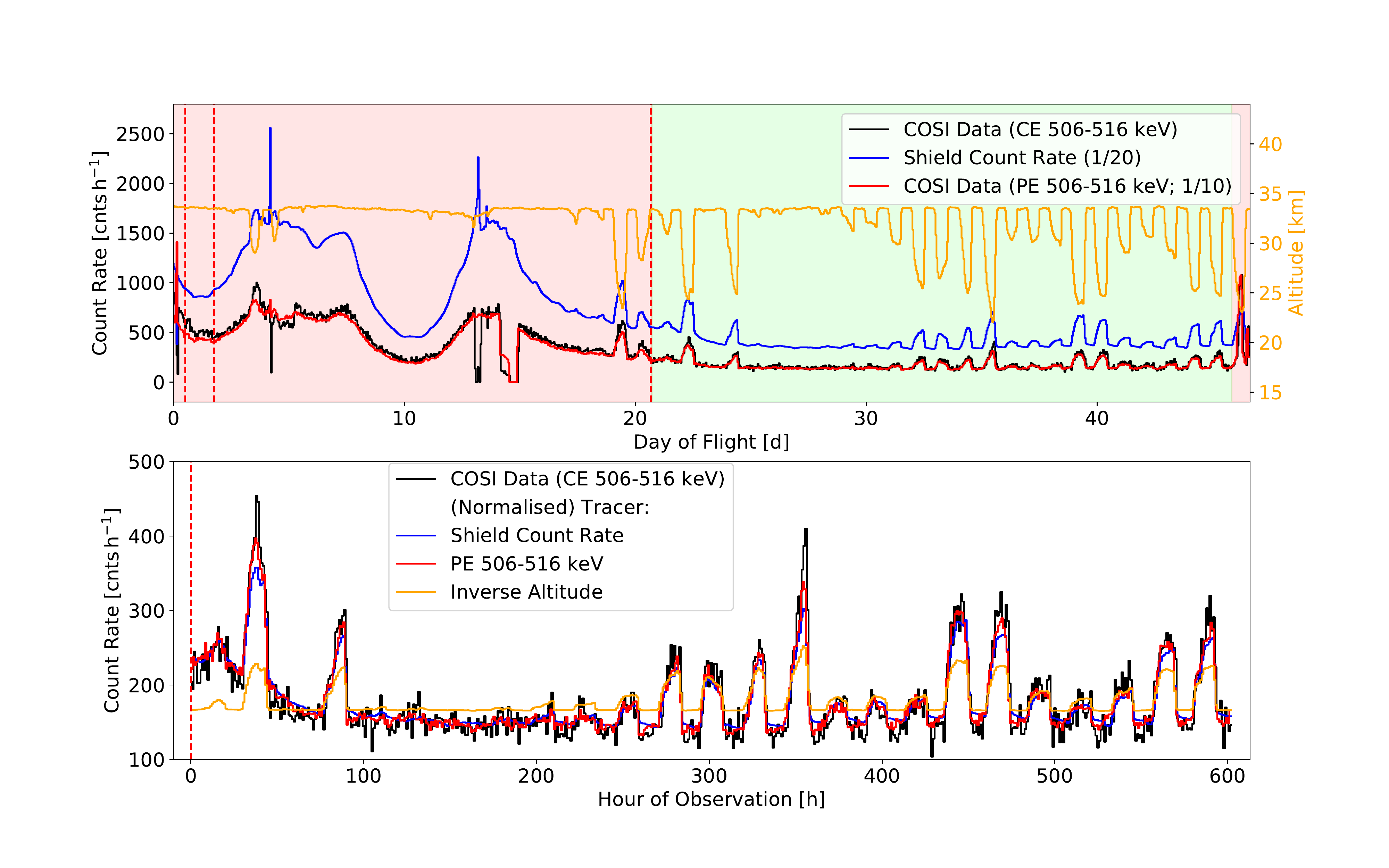}
	\caption{Measured 511\,keV Compton Event count rate (CE, black) during the 2016 COSI flight (\textit{top}), and for the chosen data set (\textit{bottom}), as a function of time. The red and green shaded areas indicate the regions of high and low background, respectively. Red dashed lines indicate times of detector failures (see also Fig.\,\ref{fig:COSI2016_flightpath}). In the top panel, also the shield count rate (blue, scaled by 1/20), the 511\,keV Photo Event rate (PE, red, scaled by 1/10), as well as the altitude (right axis, orange) is shown. The bottom panel compares qualitatively the count rate of the chosen data set (cf. Sec.\,\ref{sec:data_selection}) with potential background tracers (cf. Sec.\,\ref{sec:tracers}), normalised to the average rate.}
	\label{fig:countrate_tracers}
\end{figure*}

In order to show the unique capabilities of compact Compton telescopes, in this study we perform a rigorous imaging analysis of the 511\,keV positron annihilation line in the Milky Way, using the data from the 2016 balloon flight of COSI.
This paper is structured as follows: In Sec.\,\ref{sec:data_set}, we describe the 2016 balloon campaign, the data space intrinsic to Compton telescopes, and our specific data selection and preparation.
We show the spatial analysis of the 511\,keV line in Sec.\,\ref{sec:data_analysis}, provide our general approach for modelling the COSI data (Sec.\,\ref{sec:model_fits}), and give details about the imaging and background response of a Compton telescope in a balloon environment (Secs.\,\ref{sec:imaging_response} and \ref{sec:bg_model}).
Imaging is performed by both, an iterative deconvolution approach using a modified version of the Richardson-Lucy algorithm (Sec.\,\ref{sec:RL_deconvolution}), and in a full-forward modelling manner (Sec.\,\ref{sec:fitting_results}), based on the imaging results to identify significant structures.
Sec.\,\ref{sec:discussion} closes with a comparison to previous measurements and an outlook for future analyses.

\section{2016 campaign and data set}\label{sec:data_set}

\subsection{2016 balloon flight}\label{sec:flight}

The 46-day balloon flight of COSI in 2016 started on May 17 in Wanaka, New Zealand, and was terminated 200\,km north-west of Arequipa, Peru on July 2.
The nominal flight altitude was about 33\,km, with anomalous altitude drops related to day and night cycles (see Sec.\,\ref{sec:data_space}).
During the flight, three detectors failed, reducing the sensitivity of the instrument by $\approx 40\,\%$ (see Sec.\,\ref{sec:imaging_response}).
Because two of the malfunctions occurred in the top layer of COSI, the reduction is not proportional to the number of detectors.
The flight path of the balloon is shown in Fig.\,\ref{fig:COSI2016_flightpath}, indicating the time and position of the detector failures as well as the selected data set for our analysis (see Sec.\,\ref{sec:data_selection}).
The circumpolar winds carried the payload around Antarctica once in $\approx 14$ days before the balloon drifted towards the equator and finally landed on the west-coast of South America.
Details about the 2016 balloon flight can be found in \citet{Kierans2016_COSI} and \citet{Kierans2019_511COSI}.

The red path indicates times/regions in which the instrumental background rates were high and which are excluded in our data set (see Sec.\,\ref{sec:data_selection} for details).
In Fig.\,\ref{fig:countrate_tracers}, we show the measured count rate of 511\,keV photons (506--516\,keV), detected via multiple scatters (Compton Events (CE); black histogram) as well as complementary other rates.
The green path in Fig.\,\ref{fig:COSI2016_flightpath} coincides with the green-shaded region in Fig.\,\ref{fig:countrate_tracers}, identifying the chosen times for our data set.
During days 0--21 of the flight (red-shaded region), the 511\,keV count rate varies between $200$ and $1000$ counts per hour and no strong correlation with the flight altitude (orange, right axis) is seen.
As illustrated in Fig.\,\ref{fig:COSI2016_flightpath}, the balloon was floating at higher latitudes which influences the geomagnetic cut-off rigidity and consequently the background rate.
After day 29, frequent altitude drops lead to an increase of the CE count rate as well as the photo event rate (PE, red) and CsI shield (blue) count rate.
These nearly one-to-one correlations will be used to empirically determine the variation of the instrumental background in Sec.\,\ref{sec:tracers}, i.e. determining appropriate background tracers.
The latter are shown in the bottom panel of Fig.\,\ref{fig:countrate_tracers}, normalised to the average 511\,keV count rate during the selected data set (green-shaded region).

\subsection{COSI data space, preparation and selection}\label{sec:data_selection}

As a Compton telescope, COSI records individual triggers in the position sensitive active detector volume upon which event reconstruction is performed using the deposited energy and the kinematics of Compton scattering \citep[e.g.][]{vonBallmoos1989_ComptonTelescope,Boggs2000_EventReconstruction,Zoglauer2007_EventReconstruction}.
The stored parameters are then inherent to this measurement principle and include the total photon energy $E$, the three scattering angles, $\phi$ (Compton scattering angle; $\in [0,180^{\circ}]$), $\psi$ (polar scattering angle; $\in [0,180^{\circ}]$), and $\chi$ (azimuthal scattering angle; $\in [-180^{\circ},180^{\circ}]$), and an absolute time tag.
In addition, the aspect of COSI is saved independently as the pointing of the detector in $x$ and $z$ (optical axis) in both Galactic (longitude/latitude; $l/b$) and horizon coordinate system (specifically to perform the Earth Horizon Cut, cf. Sec.\,\ref{sec:selections}).

\newpage

\subsubsection{Binned COSI data}\label{sec:data_space}

The COSI data space therefore consists of a tag for the time and energy of each event in the three-dimensional $\{\phi \psi \chi\}$ data space.
In this work, we avoid treating each photon individually, and define a binned data space in scattering angles.
This is typically referred to as the `COMPTEL (or Compton) data space' \citep{vonBallmoos1989_ComptonTelescope,Diehl1992_CDS}.
Any narrow binning of the angles, e.g. with a bin size of $1^{\circ}$ (corresponding to $180 \times 180 \times 360 = 11,664,000$ bins), immediately results in an enormous number of data points to handle, and in fact would lead to a treatment similar to that of an unbinned analysis.
As the spatial resolution of COSI is about $5^{\circ}$, this provides a natural choice for the angular binning since we expect a signal of about $7\sigma$ \citep{Kierans2019_511COSI} to be distributed over the bulge region, thus avoiding a unmanageably large image data space.
We divide the Compton scattering angle, $\phi$, into 36 regular $5^{\circ}$ bins.
The remaining $(\psi/\chi)$-sphere is cut into 1650 irregular 2D-bins with equal solid angles \citep[cf.][]{Zoglauer2006_MEGAlib}.
The resulting $\{\phi \psi \chi\}$ data space thus contains $59,400$ scattering angle bins (see below for further reduction). 

The Ge detectors resolve an instrumental 511\,keV line with a FWHM of about 3.5\,keV.
The observed astrophysical broadening of the narrow 511\,keV component is about 2.0\,keV \citep[e.g.][]{Jean2006_511,Churazov2011_511,Siegert2019_lv511}, resulting in a combined Doppler broadening of $\approx 4$\,keV.
Thus, 99.7\,\% ($3\sigma$) of the expected counts of the 511\,keV line are included in a band of $\approx 10$\,keV.
We select only photons which fall into the energy interval $[506,516]$\,keV for our data set (one energy bin).
For a resolved COSI spectrum around the positron annihilation line, we refer to Fig.\,6.5 in \citet{Kierans2018_PhD}, and the spectral analysis in \citet{Kierans2019_511COSI}.

Since COSI is, to first order, zenith pointing, and is additionally moving around Earth, the time intervals used for the analysis should not be too long, because different exposures with and without signals will be combined together in time.
They should also not be too short as the limited number of counted photons would lead to an unnecessarily large data space.
Here, we adopt a time binning of one hour, resulting in 603 time bins of active observations, and weight different impacts on the imaging and background response accordingly within each hour (cf. Sec.\,\ref{sec:stability}).

We can further reduce the number of data space bins since many bins are never occupied either in the (selected) data set (Sec.\,\ref{sec:selections}), the background (Sec.\,\ref{sec:bg_response}) nor the imaging response (Sec.\,\ref{sec:imaging_response}).
This leads to a reduced data space, $\{\phi \psi \chi\}_R$, with $4243$ scattering angle bins. The total number of bins in the pre-defined data space is thus $4243 [\{\phi \psi \chi\}_R] \times 1 [E] \times 603 [T] = 2,558,529$.

\begin{table}
	\centering
	\begin{tabular}{lc}
		\hline
		Parameter & Selection \\
		\hline
		Energy $[\mrm{keV}]$ & $[506,516]$ \\
		Time $[\mrm{MJD}]$ & $[57545.78, 57570.86]$ \\
		Number of interactions & $[2,7]$ \\
		Interaction distance $[\mrm{cm}]$ & $>0.5$ (first 2); $>0.3$ (any) \\
		Compton Scattering Angle & $[0,60^{\circ}]$ \\
		Altitude $[\mrm{km}]$ & $[22,35]$ (all; see Appendix\,\ref{sec:appendix_reliability}) \\
		Pointing (coordinates) & full-sky (full exposure) \\
		Earth Horizon Cut & yes \\
		\hline
	\end{tabular}
	\caption{Event selections used for the 511\,keV imaging analysis.}
	\label{tab:selections}
\end{table}

\subsubsection{Event selections}\label{sec:selections}

In the above-described data space, we further select events which follow more detailed quality criteria:
While a lower altitude increases the background count rate, these drops (cf. Fig.\,\ref{fig:countrate_tracers}) happen mainly during observations of the Galactic centre, i.e. when the strongest signal is expected in 511\,keV.
We therefore do not restrict our data set to a specific altitude interval and rather use the full response (see Sec.\,\ref{sec:imaging_response}) to estimate the expected count rate (see, however, Appendix\,\ref{sec:appendix_reliability} for alternative selections as a reliability cross check).
This avoids `optimising for the signal' \citep[e.g.][pp. 79–96]{Koehler1993_priorbeliefs, Nickerson1998_confirmationbias,Pohl2004_biases} as can typically happen in background-dominated measurements with an apparently `known' outcome.

\begin{figure}[ht!]
	\centering
	\includegraphics[trim=0.0in 0.1in 0.6in 0.0in, clip=True, width=1.0\columnwidth]{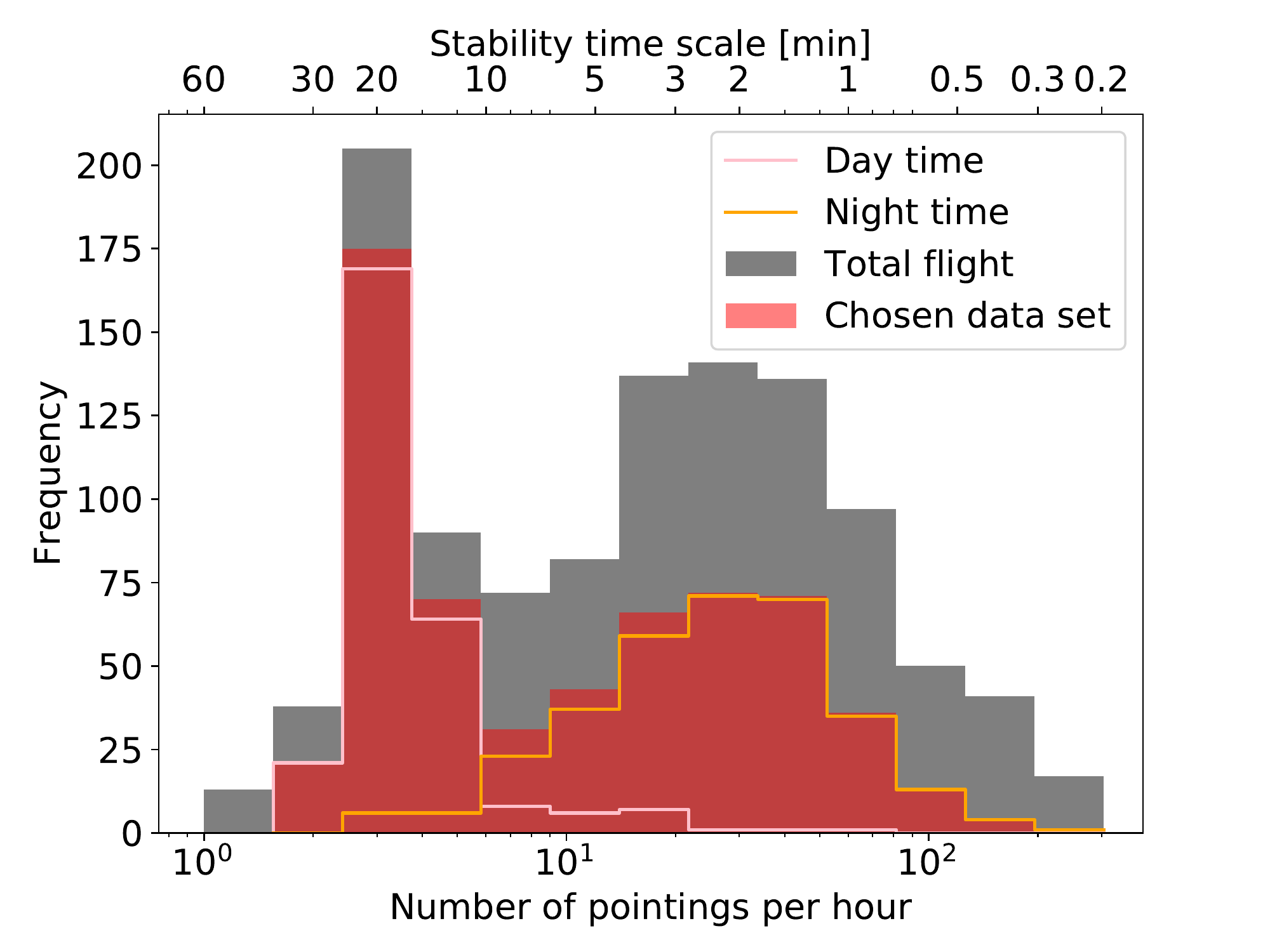}
	\caption{Number of stable pointings per hour of observation as given by the criterion in Sec.\,\ref{sec:data_space} with a $5^{\circ}$ threshold. The gray (red) shaded histogram shows the full (selected) data set. Separating the observations in day and night time explains the bi-modality of the distribution: during day times, the balloon orientation changes only every $\sim 20\,\mrm{min}$, i.e. about to the rotation velocity of Earth of $15^{\circ}\,h^{-1}$. At night, rotation, tumbling, and vast altitude changes make individual pointings unstable so that the response (see Sec.\,\ref{sec:imaging_response}) during one hour has to be re-weighted more often.}
	\label{fig:balloon_stability}
\end{figure}

We further use all exposed regions during the $603$ selected hours as this allows the background to be properly defined using regions that are expected to be empty, as well as to search for 511\,keV disk emission.
For the individual events, we only select those with a kinematic Compton reconstruction chain length (number of interactions in the detectors) of $2$ to $7$.
Events with three or more scatters provide redundant information in the reconstruction, leading to higher fraction of correctly reconstructed events \citep{Zoglauer2006_PhD}.
The angular resolution of COSI at 511\,keV is dominated by the position resolution due to the strip pitch in the Ge detectors.
As consequence, events for which the first and second interaction are farther apart have better angular resolution.
Using a minimum distances of $0.5$\,cm between the first two interactions and $0.3$\,cm between subsequent interactions inside the detectors is found to be a good compromise between improving the angular resolution and reducing the detector efficiency
The Compton scattering angle itself provides a quality measure as potential backscatters ($> 90^{\circ}$) are difficult to reconstruct.
We further select $\phi$ according to the imaging response quality for larger angles (see Sec.\,\ref{sec:imaging_response}), being less and less populated for angles larger than $60^{\circ}$.
Since the Earth Horizon Cut (see below) removes any events above $90^{\circ}$, and significantly reduces the numbers between $60^{\circ}$ and $90^{\circ}$, we restrict $\phi$ to $\leq 60^{\circ}$.
The Earth Horizon Cut rejects Compton events that, projected back onto the celestial sphere, would be intersecting with the Earth horizon.
This largely avoids albedo radiation, i.e. a physical background to our 511\,keV measurements.
The specific event selections are summarised in Tab.\,\ref{tab:selections}.
The total number of photons in our data set is then $N_{ph} = 107,880$.
Thus, only $\approx 4.2\,\%$ of the data space is populated and many bins carry zero counts.
This requires a proper statistical treatment using Poisson statistics (see Sec.\,\ref{sec:model_fits}).

\subsubsection{Balloon stability and pointing definition}\label{sec:stability}

In each of the $603$ observation hours, the balloon gondola's absolute position (aspect) is changing.
This means that either the observation direction ($z$-axis) or the detector plane ($xy$) changes from one instance in time to another.
This has to be taken into account when applying the instrument response for different times, and also within a single time bin of one hour.
We define pointings of COSI observations, i.e. over which the imaging response is applied, by a stability criterion of the gondola:
the times until the normal vectors of any instrument plane change by more than $5^{\circ}$ are accumulated and saved as weighting factors for the imaging response within individual time bins.
Such a treatment considers the steady slew of the instrument as well as intrinsic rotation and tumbling of the payload.

In total, this evaluates to $35,938$ pointings for the whole flight and $11,922$ for the $603$ one-hour time bins of our selection.
Fig.\,\ref{fig:balloon_stability} shows the distribution of pointing lengths for the complete 46-day flight (gray) as well as the chosen data set (red).
Clearly, the distribution is bi-modal, which arises from the day and night times:
during daylight, a rotator below the balloon steers the payload such that the solar panels are optimally exposed by the Sun.
This provides a smooth behaviour of the instrument aspect and is only slightly disturbed by altitude changes (first peak; pink histogram).
The stability time scale peaks at $20$\,min (corresponding to $\approx \mrm{5^{\circ}\,(20\,min)^{-1}} = \mrm{15^{\circ}\,h^{-1}}$, i.e. the rotation speed of Earth), so that only a few pointings are required to define the one-hour time bins.
At night times, the rotator is turned off and the payload more freely rotates about its zenith which leads to a stronger influence of the environment.
The stability time scale peaks around 2\,min, so that on average $\approx 30$ pointings have to be included in one hour.

\begin{figure*}[ht!]
	\centering
	\includegraphics[trim=0.5in 2.0in 0.5in 5.0in, clip=True, width=1.0\textwidth]{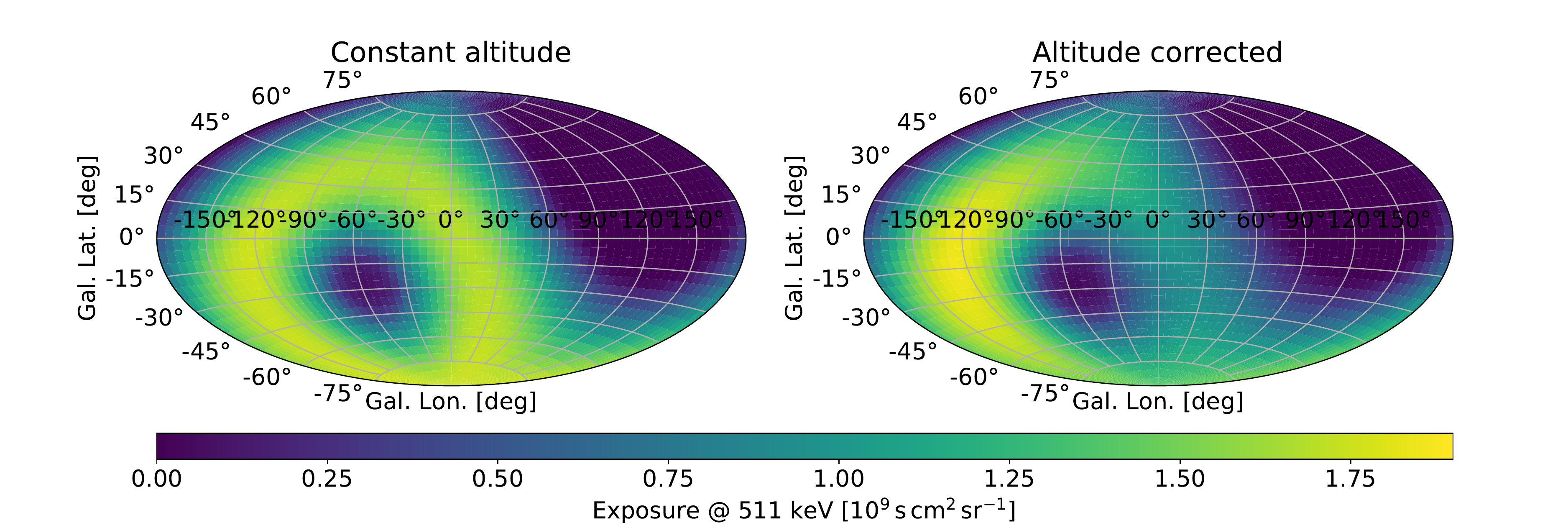}
	\caption{Exposure map of the selected data set ($T_{obs} = 603\,\mrm{h} \approx 2.2\,\mrm{Ms}$) at 511\,keV photon energy in units of $\mrm{10^{9}\,s\,cm^2\,sr^{-1}}$, assuming a constant altitude of 33\,km (left), and correcting for atmospheric absorption as a function of altitude (right). The Galactic centre is about 40\,\% less exposed when taking the altitude change into account (see also Sec.\,\ref{sec:imaging_response}).}
	\label{fig:exposure}
\end{figure*}

The distributions of $\phi$, $\psi$, and $\chi$ for each observation hour have been investigated to allow for a similar re-weighting of the background response as a function of time, altitude, and position on Earth.
These distributions are constant with respect to all observation-specific parameters, so that we can safely assume the background response to be independent of the instrument aspect.
We would expect that the background response also shows a weak dependence on the balloon altitude, but which has not been observed.
Introducing such a dependence would probably be required for longer flights when these trends become important.
We note that the amplitude of the background still shows the expected correlation with balloon altitude, which will be taken care of when defining the background model (see Secs.\,\ref{sec:bg_model} and \ref{sec:bg_response} for further details).

\section{Spatial analysis}\label{sec:data_analysis}

In this section, we will describe two approaches for inferring information about the spatial distribution of 511\,keV emission in the Galaxy from COSI data.
First, we will introduce the basic principle for full-forward modelling in the COSI-specific data space (Sec.\,\ref{sec:model_fits}), where we include the effect of the dynamic aspect of the balloon gondola in the imaging response (Sec.\,\ref{sec:imaging_response}), and the variability of the instrumental background with altitude (Sec.\,\ref{sec:bg_model}). 
For a rather model-independent approach to determine the emission morphology, we use an adapted version of the Richardson-Lucy deconvolution algorithm in Sec.\,\ref{sec:RL_deconvolution}.
This provides a baseline for the use of empirical functions in a full-forward fitting approach to reliably characterise the flux and extent of the Galactic 511\,keV emission as seen by COSI (Sec.\,\ref{sec:fitting_results}).
By using these two methods, we can cross-check different modelling assumptions and provide consistency and systematics estimates.

\subsection{General approach}\label{sec:model_fits}

We model the number of counts in a data space bin, $\left\{\phi \psi \chi t\right\}$, as a linear combination of sky model, $m_{\phi \psi \chi t}^{\mrm{SKY}}$, and background model components, $m_{\phi \psi \chi t}^{\mrm{BG}}$, such that:

\begin{widetext}
	\[
	m_{\phi \psi \chi t} = m_{\phi \psi \chi t}^{\mrm{SKY}} + m_{\phi \psi \chi t}^{\mrm{BG}} = \int_{d\Omega} \cos(b)dbdl \sum_{p_t \in t} R_{\phi \psi \chi}^{\mrm{SKY}}(Z,A,h) \cdot p_t\left((Z,A) \leftrightarrow (l,b)\right) \cdot M(l,b;\mathbf{\theta}_{s}) + R_{\phi \psi \chi}^{\mrm{BG}} \cdot T_t(\mathbf{\theta}_{b})\mrm{.}
	\tag{1} \label{eq:model_equation}
	\]
\end{widetext}

\noindent In Eq.\,(\ref{eq:model_equation}), $R_{\phi \psi \chi}^{\mrm{SKY}}(Z,A,h)$ is the imaging response of COSI as a function of zenith ($Z$), azimuth ($A$), and altitude ($h$), which is mapping the sky model, $M(l,b;\mathbf{\theta}_{s})$, to the COSI data space, $\left\{\phi \psi \chi t\right\}$, by integrating over the exposed sky region, $d\Omega$, weighted by the pointings' time, $p_t \in t$, defined in each time bin, $t$, which also links the internal zenith/azimuth coordinate system to Galactic coordinates, $(Z,A) \leftrightarrow (l,b)$.
The description of the spatial distribution of photons in image space is parametrised either by a differential flux value per individual pixel (Richardson-Lucy deconvolution; Sec.\,\ref{sec:RL_deconvolution}), or by a set of sky model parameters, $\mathbf{\theta}_{s}$, which can include the shapes, extents, and flux normalisations of a multitude of morphologies, such as individual point sources or extended emission (Sec.\,\ref{sec:fitting_results}).
The background response, $R_{\phi \psi \chi}^{\mrm{BG}}$, describes the expected distribution of photons in the data space and is constant in time and altitude.
The absolute rate of background can change with time such that its temporal variability is included by a tracer function, $T_t(\mathbf{\theta}_b)$, and parametrised by a set of background parameters, $\mathbf{\theta}_{b}$, which can include time nodes and various amplitudes (see Sec.\,\ref{sec:bg_model}).

In this way, the total model counts are predicted as parametrised through $\mathbf{\theta}_s$ and  $\mathbf{\theta}_b$, such that $m_{\phi \psi \chi t}(\mathbf{\theta}_s,\mathbf{\theta}_b)$ will be unit-less (number of photons).
Because this describes a counting experiment, the distribution of photons in each data space bin follows the Poisson statistics, and therefore the total model is determined by maximising the Poisson likelihood,

\begin{align}
	\mathscr{L}(d|m) = \prod_{{\phi \psi \chi t}} \frac{m^{d}\exp(-m)}{d!}\mrm{,}\tag{2}
	\label{eq:Poisson_likelihood}
\end{align}

\noindent with $d$ being the measured counts in each data space bin $\left\{\phi \psi \chi t\right\}$.
The general description of $M(l,b;\mathbf{\theta}_{s})$ predicts differential fluxes in units of $\mrm{ph\,cm^{-2}\,s^{-1}\,sr^{-1}}$.
Applying the imaging response, $R_{\phi \psi \chi}^{\mrm{SKY}}(Z,A,h)$ (in units of $\mrm{cm^2}$), to a sky model for a certain pointing duration, $p_t$ (in units of $s$), is computationally very expensive for a particular combination of spatial and amplitude parameters.
This would be required in each step of a likelihood maximisation.
However, the same spatial parameters (e.g. the position or the width of a 2D-Gaussian; see Sec.\,\ref{sec:2Dgaussians}) predict the same relative numbers in the $\left\{\phi \psi \chi t\right\}$ data space.
For this reason, the amplitude (i.e. flux normalisation) can be separated, as this parameter only scales the expected number in each bin up and down, but will not change the expected patterns.
This means the amplitude, $\alpha_s$, for each sky model $s$, can be handled independently of the already-`convolved sky models',

\begin{align}
	m_{\phi \psi \chi t}^{\mrm{SKY,s}} & = \int d\Omega \sum_{p_t \in t} R_{\phi \psi \chi}^{\mrm{SKY}}(Z,A,h) \cdot p_t \cdot M_s(l,b;\mathbf{\theta}_{s}) = \nonumber\\
	 & = \alpha_s \cdot \int d\Omega \sum_{p_t \in t} R_{\phi \psi \chi}^{\mrm{SKY}}(Z,A,h) \cdot p_t \cdot M_s(l,b;\mathbf{\theta}_{s}^*) = \nonumber\\
	 & = \alpha_s \cdot m_{\phi \psi \chi t}^{\mrm{SKY,s,*}} \tag{3} \mrm{.}
	 \label{eq:convolved_sky_models}
\end{align}

\noindent In Eq.\,(\ref{eq:convolved_sky_models}), the set of sky model parameters is separated into a fixed set of   parameters, $\mathbf{\theta}_{s}^*$, and the amplitude: $\mathbf{\theta}_{s} = \{\mathbf{\theta}_{s}^*, \alpha_s \}$.
In this way, for a specific (set of) model(s), $m_{\phi \psi \chi t}^{\mrm{SKY,s,*}}$ is only calculated once, and the flux determined for (a set of) pre-defined, fixed, parameters; $m_{\phi \psi \chi t}^{\mrm{SKY,s,*}}$ is termed `convolved sky model'.
This methodology will also be used when using the Richardson-Lucy deconvolution algorithm, and its modification for accelerated convergence (Sec.\,\ref{sec:arldobags}).

In contrast to the imaging response which is derived from simulations (see Sec.\,\ref{sec:imaging_response}), the background response is determined purely empirically and will therefore be treated as being unit-less.
Details about how the background modelling is approached are given in Sec.\,\ref{sec:bg_model}.

\subsection{Imaging response}\label{sec:imaging_response}

We use the Medium-Energy Gamma-ray Astronomy library \citep[MEGAlib,][]{Zoglauer2006_MEGAlib} to simulate the expected number of photons at 511\,keV as a function of the intrinsic zenith and azimuth coordinate system.
Such a simulation requires a detailed mass model of COSI, and has to take into account the three dead detectors as well as the attenuation of the atmosphere at a specific altitude.
The latter has large impact on the resulting effective area as a function of zenith because more air mass has to be passed at the same zenith angle for lower altitudes.

\begin{figure}[ht!]
	\centering
	\includegraphics[trim=0.5in 0.6in 1.0in 1.0in, clip=True, width=1.0\columnwidth]{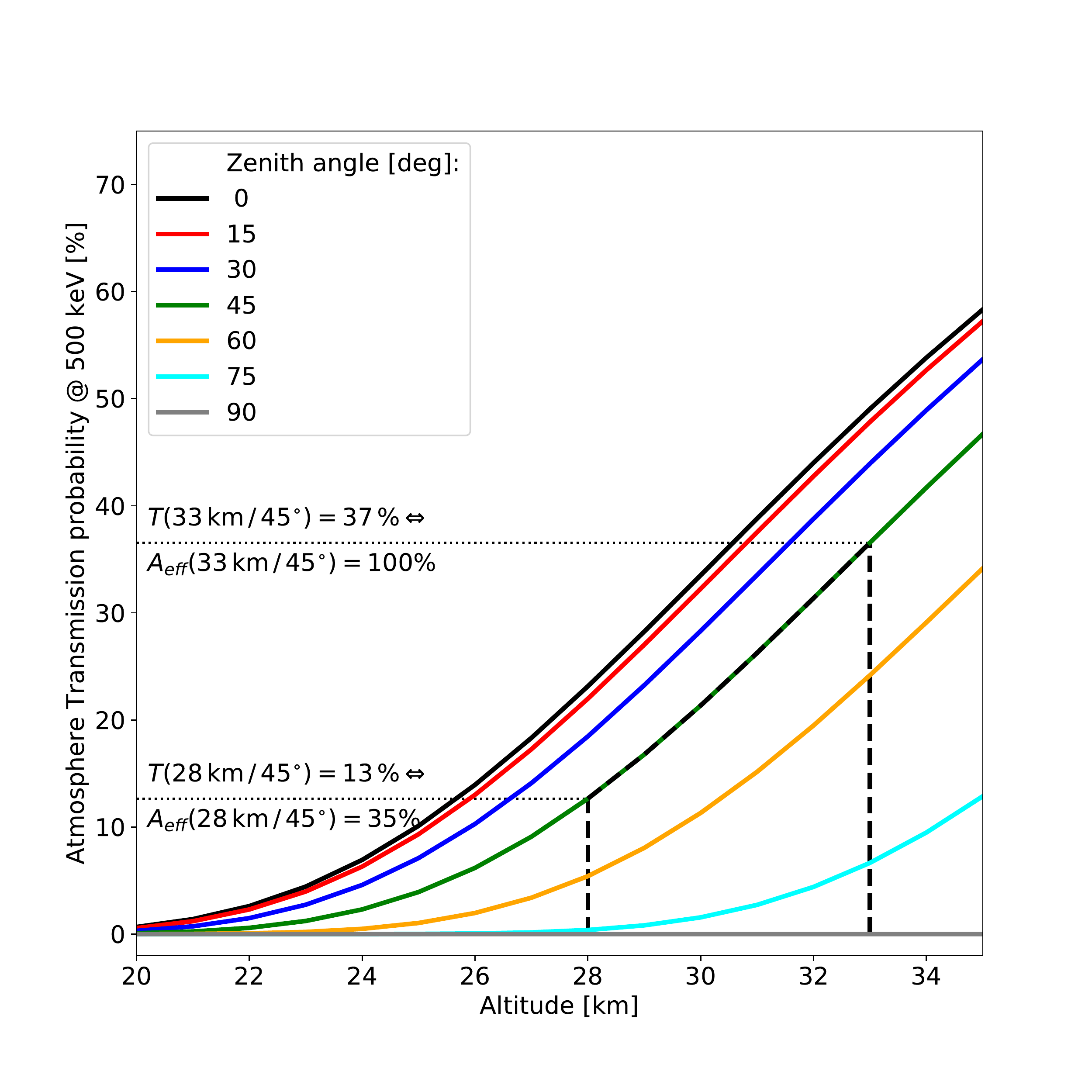}
	\caption{Renormalisation of the aspect-dependent response at 500\,keV as a function of different altitudes. The nominal response was calculated for a floating altitude of 33\,km, corresponding to a transmission probability through the atmosphere for zenith angles of $45^{\circ}$, for example, of $\approx 37\,\%$. For the same aspect angle, the resulting effective area at 28\,km altitude is reduced to 35\,\%. See text for details.}
	\label{fig:altitude_correction}
\end{figure}

The simulation setup places the mass model of COSI with $9$ functioning detectors in the centre of an isotropically emitting sphere, at a nominal altitude of $h = 33\,\mrm{km}$ (defining the transmission probabilities).
The total number of simulated photons is $\approx 2.65 \times 10^{12}$ which took about $3.5$ million CPU hours of computation time at the National Energy Research Scientific Computing Center's supercomputer \textit{Cori}.
The simulated events then pass through a well-benchmarked detector effects engine \citep{Sleator2019_COSI_DEE}, making them appear like actual data (e.g., strip numbers instead of positions, AD units instead of energy).
The simulated events then pass through the same calibration and analysis pipeline as the real data.
After event reconstruction \citep{Zoglauer2006_PhD}, the events are binned according to a pre-defined spacing in a 5-dimensional data space, defined by the zenith and azimuth angles in detector coordinates, $(Z,A)$, as well as the Compton data space, $\left\{\phi \psi \chi\right\}$.
Here, on average, a $5^{\circ}$ spacing is used.
Finally, a 5-dimensional sky response is created: $R_{\phi \psi \chi}^{\mrm{SKY}}\left(h=33\,\mrm{km};Z,A\right)$.

\begin{figure}[ht!]
	\centering
	\includegraphics[trim=0.0in 0.1in 0.5in 0.3in, clip=True, width=1.0\columnwidth]{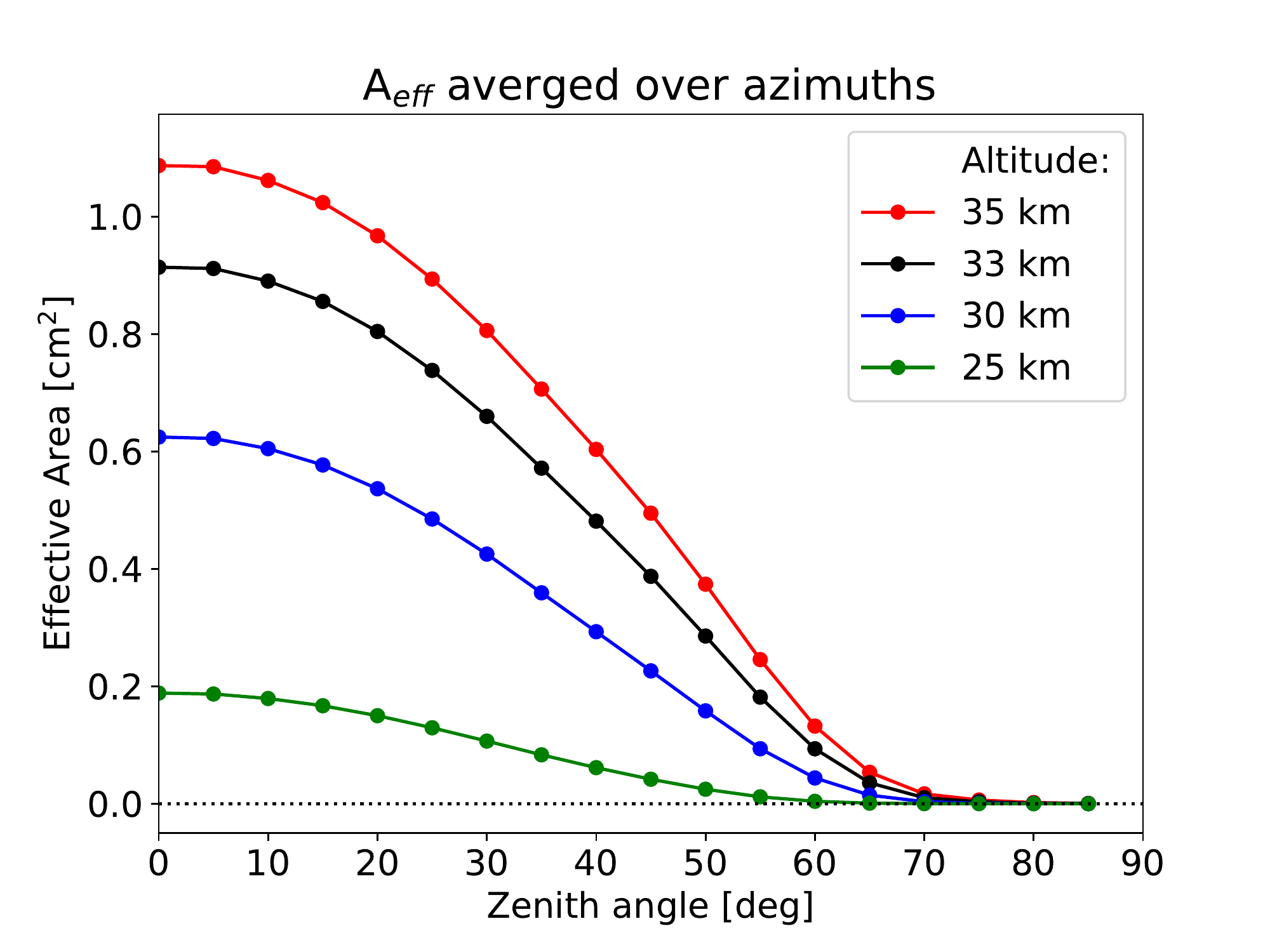}
	\caption{Absolute effective area at 511\,keV, averaged over $360^{\circ}$ of azimuths, as a function of zenith angle for different balloon altitudes. The altitude- and time-averaged effective area of the expected 511\,keV signal in the Milky Way for the chosen data set is $0.59\,\mrm{cm^{2}}.$}
	\label{fig:effective_area}
\end{figure}

Since the balloon altitude is changing between about 22\,km and 34\,km in our selected data set, the 6th dimension of altitude has to be included as well.
Instead of performing multiple simulations with ever-increasing computing time, we use the simulated response at 33\,km to build a grid of relative transmissivities for zenith and azimuth as a function of altitude.
In Fig.\,\ref{fig:altitude_correction}, the altitude-dependent atmospheric transmission probability at 500\,keV photon energies is shown for different zenith angles.
In the indicated example, the absolute transmission probability (transmissivity) at nominal altitude (33\,km) for a zenith angle of $45^{\circ}$ is 37\,\%, which corresponds to a relative effective area of 100\,\% (relative to the value at nominal altitude).
At the same zenith angle, but at a considerably lower altitude, for example 28\,km, the absolute transmission probability is only 13\,\%, for which the effective area is to be rescaled by $13\,\% \cdot 100\,\% / 37\,\% = 35\,\%$.
We create a grid of altitudes from 20 to 35\,km in 1\,km steps and zenith angles between $0$ and $90^{\circ}$ in $5^{\circ}$ steps to determine a re-normalisation for the absolute effective area of COSI around 500\,keV photon energies.
The resulting azimuth-averaged effective area is shown in Fig.\,\ref{fig:effective_area} for different altitudes.

It is evident from Fig.\,\ref{fig:effective_area} that the effective area is drastically changing with both zenith angle and altitude.
While the effective area naturally decreases with zenith due to the finite projected geometric area, the largest impact is still originating from the larger airmass that photons have to pass, reducing the effective area for larger zeniths even further.
Also above adequate flight altitudes, $\gtrsim 30\,\mrm{km}$, the effective area at zenith varies by $\approx 40\,\%$.
With $9$ functioning detectors at 33\,km altitude, COSI's effective area at zenith is about $0.91\,\mrm{cm^2}$.

\begin{figure*}[ht!]
	\centering
	\includegraphics[trim=0.5in 0.6in 0.5in 0.9in, clip=True, width=0.9\textwidth]{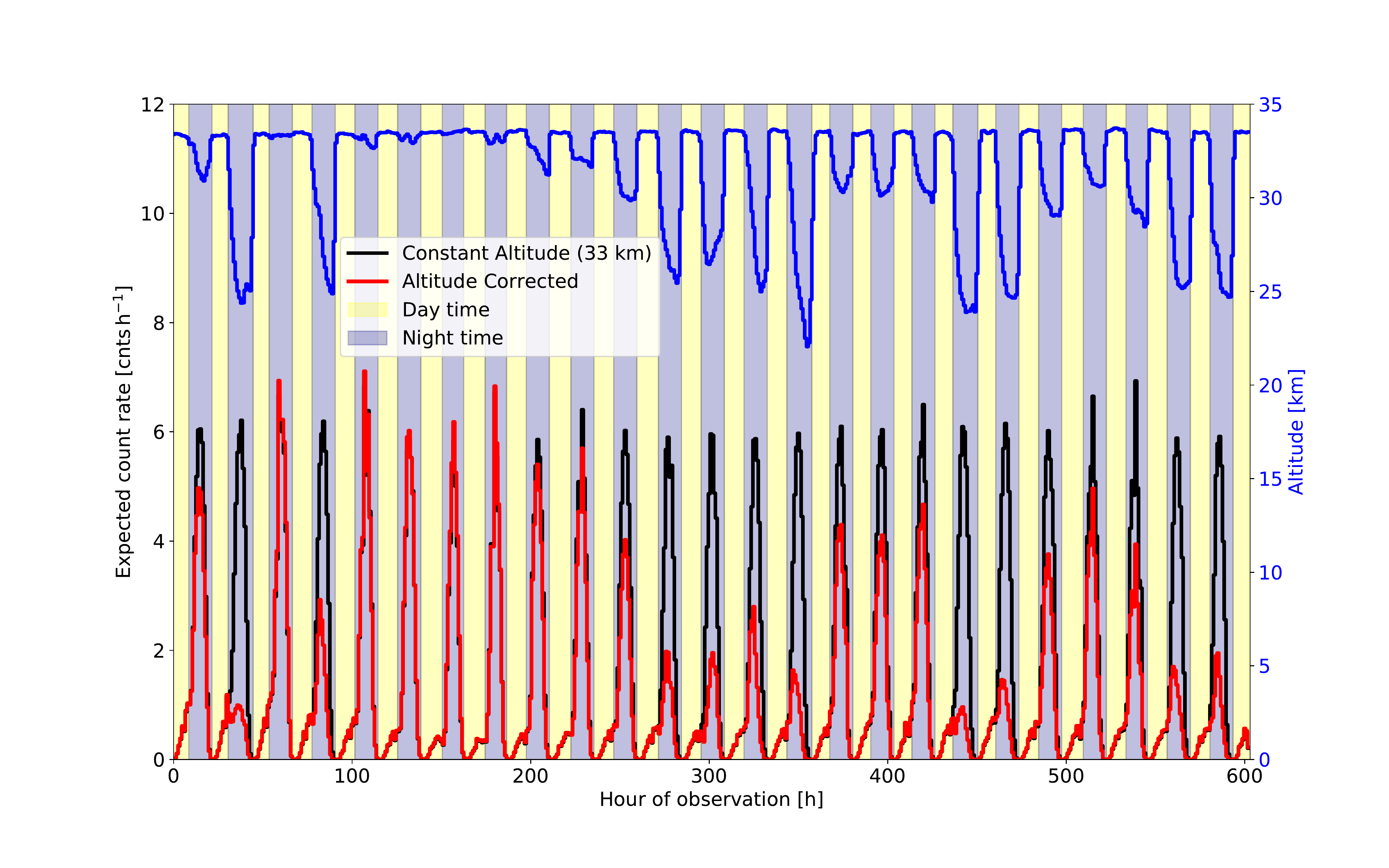}
	\caption{Expected count rate for the 511\,keV emission model of \citet{Siegert2016_511} with (red) and without (black) correction for varying balloon altitudes (right axis, blue) as a function of time. Note that each one hour time bin consists of $\approx 4000$ bins in the pre-defined (reduced) COSI $\{\phi \psi \chi\}$-data-space. Day and night times are indicated with bright and dark shading, respectively (Earth latitude/longitude dependent). The strong altitude drops mainly occur during night times, i.e. when the Galactic centre is in COSI's field of view. The total loss due to low altitudes is about 38\,\% and only affects the emission from the bulge. The expected disk emission (smaller bumps during day times) benefits from slightly higher altitudes than nominal during day times ($\approx 5\,\%$).}
	\label{fig:response_expectation}
\end{figure*}

The altitude changes are mainly connected with day and night cycles.
The strongest signal at 511\,keV is expected to come from the Galactic bulge region.
However, the bulge is mainly exposed at night, i.e. when the balloon's altitude drops;
hence, the total exposure (in units of $\mrm{cm^2\,s}$) is not uniform in COSI's field of view throughout the flight.
Fig.\,\ref{fig:exposure} shows the total exposure of the chosen data set for a constant altitude (left), and corrected for the true motion (right).
Especially at the Galactic centre, the exposure is decreased by about 40\,\%.
But since this is the region in which most of the signal is expected, and because the more-exposed regions in which no signal is expected provide a good basis for background estimates, all $603$ hours of observation after the third detector failure are kept (cf. Sec.\,\ref{sec:data_selection}).
In Fig.\,\ref{fig:response_expectation}, we show the expected number of photons from the empirical model for the 511\,keV emission as found by \citet{Siegert2016_511} (see also Sec.\,\ref{sec:fullsky}), for a constant altitude (black) and the altitude-corrected response.
Clearly, during night times (dark shaded areas), the altitude (blue) frequently drops, for which the effective area is reduced.
These times are expected to contribute most to the measured sky counts.

We want to note that assessing the quality of the response creation through simulations is difficult to benchmark under laboratory conditions for imaging diffuse emission on top of a large and varying background.
Nevertheless, \citet{Sleator2019_COSI_DEE} performed a comprehensive study of detector effects that influence how individual event messages are recorded, and the resulting spectral response and angular resolution of COSI.
Among other effects, this included charge sharing between adjacent strips, charge loss, crosstalk between electronic channels, and accurate threshold settings regarding timing and energy for veto systems.
These effects are taken into account in the imaging response creation.

\subsection{Background modelling}\label{sec:bg_model}

The instrumental background in soft \gr telescopes is the dominant contributor to the measured count rate.
A rough prediction of the expected background spectrum as well as its intensity can be made from expensive simulations using the full mass model of both the payload and the mount, and the complete environmental conditions.

In space, a large portion of the instrumental background comes from the interaction of primary cosmic-ray and solar particles with the satellite and instrument materials \citep[e.g.][]{Gehrels1985_balloonBG,Boggs2002_SPIBG,Jean2003_SPIBG,Cumani2019_MeVBG}.
Secondary, then lower-energy, particles ($\sim \mrm{MeV}$) lead either to nuclear excitations followed by de-excitation through the emission of {\grs}, or other nuclear reactions, building short- and long-lived radioactive nuclei, which then also emit \grs after having decayed (activation, radioactive build-up).
This constitutes a family of prompt and delayed \gr line emission.
A dominant instrumental continuum background is created by $\beta$-particles depositing their energy inside detectors, or electromagnetic cascades induced by high-energy cosmic-rays.
Because the general cosmic-ray flux at Earth and consequently the instrumental background rate depends strongly on the 11-year solar cycle, and furthermore on the unpredictable occurrence of solar flares, a physical background model, applicable for each time and position, is not feasible \citep{Diehl2018_BGRDB,Siegert2019_SPIBG}.

The problem of instrumental background is complicated even more in a typical balloon environment:
Even though the atmosphere becomes more transparent at higher altitudes and for larger energies, cosmic-ray particles interact with the atmosphere and create a strong soft $\gamma$-ray continuum, i.e. the atmosphere is shining in hard X-rays and \grs \citep[Earth albedo;][]{Cumani2019_MeVBG}.
Furthermore, the bremsstrahlung from secondary electrons, for example, depends on the exact altitude of the payload (density of air, passed airmass, zenith/azimuth) and the position on Earth (geomagnetic cutoff).
In both space and the atmosphere, primary emission photons, such as from the Galactic plane or extragalactic background light, will lead to downscattered \gr photons that might be seen as instrumental background.

Attempts to model the expected background behaviour, especially at 511\,keV photon energies, typically lead to a robust order of magnitude estimate.
Proposing an absolute number of background counts for each observation, even given all appropriate environmental conditions and instrument-specific properties, would still require an uncertainty attached to these model predictions.
These are difficult to determine.
The description of the low-energy background ($\lesssim 10$\,MeV) at balloon altitudes, and in particular for the 511\,keV line, by \citet{Ling1975_MeVBG} is nevertheless useful to perform simulations and assess the adequacy of background modelling and parameter inference.
This model was later also shown to provide a good description of one of the first measurements of atmospheric 511\,keV \grs with a balloon-borne Ge(Li) spectrometer \citep{Ling1977_511keV_atmosphericBG}.
Alternatively, calculations for the atmospheric cosmic-ray spectrum and the resulting electromagnetic emission as a function of longitude, latitude, and altitude are available from \citet{Sato2016_EXPACS}.
These predictions are based on least-square fits to smooth analytical functions to describe the cosmic-ray spectra, and might not capture the required flexibility of the actual measurement.
Assessing the suitability of absolute background models in a changing balloon environment is beyond the scope of this paper.

For these reasons, we build an empirical, three-component, background model, that is parametrised in variability and amplitude, in order to be fitted simultaneously with a model or along iterative image deconvolutions to describe the celestial emission.
The expected number of background photons in the COSI data space $\left\{\phi \psi \chi t\right\}$ is consequently modelled as
\begin{align}
	m_{\phi \psi \chi t}^{\mrm{BG}} & = R_{\phi \psi \chi}^{\mrm{BG}} \cdot T_t(\mathbf{\theta}_{b}) = \tag{4} \\
	& = R_{\phi \psi \chi}^{\mrm{BG}} \cdot \sum_{b \leftarrow (b_i,b_f) \in \mathscr{B}} \beta_b \cdot \Theta(t-b_i) \cdot T_t \cdot \Theta(b_f-t) = \nonumber\\
	& = R_{\phi \psi \chi}^{\mrm{BG}} \cdot \sum_{b \leftarrow (b_i,b_f) \in \mathscr{B}} \beta_b \cdot \mathscr{R}(t,b_i,b_f) \cdot T_t \mrm{,}\nonumber
	\label{eq:bg_model}
\end{align}

\noindent where $R_{\phi \psi \chi}^{\mrm{BG}}$ is the background response (Sec.\,\ref{sec:bg_response}), $T_t$ is a tracer function (Sec.\,\ref{sec:tracers}) which provides a first-order background variability estimate, and $\mathscr{B}$ is a set of time nodes (Sec.\,\ref{sec:rescaling}) which sub-divides the tracer function into a pre-defined number of subsets with amplitudes, $\beta_b$, for each time interval between two time nodes $b_i$ and $b_f$ with $b_i < b_f$.
Those are then fitted simultaneously with the sky model amplitude $\alpha_s$ (cf. Eq.\,(\ref{eq:convolved_sky_models})).
Cutting the tracer function into smaller portions allows for a second-order correction to the background variability, as it may take uncaptured variations into account.
The set of background parameters is $\theta_b = \left\{\beta_b,\mathscr{B}\right\}$, and $\Theta$ is the heaviside function, such that $\Theta(t-b_i) \cdot \Theta(b_f-t) = \mathscr{R}(t,b_i,b_f)$ is the rectangle (boxcar) function that returns $1$ for $b_i \leq t \leq b_f$ and $0$ otherwise.
This then defines a linear combination of $|\mathscr{B}|$ background models, with a fixed relative variation between each starting and ending time node, and zero otherwise.
The covariance between these individual blocks is naturally low, and mainly determined by the contribution of the sky emission in each block.

\subsubsection{Finding a good background model}\label{sec:chosen_bg}

We evaluate the performance of different background response, tracer, and time-node combinations by performing the above-described maximum likelihood fits for all cases.
We choose several combinations among a large number of possibilities which appear most plausible as background response (Sec.\,\ref{sec:bg_response}), background tracer (Sec.\,\ref{sec:tracers}), and background re-scaling time nodes (Sec.\,\ref{sec:rescaling}) to explore our background model.
Even though the background dominates the signal in any case, we require an optimisation of the model accounting for both background and sky components.
For this, we include best-fit 511\,keV sky model by \citet{Siegert2016_511}, and allow the sky amplitude to change.
The choice of this model compared to other models, for example the full-sky model by \citet{Skinner2014_511} or a simple 2D-Gaussian to only represent the bulge, has no influence on the derived background model parameters.
This is reasonable since the instrumental background is anyway dominating the total signal and any first-order image proposition is re-scaled to the actual number of counts in the chosen COSI data set by our fitting approach.

Since the likelihood naturally increases by introducing more parameters (`fits better'), we make use of the Akaike Information Criterion \citep[AIC;][]{Akaike1974_AIC,Burnham2004_AICBIC,Burnham2004_ModelSelectionBook} which penalises `over-fits' by taking into account the number of fitted parameters, $n_{par}$, such that

\begin{align}
	\mrm{AIC} = 2 n_{par} - 2 \mathscr{L}( \hat{\theta}_s,\hat{\theta}_b )\mrm{,}\tag{5}
\end{align}

\noindent where $\mathscr{L}( \hat{\theta}_s,\hat{\theta}_b )$ is the likelihood of Eq.\,(\ref{eq:Poisson_likelihood}), evaluated at the best-fit parameters, $\hat{\theta}_s$ and $\hat{\theta}_b$, for sky and background model, respectively.

 In general, the lower the AIC, the `better' the model.
 We note that the AIC is not an absolute `goodness-of-fit' criterion, but allows for a restricted set of tested models to identify the most probable \citep{Burnham2004_ModelSelectionBook}.
 Since the data set is very sparsely populated, any use of an approximate $\chi^2$ goodness-of-fit measure will be flawed.
 Instead, we will use posterior predictive checks \citep[PPCs;][]{Guttman1967_PPC,Rubin1981_PPC,Rubin1984_PPC,Gelman1996_PPC} to evaluate the adequacy of our fits (see Sec.\,\ref{sec:results} for further details).
 In the following, we describe different parts of our background model setup in more detail.

\subsubsection{Background response}\label{sec:bg_response}

The background response, $R_{\phi \psi \chi}^{\mrm{BG}}$, is not uniquely defined.
In general, it provides an expected number of counts in the $\left\{\phi \psi \chi\right\}$ data space, which should be independent of time.
This does not mean that the amplitude of the background is constant in time, but the appearance in the COSI data space is\footnote{Note that it will also have a dependence on energy. Since we are only taking 511\,keV photons into account, we omit the dimension of energy.}.
An exhaustive simulation using the complete mass model could potentially provide a first-order background response, however the true environment, conditions, and circumstances will alter this expected behaviour.
As these parameters are constantly changing, determining an absolute background response for each instance in time through simulations is infeasible.
For these reasons, we infer a background response empirically from the data:

Order-of-magnitude simulations show that the expected instrumental background compared to the 511\,keV sky signal is about a factor of 100.
Thus, integrating the measured count rate over long times, i.e. different aspect angles and altitudes, will smear out any contribution of the sky from which a background response can be created.
Any background-dominated measurement can thus be used to define a response empirically via

\begin{align}
	R_{\phi \psi \chi}^{\mrm{BG}} = \sum_t \sum_{e \in \mathscr{E}} d_{\phi \psi \chi t e}\mrm{.}\tag{6}
	\label{eq:bg_response}
\end{align}

\noindent In Eq.\,(\ref{eq:bg_response}), $d_{\phi \psi \chi t e}$ describes the data, i.e. photons with their identifiers $\phi$, $\psi$, $\chi$ in the instrument-specific data space, the time (of arrival) $t$, as well as the photons' reconstructed energy $e$.
The sum over all times and a selected energy interval, $\mathscr{E}$, sorts each measured photon in the appropriate $\left\{\phi \psi \chi\right\}$-bin.
We normalise any such-constructed background response to $1.0$.

The energy interval $\mathscr{E}$ has to be chosen such that 1) there is enough statistics available for the background response to predict the relative number of counts in each $\left\{\phi \psi \chi\right\}$-bin, 2) that the correct processes in the instrument that lead to the 511\,keV are presented, and 3) that possible contaminations of sky emission are either completely smeared out or masked.
We construct a total of eight background responses from different energy bands, listed in Tab.\,\ref{tab:response_energy_bands}, to determine the best representation of our data, which always includes a possible sky contribution.

\begin{table}
	\centering
	\begin{tabular}{cc}
		\hline
		Energy band $\mrm{[keV]}$ & Comments \\
		\hline
		$[506,516]$ &  Line only$^*$ \\
		$[460,560]$ &  Line + continuum\\
		$[460,500]$ &  Adjacent low-energy\\
		$[520,560]$ &  Adjacent high-energy\\
		$[375,500]$ &  Adjacent low-energy, broad\\
		$[520,645]$ &  Adjacent high-energy, broad\\
		$[460,500]$ \& $[520,560]$ &  Adjacent continuum, no line\\
		$[375,500]$ \& $[520,645]$ &  Adjacent continuum, broad, no line\\
		\hline
	\end{tabular}
	\caption{Energy bands for background response creation. $^*$\,Best-fit background response that is used throughout this work.}
	\label{tab:response_energy_bands}
\end{table}

This approach is similar to the empirical background modelling by \citet{Siegert2019_SPIBG} for the SPI telescope, but in the instrument-specific data space of COSI, $\left\{\phi \psi \chi\right\}$, instead of SPI's 19 Ge detectors shadowed by a coded mask.
We note that this approach of defining a background response can be refined even further by separating different (physical) processes inside the instrument, for example distinguishing between the 511\,keV line and its underlying continuum, or also for smaller energy bin sizes.
Such an elaboration, however, requires a lot of statistics in the individually-defined data space bins and might be unreliable for the current COSI data set.
A running average across energies or using general linearised models might be used in future background response generations for fine spectroscopy.

While there is also a dependence on the other background parameters, such as the chosen tracer or the additional background time nodes (see next sections), using the energy band $[506,516]$\,keV for creating the background response provides the best fits compared to all other cases (see Appendix Fig.\,\ref{fig:allBGresponses_AIC_performace}).

\subsubsection{Variability tracer}\label{sec:tracers}

The intrinsic variability of the above-mentioned processes that lead to instrumental background radiation cannot be predicted from physically-motivated models.
For this reason, tracers of this variability are determined.
These may be any function in time that could be related to the background-generating processes, for example on-board or external monitors measuring the cosmic-ray flux, a voltage-meter, the CsI veto-shield count rate, or the balloon altitude.
As a further step, these functions may be orthogonalised and combined with different weightings to capture additional variability \citep[e.g.][]{spiorthomodel}.
Alternatively, one `best-performing' tracer function may be cut further as depending on time or, for example, based on the intrinsic variability of the measured count rate (see Sec.\,\ref{sec:rescaling}).

In this study, we use three background tracer function which are supposedly closely related to the measured Compton event rate at 511\,keV:
the CsI shield rate ($\mrm{SR}$), the (inverse of) the balloon altitude ($h^{-1}$), and the photoabsorption event count rate in the analysed band between 506 and 516\,keV ($\mathrm{PE}$).

The shield rate provides a well-sampled, i.e. high statistics, general trend of any possible process that might lead to background emission.
The shield is sensitive to energies $\gtrsim 80$\,keV \citep{Kierans2018_PhD,Sleator2019_PhD}, but with no energy information, it also counts a large number of events which are unrelated to the specific 511\,keV range.
The $\gamma$-ray background is higher at lower altitudes, and particularly for 511\,keV, lower altitudes result in more cosmic-ray particle showers which include $\beta$-particles and secondary decay positrons.
Therefore, the inverse of the altitude may be an appropriate tracer for 511\,keV.
While the 511\,keV PEs also include photons from the expected sky emission, the total contribution to the count rate is less than 0.1\,\%.
Thus, as the 511\,keV PE rate is about ten times larger than the 511\,keV CE rate, these single site interactions might provide a sufficient tracer of the multiple-site events.

For a zero-order estimate of the predictability of any tracer, we calculate the Pearson correlation coefficient, $\rho(\mathrm{CE,X})$, between the measured 511\,keV Compton events per hour and any tracer ($\mathrm{X}$).
The strongest correlation is found between CE and PE with $\rho(\mathrm{CE,PE}) = 0.958$, followed by $\rho(\mathrm{CE,SR}) = 0.948$ for the CsI shield rate, and $\rho(\mathrm{CE,h^{-1}}) = 0.862$.
While all chosen tracers strongly correlate with the CE count rate, this still should be taken as only an indication for a possible tracer, because there are also photons from the sky included in the CEs (and PEs) which might also be correlated with these functions.

The number of fitted background parameters, and how they are set (time nodes) in addition to a contribution from the sky, influences the fit adequacy. Nevertheless, the PE tracer performs on average better than the other two (see also Fig.\,\ref{fig:allBGs_AIC_performace}).

\subsubsection{Amplitude renormalisation}\label{sec:rescaling}

A function which could predict only the background variability and which is orthogonal to any celestial emission would require only one parameter $\beta_b$ in Eq.\,(\ref{eq:bg_model}), i.e. one set of time nodes before the first and after the last time bin of our observations.
We can, however, not be sure that any of the tracer functions behaves in this desired way in the first place, for which reason we define three different possibilities to set varying time nodes to re-scale the background tracers.

Such a re-scaling is generally useful when the full sky is included in the data set and the true emission is unknown.
Because COSI has a $\sim 1\pi$ field of view and is not performing targeted observations, i.e. the exposure changes smoothly with time, any point source location will not be visible at all times.
This means, any non-perfect background model tracer which intends to describe the pointing-to-pointing variation (or here hour-to-hour variation), will over-predict the number of background counts whenever the source is not in the field of view.
For this reason, it might be useful to set time nodes for the background to re-scale (introduce another background parameter) whenever the source is in the field of view.
However, when either the position of the source is not known, or the emission is of general diffuse nature with unknown extents, a more general approach to set these time nodes has to be chosen.

\begin{figure}[ht!]
	\centering
	\includegraphics[trim=0.5in 0.2in 1.2in 0.2in, clip=True, width=1.0\columnwidth]{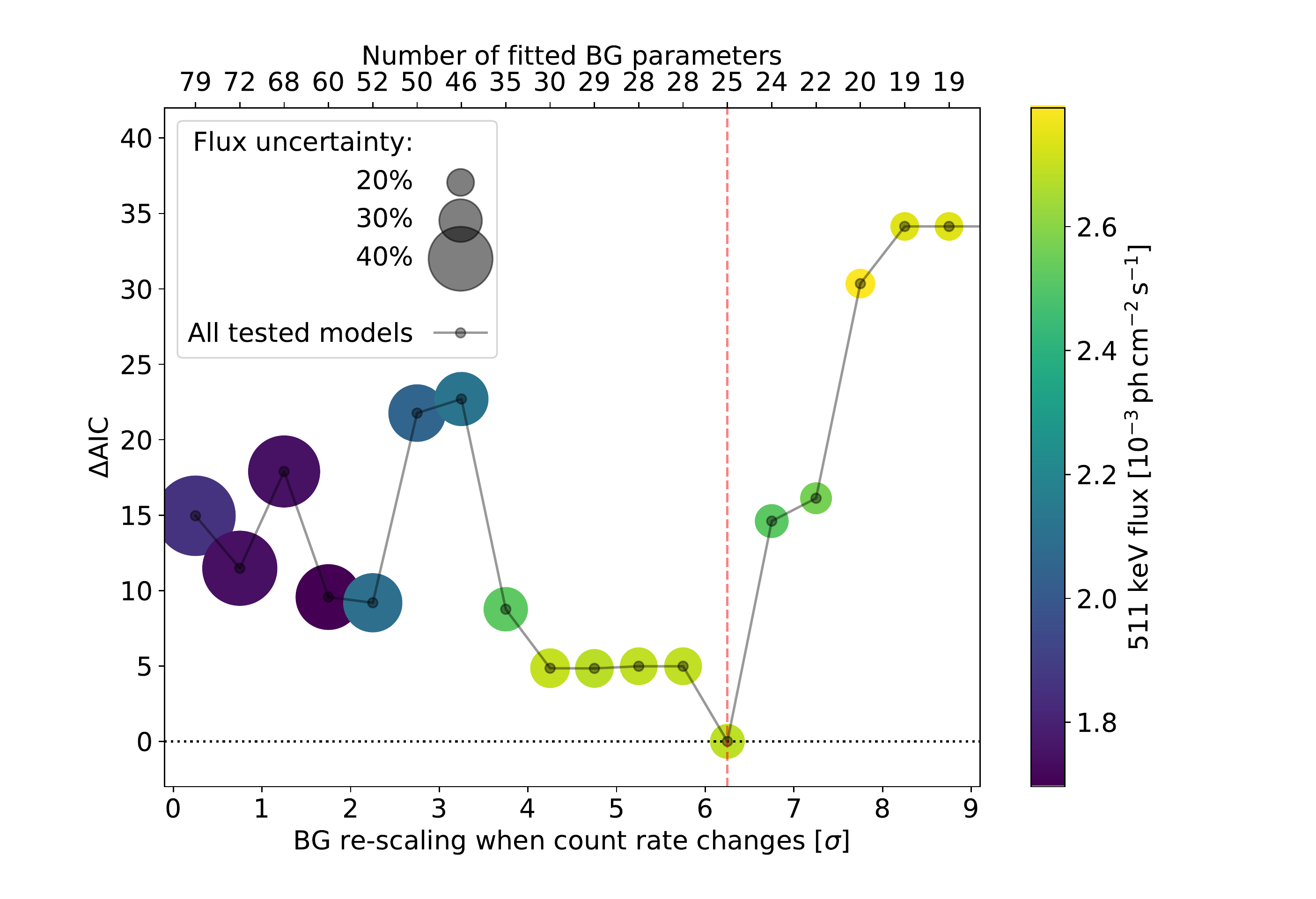}
	\caption{Performance of the background model combination: PE tracer, $506$--$516$\,keV BG response, BG amplitude re-normalisation using Bayesian block. The minimum AIC indicates when the optimal number of BG parameters (top axis) is reached, avoiding at the same time `bad fits' (too few parameters) and `over-fitting' (too many parameters). For a threshold of $\approx 6\sigma$ in the change of the measured 511\,keV count rate (cf. Fig.\,\ref{fig:countrate_tracers}), the optimum is found by using 25 BG parameters (red dashed line). The fitted sky model fluxes are colour-coded with their estimated uncertainties shown by the size of the symbols. See text for more details.}
	\label{fig:PEtracer_AIC_performance}
\end{figure}

As it can be assumed that the background changes are not traced completely from the function $T_t$ alone, a natural choice comes from the time dimension.
We divide the 603 observation hours in equidistant time intervals, ranging between 1 time interval (i.e. $\theta_b = \left\{\beta_1, \{0, 603\} \right\}$, thus 1 background parameter) to 603 time intervals (i.e. $\theta_b = \left\{\beta_1, \{0, 1\}, \beta_2, \{1, 2\}, \dots, \beta_{603}, \{602, 603\} \right\}$, thus 603 background parameters).
This defines 48 different cases.

The altitude can serve as a second-order predictor for when the background should be re-scaled.
As the altitude changes between 22 and 34\,km, we define time nodes whenever the balloon crosses a certain mark, here in unit steps of 1\,km.
This defines 12 different cases with a number of background parameters between 2 and 48, now set at times according to the altitude crossings.
This means even though the number of background parameters in the time interval case and in the altitude case can be equal, the resulting likelihood might be different.

\begin{figure*}[ht!]
	\centering
	\includegraphics[trim=0.7in 0.3in 1.3in 1.0in, clip=True, width=0.8\textwidth]{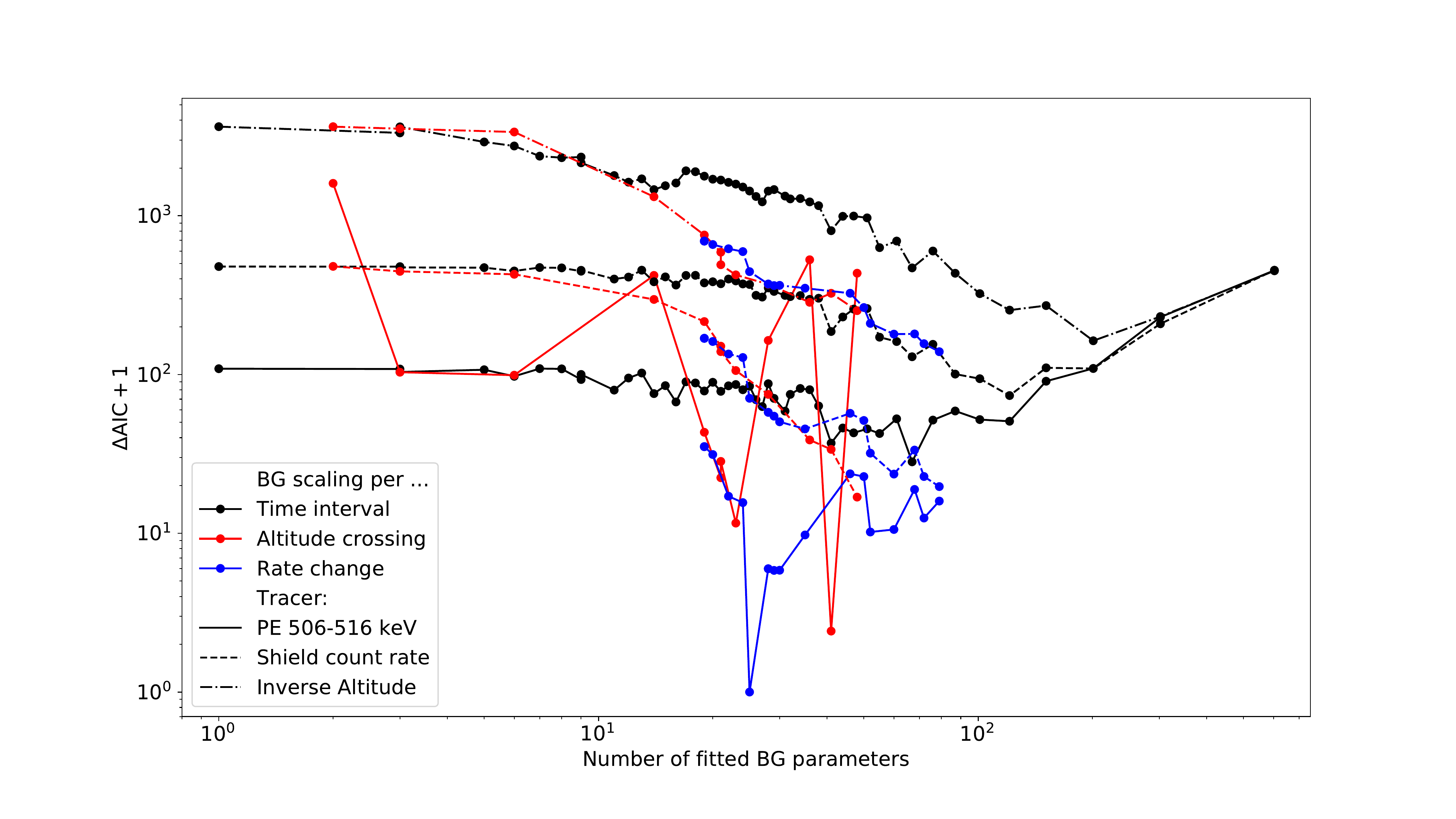}
	\caption{Performance of all background model combinations using the $506$--$516$\,keV BG response. Clearly, choosing to re-scale the BG amplitude at time nodes which correspond to strong changes in the count rate (Bayesian blocks, blue) performs best. The background is also adequately determined by using changes with altitude (red), however requiring about twice the number of parameters. Equidistant time intervals (black) show a smoother behaviour, but in general perform worse. The trend among different tracers is clear, with the 511\,keV PE performing best (solid lines), followed by the shield rate (dashed lines), and the altitude (dash-dotted lines). A summary for different choices of the BG response is given in the Appendix (Fig.\,\ref{fig:allBGresponses_AIC_performace}).}
	\label{fig:allBGs_AIC_performace}
\end{figure*}

As a third alternative to when to set additional time nodes, we use Scargle's Bayesian blocks \citep{Scargle1998_BayesianBlocks,Scargle2012_BayesianBlocks}.
This methods determines `change points' of a count rate according to a false alarm probability threshold.
We define $20$ different thresholds, $\tau$, for the Bayesian block algorithm, according to a survival probability, $S(\tau) = 1 - 2\cdot \int_{0}^{\tau} dx \mathscr{N}_{x}(0,1)$, of the standard normal distribution, $\mathscr{N}_{x}(0,1)$, between $\tau = 0.25\sigma$ ($S(0.25) \approx 80\,\%$) and $\tau = 9.75\sigma$ ($S(9.75) \approx 1.8 \times 10^{-22}$) in $\Delta \tau = 0.5\sigma$ steps.
Since different thresholds can lead to the same change points, this defines 16 unique cases.

In Fig.\,\ref{fig:PEtracer_AIC_performance}, we show the performance of the best-fitting background model combination:
506--516\,keV background response, PE tracer, Bayesian block re-scaling time nodes.
Clearly, the more background parameters (top axis, right to left) are included in the fit, the better the resulting likelihood (AIC).
After including more than $25$ background parameters ($\tau = 6.25\sigma$), however, the large number of fitted parameters is penalised by the AIC.
We note that for smaller thresholds (and in general for larger number of parameters), the AIC is not a smooth function, and also the resulting flux (colour-coded in Fig.\,\ref{fig:PEtracer_AIC_performance}) is not directly related to $n_{par}$.
The uncertainties on the flux naturally increase with the number of fitted parameters.

A summary of all background model combinations using the 506--516\,keV response is shown in Fig.\,\ref{fig:allBGs_AIC_performace}.
The PE tracer (solid lines) performs best, independent of the chosen background re-scaling time nodes.
This is reassuring that our methodology is consistent.
Depending on the time nodes set, the AIC minimum is found in the range between $25$ and $64$ background parameters.
For other tracers, the minima move to a larger number of parameters.
This evaluation has been performed with the full-sky 511\,keV model from \citet{Siegert2016_511}.
We again note that different sky models in this procedure, for example using only a 2D-Gaussian component to represent the bulge, alter the absolute likelihood values.
Nevertheless, the number of background parameters that are required in this data set are consistently found between $25$ and $64$.
This appears reasonable since the celestial contribution is always small and the data set is dominated by instrumental background.
From using different background model combinations (cf. Fig.\,\ref{fig:allBGs_AIC_performace} and Appendix Fig.\,\ref{fig:allBGresponses_AIC_performace}), then with also more parameters, we estimate a systematic uncertainty in our derived flux values of 30\,\%.
We note that this is not, and can never be, a full exploration of all\footnote{For the 603 time bins in our data set, the number of all possible, independent, background time nodes is $\approx 10^{181.22}$. The number of possible background tracers is infinite, and hence describes an open set, for which no `absolute best fit' can be found.} possibilities to set time nodes and to choose among tracers.
We instead chose among a plausible set of combinations and investigated which produces the most probable outcome, always including a first-order sky model.

\section{Imaging}\label{sec:results}

In Secs.\,\ref{sec:model_fits} and \ref{sec:imaging_response}, we introduced the imaging response in general, and as applied to our specific data set.
For a fixed set of observations, as used here for the 603 hours of 511\,keV measurements, the sum $\sum_{p_t \in t} R_{\phi \psi \chi}^{\mrm{SKY}}(Z,A,h) \cdot p_t$ from Eq.\,(\ref{eq:convolved_sky_models}) can be isolated and work as a data-set-specific response, $R_{\phi \psi \chi t}^{\mrm{SKY}}(Z,A,h)$.
This response then carries entries for the $4243$ non-zero bins in the COSI data space $\{ \phi \psi \chi\}$, times $603$ entries in the time domain, times the chosen zenith/azimuth-binning, here $36 \times 72 = 2592$.
The response thus requires at least an allocation of 53\,GB memory alone.

We use this response to 1) perform an image reconstruction using a modified version of the Richardson-Lucy algorithm (Sec.\,\ref{sec:RL_deconvolution}), as well as 2) calculate a set of empirical functions to describe the 511\,keV sky as seen with COSI (Sec.\,\ref{sec:fitting_results}).

\subsection{Richardson-Lucy deconvolution}\label{sec:RL_deconvolution}

\begin{figure*}[ht!]
	\centering
	\includegraphics[trim=0.5in 0.2in 1.2in 0.5in, clip=True, width=0.75\textwidth]{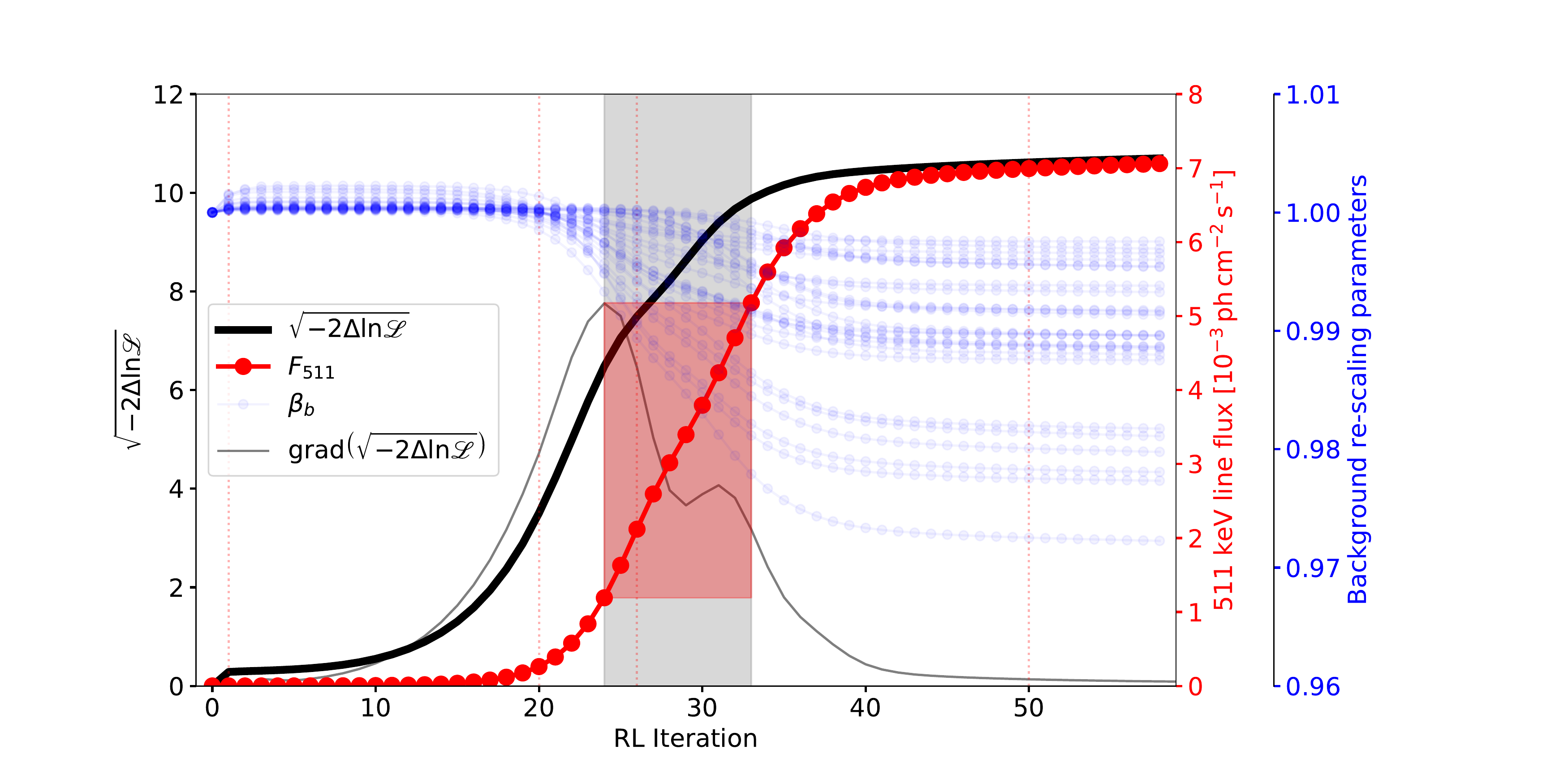}
	\caption{Properties of the modified Richardson-Lucy algorithm for COSI 511\,keV data as varying with iteration. The black curve shows the test statistics of the current image proposal vs. a background-only description of the data. The first sharp step is due to the large acceleration parameter found for the first iteration. Without the acceleration parameter, the first sharp step would typically take several tens to hundreds of iterations. The total map-integrated 511\,keV flux is shown in red (first right axis), and the background parameters for each step in blue (second right axis). The gradient of the used test statistics (gray, arbitrary units) can be used to define a region of iterations that adequately describe the 511\,keV data, defined by the first inflection point towards positive curvature and largest positive curvature before converging to the noise-dominated maximum likelihood solution. Iteration $26$ is the one that represents the first maximum positive curvature. See text for details.}
	\label{fig:RL_params}
\end{figure*}

In order to investigate the celestial contributions of the current data set without relying on a priori assumptions, we perform an iterative image reconstruction using the Richardson-Lucy deconvolution technique \citep{Richardson1972_RichardsonLucy,Lucy1974_RichardsonLucy}.
This algorithm has been successfully used in MeV \gr{} astrophysics \citep{Knoedlseder1996_COMPTELimaging,Knoedlseder1999_26AlCOMPTEL,Knoedlseder2005_511}, and can provide a less-biased picture of the underlying morphology.
It might further reveal structures, shapes, and regions which might not be tested by a pure empirical model-fitting approach.
We note, however, that this method cannot replace physical modelling of the 511\,keV, and individual features should not be over-interpreted.
In particular, we expect on the order of $10^3$ celestial 511\,keV photons \citep[cf.][]{Kierans2019_511COSI}, which would be distributed over the number of pixels (here: $2,592$ $5^{\circ} \times 5^{\circ}$ pixels, i.e. degrees of freedom).
With an expected significance of about $7\sigma$ from COSI data \citep{Kierans2019_511COSI}, only about $16$ (\textit{sic!}) significant ($3\sigma$) pixels would be present.

The general algorithm has been proven to converge to the maximum likelihood solution of the problem \citep{Shepp1982_RL}, which however tends to find noise peaks in the background-dominated data of MeV instruments \citep[cf.][]{Knoedlseder1999_26AlCOMPTEL}.
The basic version the Richardson-Lucy algorithm is described by the iterative update of an initial image, typically set to an isotropic low flux map, by forward and backward application of the response, such that

\begin{align}
M^{k+1}_j = M^k_j + \delta M^k_j = M^k_j + M^k_j \left( \frac{\sum_{i} \left( \frac{d_{i}}{\epsilon_{i}^k} - 1 \right) R_{ij}}{\sum_{i} R_{ij}} \right)\mrm{.} \tag{7}
\label{eq:RL_deconv_v1}
\end{align}

\noindent In Eq.\,(\ref{eq:RL_deconv_v1}), $M^k_j$ is the k-th image (`map', with image space indexed by $j$) proposal, and iteratively updated by $\delta M^k_j$, in which the observation specific response, $R_{ij}$ (with data space indexed by $i = \{\phi \psi \chi t\}$), is applied to an expectation, $\epsilon^k_i = \sum_j R_{ij} M^k_j + \epsilon_i^{\mrm{BG}}$, given the data set $d_i$.
The expected number of background counts is $\epsilon_i^{\mrm{BG}}$.
The application of the imaging response from image space $j$ into the data space $i$ would be forward folding (how does the instrument see an image), whereas the application from data space into image space would be equivalent to a backward projection of (all) data space counts onto the sphere of the sky.
The latter also includes the background photons (whose absolute portions are fixed in the standard algorithm, Eq.\,(\ref{eq:RL_deconv_v1})), so that a single back-projection would merely show the instrument itself.
For this reason, the total expectation in the data space has to be updated in several iterations.
We note that a back-projection of residual counts, for example from model fitting (Sec.\,\ref{sec:model_fits}), might identify hot spots in the image dimension which are not captured by the used sky models.

Since the standard Richardson-Lucy algorithm, Eq.\,(\ref{eq:RL_deconv_v1}), typically uses a fixed background model, and because the delta-image is typically updating only in marginal steps which makes the algorithm very slow \citep[i.e. low flux differences in specific regions; cf.][]{Kaufman1987_RL,Lucy1992_RL}, we modify the standard algorithm to also take into account the uncertain background.

\subsubsection{Description of algorithm used}\label{sec:arldobags}

It was shown in the case of MeV \gr{} imaging \citep{Knoedlseder1999_26AlCOMPTEL}, that the standard Richardson-Lucy algorithm can be accelerated by applying a multiplicative factor, $\lambda^k$, to the delta image in each iteration, which will be determined by a maximum likelihood fit (cf. Sec.\,\ref{sec:model_fits}).
It must be guaranteed that each image pixel $j$ of the (k+1)-th iteration is still positive, for which reason $\lambda^k$ is constrained to $\lambda^k > -M^k_j / \delta M^k_j$.
Note that the delta image can and must contain negative pixels.

Weakly exposed regions in the data sets of Poisson count limited experiments, such as COSI, are prone to artefacts as individual fluctuations in the very sparsely populated data space can lead to unnatural high expectations $\epsilon_i$.
For this reason, we apply the noise damping approach from \citet{Knoedlseder2005_511}, and introduce a factor $w_j = \sqrt{\sum_i R_{ij}}$ for weighting the delta image, and apply a $2.5^{\circ}$ Gaussian filter to reduce the effective number of degrees of freedom in the image reconstruction.

\begin{figure}[ht!]
	\centering
	\includegraphics[trim=1.0in 2.0in 1.0in 1.0in, clip=True, width=1.0\columnwidth]{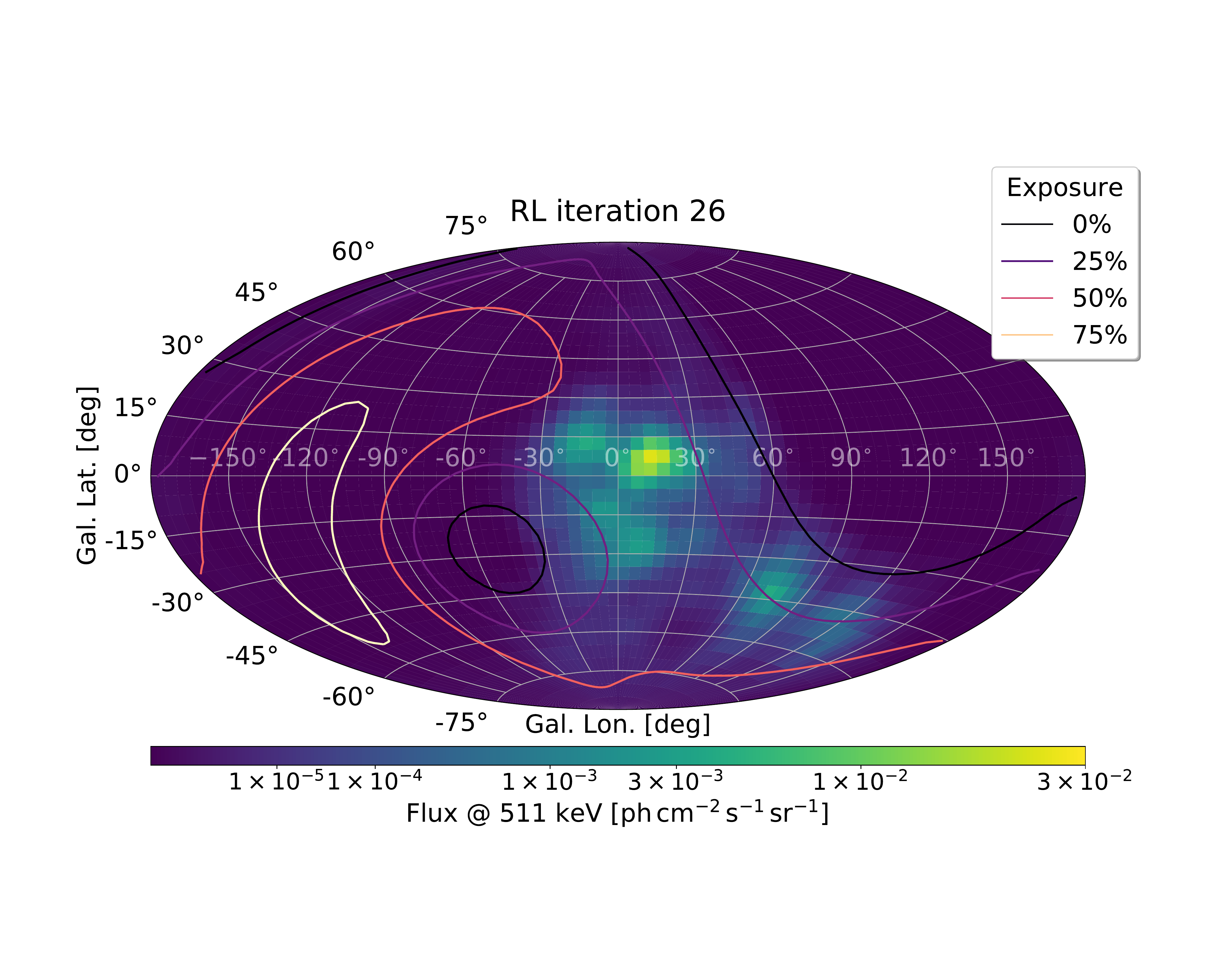}
	\caption{Iteration $26$ of our modified version of the Richardson-Lucy deconvolution algorithm, Eq.\,(\ref{eq:RL_deconv_v2}), together with the exposure regions, including $0\,\%$ (black contours), $25\,\%$ (purple), $50\,\%$ (red), and more than $75\,\%$ (white) with respect to the maximum exposure. Iteration $26$ represents the case at which the likelihood ratio function (with respect to a background-only fit) shows the largest positive curvature (cf. Fig.\,\ref{fig:RL_params}), typically chosen as the best trade-off between granularity of the map and likelihood, with a total integrated 511\,keV flux of $2.1 \times 10^{-3}\,\mrm{ph\,cm^{-2}\,s^{-1}}$. See text for further discussion.}
	\label{fig:RL_maps}
\end{figure}

A third modification to the Richardson-Lucy algorithm is provided in this work by allowing the background to vary between iterations:
A fixed background model expectation, $\epsilon_i^{\mrm{BG}}$, for example from an acceptable maximum likelihood fit using a first-order sky model, will result in a reconstruction that strongly depends on this first image and the resulting total number of background photons in each data space bin.
Consequently, the reconstruction will be a distortion of the best-fit maximum likelihood solution image, and introduces some granularity, but which may just be `filling the residuals' with sky emission.
Such an approach is naturally flawed because only specific data space bins may be re-populated due to the forward application of the response, as the background is fixed.
This is equivalent to subtracting a background model, and neglecting to consider that this model also carries its own uncertainties.
In our modified algorithm, we re-determine the $25$ background re-scaling parameters, $\beta_b^k$, in each iteration, together with the acceleration parameter $\lambda^k$, so that the updated image is built from how much background is required to explain the data - and not assuming it in the first place.

Finally, our full modified Richardson-Lucy deconvolution version is written

\begin{align}
M_j^{k+1} & = M_j^k + \lambda^k \left[w_j M_j^k \left( \frac{\sum_{i} \left( \frac{d_{i}}{\epsilon_{i}^k} - 1 \right) R_{ij}}{\sum_{i} R_{ij}} \right) \right]_{2.5^{\circ}} \tag{8} \\
\label{eq:RL_deconv_v2}
\mrm{with} &  \nonumber\\
\epsilon^k_i & = \sum_j R_{ij} M^k_j + \sum_{b \in \hat{\mathscr{B}}} \beta_b^k  \hat{R}_{i}^{\mrm{BG}} \mrm{,} \tag{9}
\end{align}

\noindent where $\hat{R}_{i}^{\mrm{BG}}$ is the best-fit background model response from Sec.\,\ref{sec:bg_response}, together with set $\hat{\mathscr{B}}$, containing the $25$ required time intervals to guarantee an adequate fit.

\subsubsection{Images}\label{sec:RLimages}

The general problem with any such iterative procedure is to find when to stop the algorithm, or determine which image to pick as best representing the data.
In fact, there are no definite answers to these questions, as also each solution is in itself uncertain and just represents one realisation of the set of parameters.
We use the gradient of the shape of the test statistics, $\sqrt{-2\Delta\ln\mathscr{L}}$, between the current image proposal and a background-only description (iteration $0$) to identify plausible iterations that describe the COSI 511\,keV data adequately (see Fig.\,\ref{fig:RL_params}).
In the case of priors that set the correlations lengths of the pixels, for example, to regularise the frequency of noise in the Poisson count dominated data, \citet{Allain2006_gammaimaging} used a trade-off between the likelihood and the prior to extract an adequate solution (`\textit{L-curve}').
Our regularisation is approximately given by the Gaussian smoothing kernel and thus constant.
This means the inflection points of the likelihood function alone provide a first-order criterion.
We find that iteration $24$ is the first inflection point, followed by iteration $26$ showing the largest positive curvature.
Another inflection point is found at iteration $28$, and the last largest positive curvature until convergence to the maximum likelihood (noise-dominated) solution at iteration $33$ (see Fig.\,\ref{fig:RL_params}). 

Thus, iteration $26$ provides a map with a compromise between noise and granularity.
We show iteration $26$ of the modified Richardson-Lucy algorithm in Fig.\,\ref{fig:RL_maps}.
Clearly, there is emission around the centre of the Galaxy which is also found to be uncorrelated with the exposure map (contours).
This is reassuring that the algorithm works as expected.
We note that beyond iteration $33$, the low-frequency noise takes over and can enhance individual emission features, especially in regions with $\lesssim 25\,\%$ of exposure.
Between iterations $24$ and $33$, the total 511\,keV flux varies between $1.1$ and $5.1 \times 10^{-3}\,\mrm{ph\,cm^{-2}\,s^{-1}}$, with iteration $26$ showing $2.1 \times 10^{-3}\,\mrm{ph\,cm^{-2}\,s^{-1}}$.
The integrated flux in the central region of the reconstructed image (angular radius $\leq 40^{\circ}$) is $1.9 \times 10^{-3}\,\mrm{ph\,cm^{-2}\,s^{-1}}$.
All these values are consistent with previous measurements, considering the full sky.

We want to remind that individual emission features should not be over-interpreted, in particular because the significance of the full-sky emission in this data set is $\gtrsim 7\sigma$ (cf. Fig.\,\ref{fig:RL_params}), distributed over hundreds\footnote{By smoothing the delta images, the effective number of degrees of freedom (data points) will be smaller than the number of used pixels.} of pixels.
For example, the apparently-bright spots around $b \approx -45^{\circ}$, $60^{\circ} \lesssim l \lesssim 120^{\circ}$, are very close to the completely unexposed regions of the sky (inside black contours), and therefore these might only be image artefacts.
This may then also be related to the reliability of the imaging response for larger zenith angles ($\gtrsim 45^{\circ}$) and the statistics in the response-generating simulation.
Additionally, a stronger than expected dependence on the altitude may lead to skewed correction factors, again especially at large zenith angles.

The high-latitude features only appear in later iterations, whereas the central component is immediately present after only a few iterations.
Likewise, the three distinct emission features around the Galactic centre are probably due to the reconstruction method itself, favouring distinct emission spots rather than correlated pixels, and it should be considered only as describing the general extent of the emission.
Nonetheless, it appears here that the emission is more extended than what was found in earlier measurements.
We investigate the emission extent in more detail in Sec.\,\ref{sec:fitting_results} using empirical emission templates to obtain uncertainties on the parameters that describe the morphology.
Finally, we note that any such reconstruction always depends on the choice of the background model.

From the image deconvolution, we find no evidence of a 511\,keV disk.
Such a feature would only be visible for negative longitudes as COSI's exposure is restricted to $l \lesssim 60^{\circ}$.
We further discuss reliability of the image reconstruction in Appendix\,\ref{sec:appendix_reliability} and provide examples of simulated data sets including different flux levels in Appendix\,\ref{sec:appendix_simulations}.

\subsection{Model fitting}\label{sec:fitting_results}

As described above, the Richardson-Lucy algorithm is prone to produce noise peaks in individual, also low-exposure regions, for later iterations.
These could be alleviated by the usage of more elaborate image reconstruction methods.
For example, the Multiresolution Regularized Expectation Maximization (MREM) method, which is based on the Richardson-Lucy algorithm, tries to damp the low-frequency noise in the delta images through wavelet thresholding \citep[see, e.g., ][for an application to COMPTEL \nuc{Al}{26} data]{Knoedlseder1999_26AlCOMPTEL}.
Alternatively, the Maximum Entropy method applies a prior in image space, measuring entropy of an image proposition by its distance to a default image, and thus counteracting the likelihood solution \citep[e.g., ][]{Knoedlseder1996_COMPTELimaging}.
Because the signal strength in our current data set is in general very low, we want to quantify the current findings by a more restrictive approach.
Using pre-defined templates that are parametrised by only a few parameters, we can provide a robust estimate of the emission parameters and furthermore compare to previous findings.

Consequently, the final equation to describe the COSI 511\,keV data is assuming already convolved sky models, Eq.\,(\ref{eq:convolved_sky_models}), marked by an asterisk ($^*$), as well as the best-fitting background response from Sec.\,\ref{sec:bg_model}, marked by a hat ($\,\hat{}\,$; still with free amplitude parameters).
Thus,

\begin{align}
	m_{\phi \psi \chi t} & = \sum_{s \in \mathscr{M}} \alpha_s \cdot m_{\phi \psi \chi t}^{\mrm{SKY,s,*}} + \nonumber\\
	& + \sum_{b \leftarrow (b_i,b_f) \in \hat{\mathscr{B}}} \beta_b \cdot \hat{R}_{\phi \psi \chi}^{\mrm{BG}} \cdot \mathscr{R}(t,b_i,b_f) \cdot \mrm{PE}_t \tag{10}
	\label{eq:model_fitting_model}
\end{align}

\noindent includes the scaling parameters $\alpha_s$ (solid-angle integrated sky flux) for each map in the set of chosen sky models to be fitted, $\mathscr{M}$, and the $25$ background model re-scaling parameters, $\beta_b$, whose associated time nodes, $(b_i,b_f) \in \hat{\mathscr{B}}$, had been calculated by the Bayesian block algorithm with a change point threshold of $\hat{\tau} = 6.25\sigma$.
For a single sky map, the total number of fitted parameters is thus $26$.

For an intuitive check of the absolute values of resulting background parameters, we normalise each background block (time span) to the number of measured counts inside this block.
This leads to a background re-scaling parameter of $\beta_b = 1.0$ if the contribution of celestial emission is zero, and should be $<1.0$ if the sky response suggests a contribution different from zero.
Thus, if an expected signal is visible throughout all exposures, the background parameters should all deviate from $1.0$ (=background-only) in the fit.

We include priors for the sky and background scaling parameters, based on our image reconstruction (Sec.\,\ref{sec:RLimages}), previous measurements with SPI and other instruments, and the expected contribution of background counts to the total signal, such that

\begin{align}
	\pi( \theta | d_{\phi \psi \chi t}) \propto \mathscr{L}(d_{\phi \psi \chi t} | \theta) \pi(\theta) \tag{11}
	\label{eq:bayes_posterior}
\end{align}

\noindent is the joint posterior distribution of all parameters.
We sample the posterior by using the No U-Turn Sampler \citep[NUTS;][]{Hoffman2011_NUTS,Hoffman2014_NUTS} built in Stan \citep{Carpenter2017_stan}.
In Eq.\,(\ref{eq:bayes_posterior}), $\mathscr{L}(d_{\phi \psi \chi t} | \theta)$ is the likelihood given in Eq.\,(\ref{eq:Poisson_likelihood}), and $\pi(\theta)$ are the prior distributions.

In previous studies, the 511\,keV line flux in the bulge region of Galaxy was consistently found to be of the order of $\sim 10^{-3}\,\mrm{ph\,cm^{-2}\,s^{-1}}$.
The full sky emission is less well-determined, and the fluxes range between $1.7$ to $3.5 \times 10^{-3}\,\mrm{ph\,cm^{-2}\,s^{-1}}$ \citep[e.g.][]{Knoedlseder2005_511,Skinner2014_511,Siegert2016_511}, with a tendency for higher fluxes with increasing exposure and INTEGRAL/SPI observations of the Milky Way disk and higher latitudes.
We note that the absolute flux values from OSSE can be considerably smaller \citep[][]{Purcell1997_511}.
We choose a prior on the 511\,keV line flux that is normalised to the flux of the convolved sky map in each case with $F_{511}^{\mathrm{bulge}} = 10^{-3}\,\mrm{ph\,cm^{-2}\,s^{-1}}$ for bulge-only maps (Sec.\,\ref{sec:2Dgaussians}), and varying for the full-sky maps ($F_{511}^{\mathrm{full\,sky}}$, Sec\,\ref{sec:fullsky}).
In this way, we can set a truncated normal prior for the sky amplitude $\alpha \sim \mathscr{N}_{\alpha>0}(\mu = 1, \sigma = 2/3)$.
The choice of the large prior width originates from the unknown systematics in the response creation of COSI, for which the absolute efficiency at 511\,keV can easily be off by several tens of per cent.
In general in this study, for the sky model flux, the prior functions as a scale to the problem, and forces positivity to the signals.
\citet{Kierans2019_511COSI} extracted a positron annihilation spectrum with COSI from the central $16^{\circ}$ around the Galactic centre, and found a 511\,keV flux of $(3.9 \pm 0.4) \times 10^{-3}\,\mrm{ph\,cm^{-2}\,s^{-1}}$.
Note that while the absolute flux of the 511\,keV bulge emission has been measured by several balloon-borne and satellite-based instruments \citep[e.g.][and references therein]{Purcell1997_511,Prantzos2011_511}, the different systematics inherent to each measurement can lead to several tens of per cent of additional margin.
Based on the Richardson-Lucy image deconvolution, we find a 511\,keV flux of $\approx 2 \times 10^{-3}\,\mrm{ph\,cm^{-2}\,s^{-1}}$ (Sec.\,\ref{sec:RLimages}).
Consequently, within $3\sigma$, the prior width provides a factor of $2$ uncertainty in the absolute measurement.

For the background, the choice of the priors is set to a truncated normal distribution, $\beta_b \sim \mathscr{N}_{\beta>0}(\mu = 1, \sigma = 0.1)$, for each block $b$, as large variations between different blocks would be unexpected (as already normalised to the count rate), and still provide enough leverage for the sky contribution to exceed previous expectations.
We note that the truncation for positive fluxes (or background contributions) does not prevent the fit to reach zero sky flux, and furthermore restricts the signal to be positive, as naturally expected from an emission process.

To address the adequacy of our fits, we use posterior predictive checks (PPCs).
The posterior predictive distribution is given by

\begin{align}
	\pi(\tilde{d}_{\phi \psi \chi t} | d_{\phi \psi \chi t}) = \int d\hat{\theta} \mathscr{L}(\tilde{d}_{\phi \psi \chi t} | \hat{\theta}, d_{\phi \psi \chi t}) \pi( \hat{\theta} | d_{\phi \psi \chi t})\mrm{,} \tag{12}
	\label{eq:PPC}
\end{align}

\noindent where $\pi( \hat{\theta} | d_{\phi \psi \chi t}) \propto \mathscr{L}(d_{\phi \psi \chi t} | \hat{\theta}) \pi(\hat{\theta})$ is the joint posterior distribution of all fitted parameters $\hat{\theta}$, and $\mathscr{L}(\tilde{d}_{\phi \psi \chi t} | \hat{\theta}, d_{\phi \psi \chi t})$ would be the predictive distribution of `replicated' (simulated) data, $\tilde{d}_{\phi \psi \chi t}$, from the inferred parameters, given the original data set \citep{Guttman1967_PPC,Rubin1981_PPC,Rubin1984_PPC,Gelman1996_PPC}.
In this way, the data generating process is used to predict future data in the same data space, which can then be compared to the current data set, and possibly uncover systematic deviations in the assumed model.
While the PPC provides a probability distribution for each data point in the complete $\{\phi \psi \chi t\}$ data space, a comparison in summary-statistics is found sufficient \citep{Gabry2019_BayesianWorkflow}, as the behaviour should change, if at all, smoothly in either dimension.
In this study, we use the partial sums over either dimension of the data space to compare with the PPC.
For example, $\tilde{d}_{\phi} = \sum_{\psi \chi t} \tilde{d}_{\phi \psi \chi t}$ describes the time-averaged distribution of Compton scattering angles for all polar and azimuth scattering angles.

\subsubsection{Empirical description}\label{sec:2Dgaussians}

Similar to previous studies, we intend to describe the diffuse 511\,keV line emission and flux empirically by a number of 2D-Gaussian functions.
Here, we use a grid of asymmetric 2D Gaussians, $G(l,b;\sigma_l,\sigma_b)$ with longitudinal width $\sigma_l$, latitudinal width $\sigma_b$, normalised to a full-sky-integrated line flux $F_{511}$, and centred on the Galactic centre $(l_0/b_0) = (0/0)$:

\begin{align}
	G(l,b;\sigma_l,\sigma_b) = \frac{F_{511}}{2 \pi \sigma_l \sigma_b} \exp\left( -\frac{1}{2} \left[ \frac{l^2}{\sigma_l^2} + \frac{b^2}{\sigma_b^2} \right] \right)\mrm{.} \tag{13}
	\label{eq:2DGaussians}
\end{align}

\begin{figure}[ht!]
	\centering
	\includegraphics[trim=0.1in 0.1in 0.8in 0.6in, clip=True, width=1.0\columnwidth]{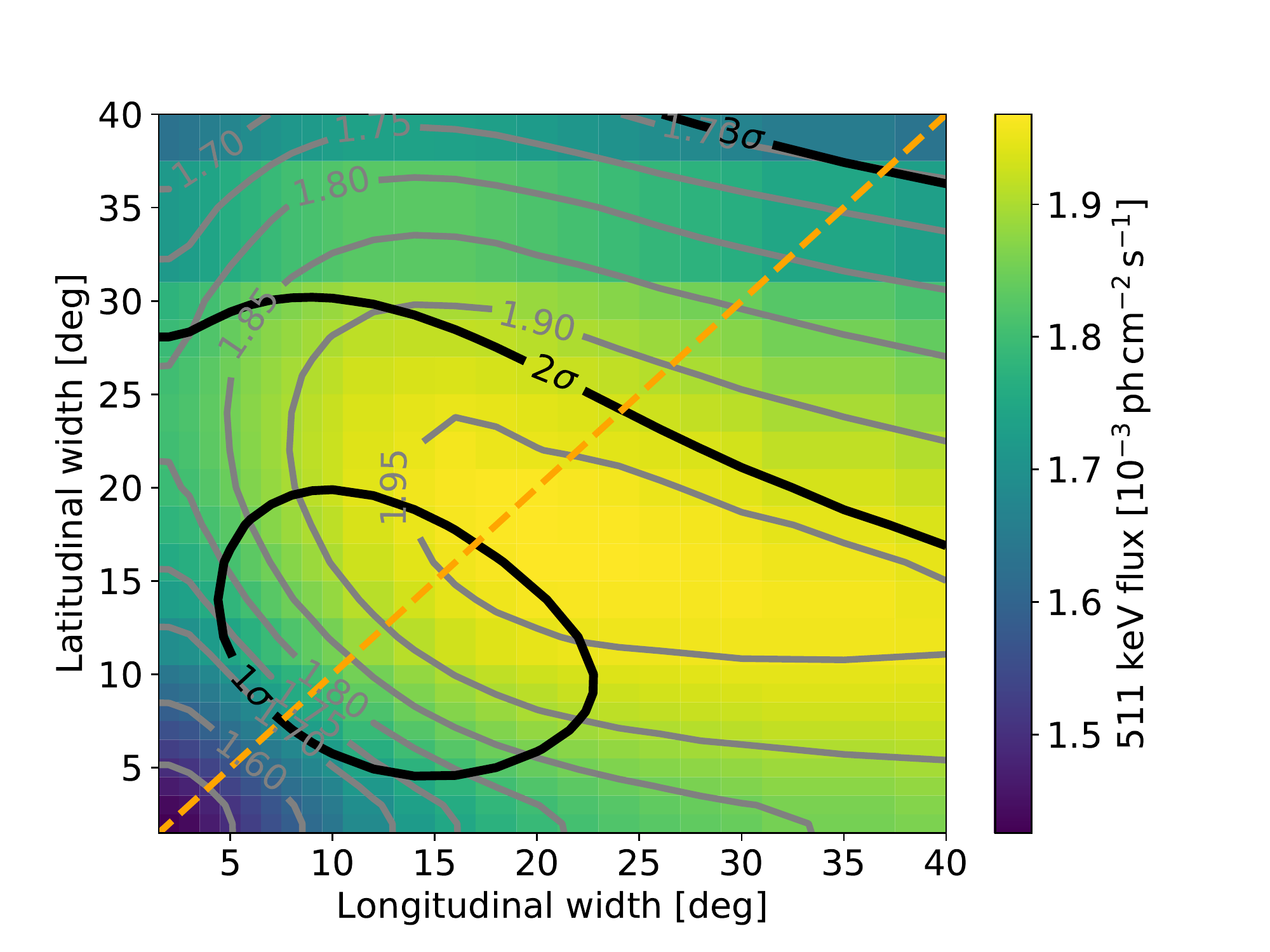}
	\caption{Likelihood profile (black contours) of the grid search in Sec.\,\ref{sec:2Dgaussians} as a function of 2D-Gaussian widths in longitude ($\sigma_l$) and latitude ($\sigma_b$), centred on $(l_0/b_0) = (0/0)$. The likelihood is maximised for $\sigma_l \approx \sigma_b = 12^{+8}_{-5}\,\mrm{deg}$, resulting in a best-fit flux (colour-coding, gray contours) of $(1.90 \pm 0.45) \times 10^{-3}\,\mrm{ph\,cm^{-2}\,s^{-1}}$.}
	\label{fig:likelihood_2DGaussians}
\end{figure}

\noindent Our chosen grid is equally-spaced in $\sigma_l$ and $\sigma_b$, respectively, from $1^{\circ}$ to $10^{\circ}$ in $1^{\circ}$ steps, to $30^{\circ}$ in $2^{\circ}$ steps, and to $40^{\circ}$ in $5^{\circ}$ steps.
This defines a grid of $22 \times 22 = 484$ individually-tested maps.
In each fit, the sky amplitude $\alpha$ and the background parameters $\beta_b$ are re-determined simultaneously.
This results in a likelihood profile as shown in Fig.\,\ref{fig:likelihood_2DGaussians}.

The 511\,keV emission in the Galactic bulge region is found to be diffuse, as the the width parameters favour values larger\footnote{We note that diffuse emission on smaller scales than the angular resolution \textit{can} be fitted, as the imaging response broadens any emission profile by its $5^{\circ}$ resolution.
Only a point source would appear as a $5^{\circ}$ emission feature;
a 2D Gaussian with a FWHM of $2^{\circ}$, for example, would be seen as a $\approx 5.4^{\circ}$ emission feature, approximated by Gaussian quadrature.
With high enough statistics, such a deviation can be identified.} than COSI's spatial resolution of $\approx 5^{\circ}$.
There is no strong asymmetry found in the shape of the 2D profiles, so that we reduce the asymmetric Gaussian function to a symmetric one, $\sigma_l = \sigma_b \equiv \sigma_{sym}$ (dashed orange line in Fig.\,\ref{fig:likelihood_2DGaussians}), and find a best-fit extent of $\hat{\sigma}_{sym} = 12_{-5}^{+8}\,\mrm{deg}$.
This value is in agreement with the extension $\approx 14^{\circ}$ estimated in \cite{Kierans2019_511COSI}.
We refer to this best-fit model as $\mrm{G_{12}}$ throughout the next sections.
At this point in the grid, the fitted 511\,keV flux is determined to be $(1.90 \pm 0.45) \times 10^{-3}\,\mrm{ph\,cm^{-2}\,s^{-1}}$.
Note that there is a dependence of the extent on the flux uncertainties.
This can be estimated from the tangents of flux contours in Fig.\,\ref{fig:likelihood_2DGaussians} with the $\Delta \mathscr{L} = 1\sigma$ contours \citep[cf., for example, Appendix A.1 in][]{Siegert2016_511} and results in an uncertainty of $(_{-0.20}^{+0.07}) \times 10^{-3}\,\mrm{ph\,cm^{-2}\,s^{-1}}$.
These values are also consistent with our findings from the image reconstruction, Sec.\,\ref{sec:RLimages}, in both flux and extent, again reassuring that our formalism is robust.
We use simulations to assess the reliability of this likelihood profile and its accuracy.
The results of these simulations are shown in Appendix\,\ref{sec:appendix_simulations} for different 511\,keV fluxes.
In addition, we can now quantify the emission uncertainties and find that emission features beyond $\approx 40^{\circ}$ are not required to explain the COSI 511\,keV data.
We discuss possible differences with literature values and implications in Sec.\,\ref{sec:discussion}.

\subsubsection{Full-sky emission}\label{sec:fullsky}

Previous studies \citep[e.g.,][]{Purcell1997_511,Knoedlseder2005_511,Weidenspointner2008_511b,Bouchet2010_511} suggested that there is an additional disk-like structure in the Galactic-wide 511\,keV emission beyond what would typically be attributed to the Galactic bulge or bar.
Observations with more exposure confirmed the longitudinally-extended morphology \citep{Skinner2014_511,Siegert2016_511}, and opened a discussion if this component is really the Galactic thin or thick disk, or if the component actually points towards a more halo-like structure.

\begin{figure*}[ht!]
	\centering
	\includegraphics[trim=1.0in 2.0in 1.0in 1.0in, clip=True, width=1.0\textwidth]{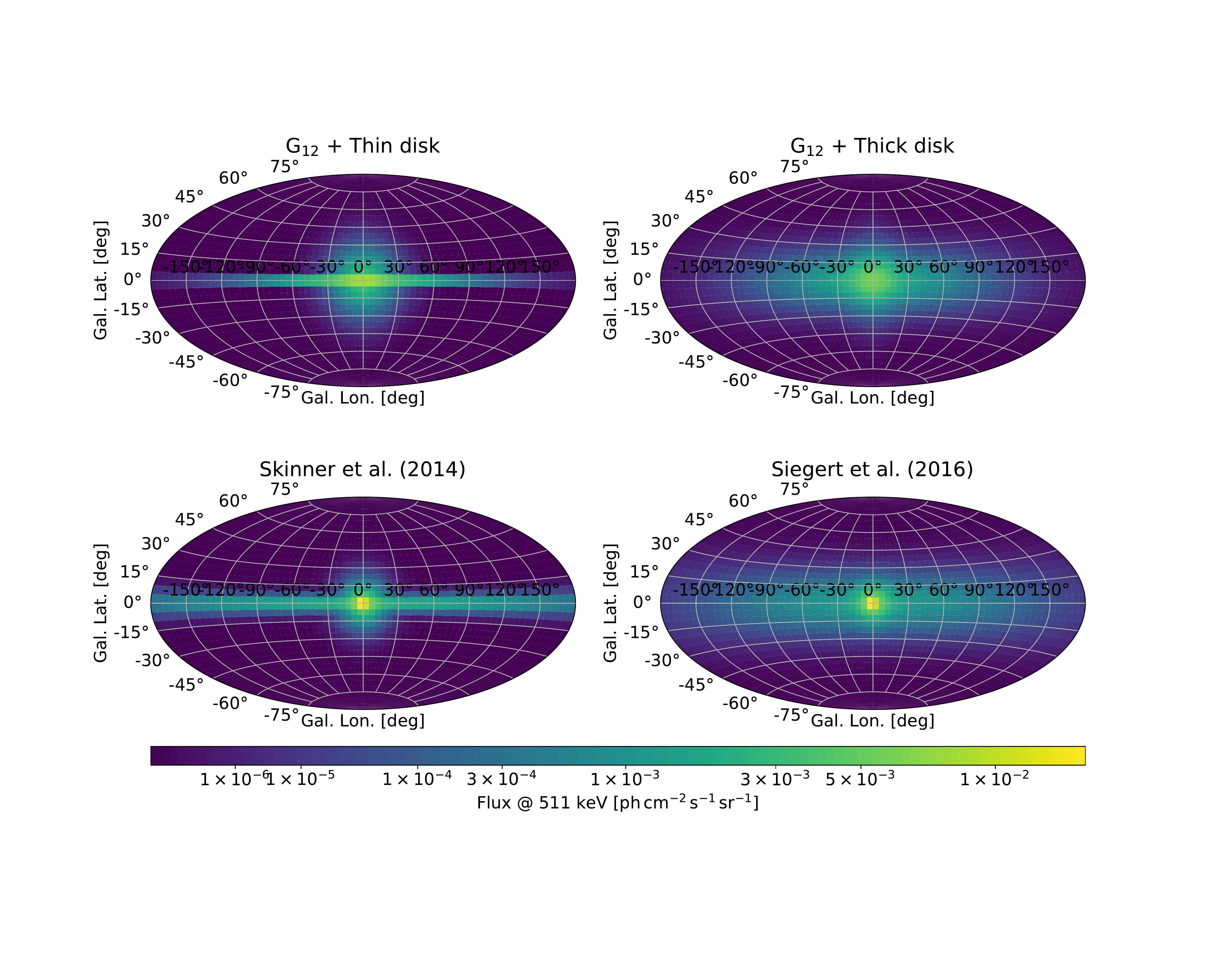}
	\caption{Tested full-sky emission morphologies. The top panel shows the best-fit 2D-Gaussian ($\mrm{G_{12}}$) from Sec\,\ref{sec:2Dgaussians}, Fig.\,\ref{fig:likelihood_2DGaussians} with symmetric width of $12^{\circ}$, plus thin disk (left, vertical extent $2^{\circ}$) and a thick disk (right, vertical extent $10^{\circ}$). The bottom panels show the multi-component models as found by \citet[][, left]{Skinner2014_511} and \citet[][, right]{Siegert2016_511}.}
	\label{fig:tested_disk_maps}
\end{figure*}

With the ultimately more-definite imaging response of Compton telescopes, this question can be investigated further.
We use different model components, either in addition to the best-fit bulge from Sec.\,\ref{sec:2Dgaussians}, or complete full-sky descriptions from the recent literature, to investigate whether there is a disk-like component present in COSI 511\,keV data.

We use the model $\mrm{G_{12}}$ and add a second disk-like component to the fit, modelled as additional 2D Gaussian with either a small \citep[$\sigma_b = 2^{\circ }$; cf.][]{Skinner2014_511} or a large \citep[$\sigma_b = 10^{\circ }$; cf.][]{Siegert2016_511} latitudinal extent, and longitudinal extend of $\sigma_l = 40^{\circ}$ in both cases.
The quoted disk-fluxes in the literature range between $0.0$ and $2.9 \times 10^{-3}\,\mrm{ph\,cm^{-2}\,s^{-1}}$, depending on the instrument and the total exposure \citep[e.g.][]{Purcell1997_511,Prantzos2011_511,Siegert2019_lv511}.
Later observations with more than ten years of INTEGRAL/SPI exposure consistently found a disk-like component, with flux values in the range $1.0$--$2.0 \times 10^{-3}\,\mrm{ph\,cm^{-2}\,s^{-1}}$, for which reason we set the normalisation of any such second component to $1.5 \times 10^{-3}\,\mrm{ph\,cm^{-2}\,s^{-1}}$, and use again a truncated normal prior of $\alpha_{\mrm{DISK}} \sim \mathscr{N}_{\alpha>0}(\mu = 1, \sigma = 2/3)$.

\begin{figure*}[ht!]
	\centering
	\includegraphics[trim=1.0in 0.5in 1.4in 1.0in, clip=True, width=1.0\textwidth]{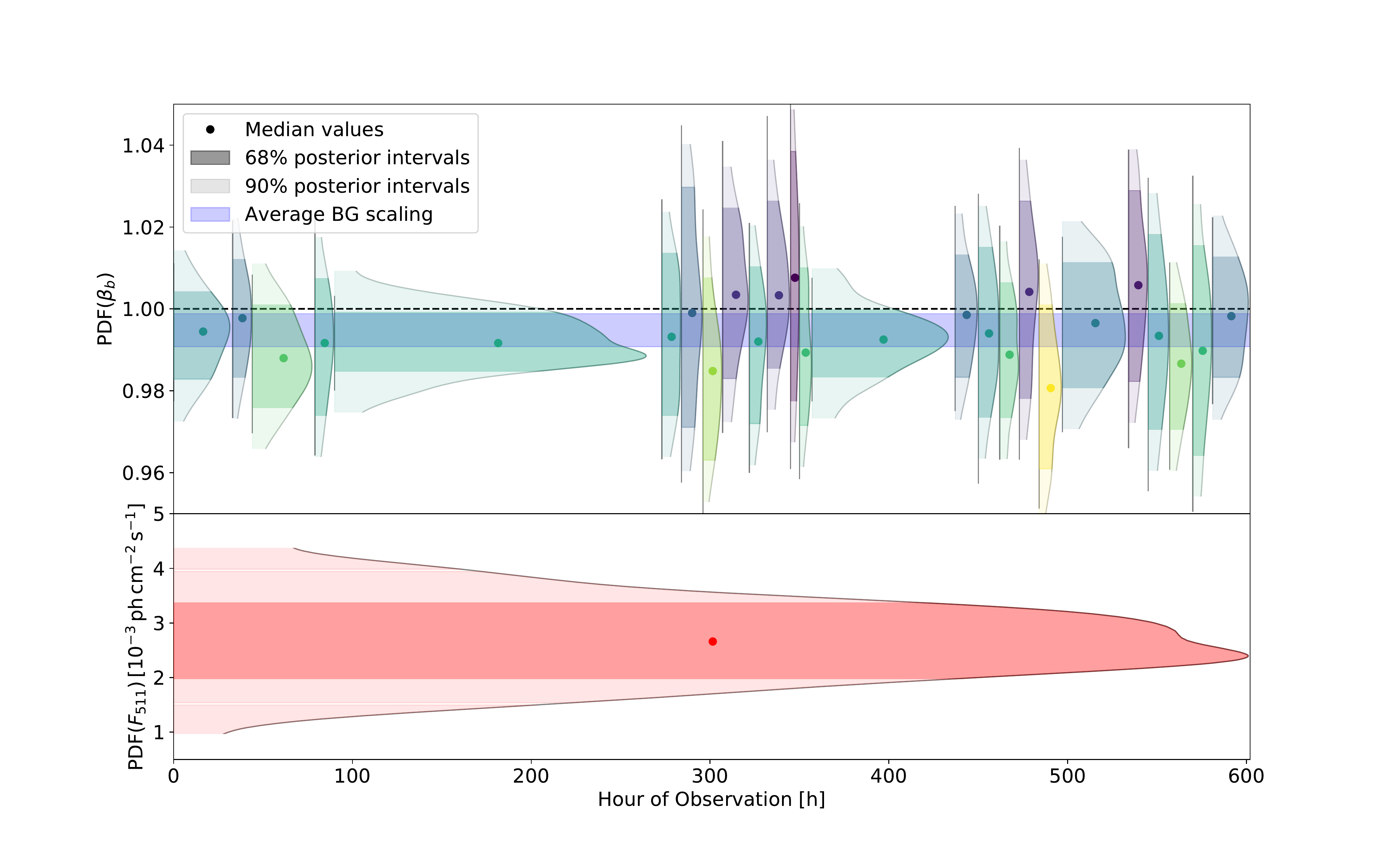}
	\caption{Posterior probability distributions of the background re-scaling parameters, $\beta_b$ (top), and the resulting 511\,keV flux, $F_{511}$ (bottom), assuming the \cite{Siegert2016_511} model. The horizontal width of each posterior resembles the time (x-axis) for which a specific parameter is active. The vertical width includes the 68\,\% (dark shaded, $\approx 1\sigma$) and 90\,\% (light shaded, $\approx 1.7\sigma$) percentiles of the sampled distributions. The colours indicate how far the median of the $\beta_b$-values deviates from a background-only time interval (i.e. $1.0$). See text for further detail.}
	\label{fig:posteriors_siegert2016_fit}
\end{figure*}

We find no significant excess in either combination, and provide a $3\sigma$ upper limit on the 511\,keV flux of $<3.1$ and $<4.3 \times 10^{-3}\,\mrm{ph\,cm^{-2}\,s^{-1}}$ for the thin and thick disk, respectively.
Note that the 99.85\,\% percentile (one-sided $3\sigma$-bound) of the chosen prior includes $4.5 \times 10^{-3}\,\mrm{ph\,cm^{-2}\,s^{-1}}$, showing that the upper limits are dominated by the contributions from the likelihood.
We find that including a second component reduces the flux of the central bulge component by $\approx 25\,\%$ in each case, which points to a non-zero contribution of a disk-like component.
This is reassuring as the bulge flux now appears closer to literature values.
We can, however, not claim a detection of an additional component beyond the Galactic bulge, as described by the $\mrm{G_{12}}$ model.

For additional full-sky model tests, we use the four-component models by \citet{Skinner2014_511} and \citet{Siegert2016_511} with fixed relative amplitudes to have complete descriptions across the entire sky.
The number of sky model parameters for these tests is thus reduced to one again.
This provides an estimate of the total Galactic 511\,keV flux as seen by COSI during its 2016 flight.
In Fig.\,\ref{fig:tested_disk_maps}, a summary of the used full-sky models is shown.

\begin{figure*}[ht!]
	\centering
	\includegraphics[trim=1.0in 0.7in 1.0in 1.2in, clip=True, width=1.0\textwidth]{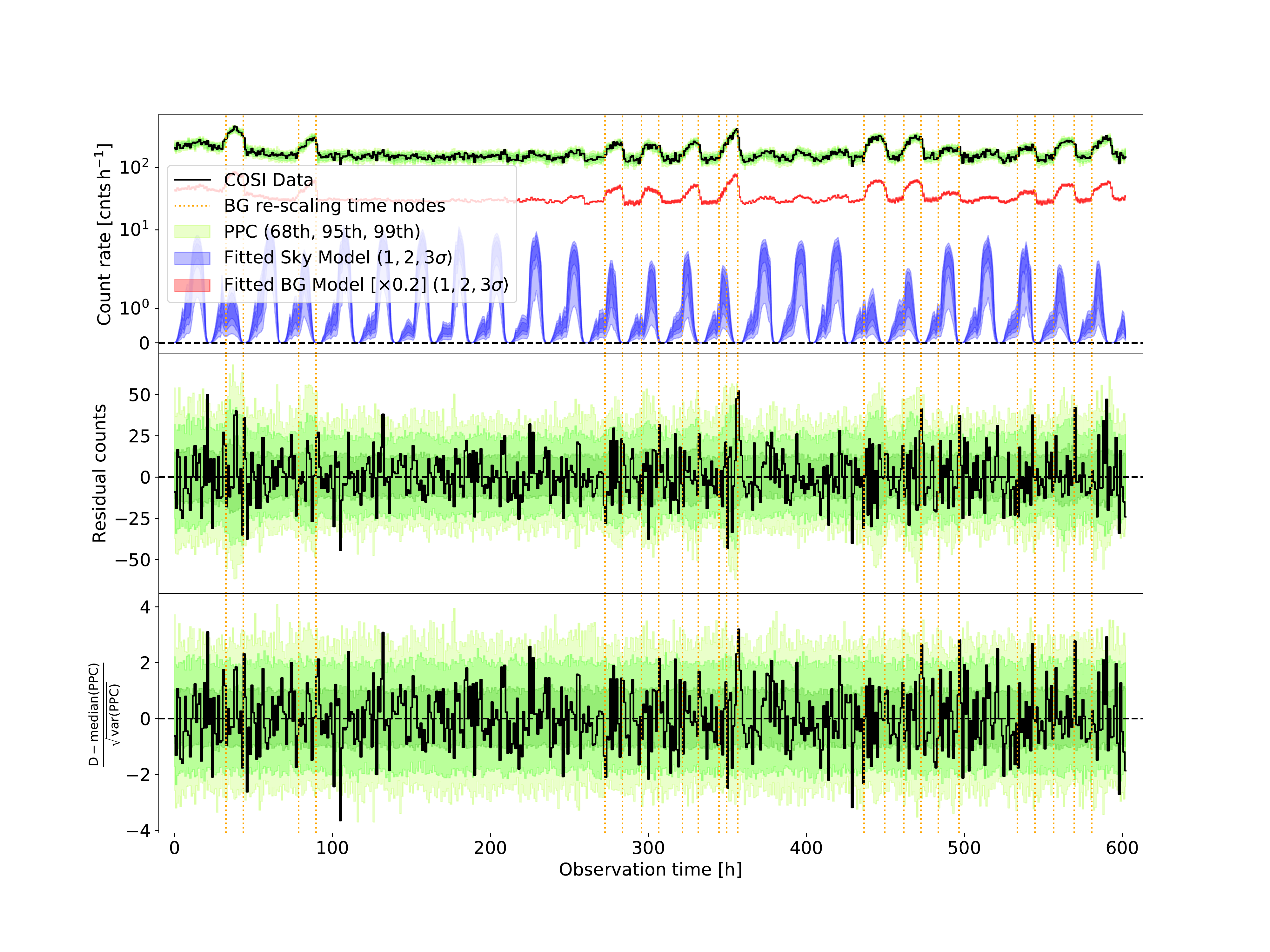}
	\caption{Model fit quality (summary statistics) in the time domain: Shown are the COSI data (black histogram), together with the 68th, 95th, and 99th percentile of the PPC (green bands, cf. Eq.\,(\ref{eq:PPC})), summed over the COSI data space $\{\phi \psi \chi \}$, for each time bin, together with the fitted background model (red; for illustration purpose scaled by $0.2$) and sky model (blue). From \textit{top} to \textit{bottom}, the data space, the absolute residuals, and the z-scores are shown. The chosen background time nodes are indicated by vertical lines. Clearly, the fit appears adequate as only 12 out of 603 time bins fall outside the 99th percentile of the PPC. See Fig.\,\ref{fig:PPC_phi_siegert2016} (Appendix) for additional summary statistics in other data space dimensions.}
	\label{fig:PPC_time_siegert2016}
\end{figure*}

For the \citet{Siegert2016_511} model, we show exemplarily the posterior distributions of the fitted sky model scaling parameter as well as the $25$ background scaling parameters as they vary with time in Fig.\,\ref{fig:posteriors_siegert2016_fit}.
The full-sky 511\,keV flux is found to be $(2.7 \pm 0.7) \times 10^{-3}\,\mrm{ph\,cm^{-2}\,s^{-1}}$ (bottom panel), consistent with previously-found estimates.
The posterior distributions for the background re-scaling parameters are shown in the top panel, each of them shown according to the time intervals between which the Bayesian block method set the time nodes (cf. Sec.\,\ref{sec:rescaling} and Fig.\,\ref{fig:PPC_time_siegert2016}).
If there was no sky contribution present, i.e. $F_{511} = 0$, the $\beta_b$-values should consistently scatter around $1.0$.
While each fitted background parameter is individually consistent with $1.0$, there is a clear trend for a reduced background level during the $603$ observation hours, as indicated by the blue shaded band. 

In Fig.\,\ref{fig:PPC_time_siegert2016}, we show the PPC of this model in the time domain for a fit quality check (additional PPCs in the remaining COSI data space, i.e. $\{\phi \psi \chi\}$, are shown in Appendix \ref{sec:appendix_figure}).
The top panel shows the COSI 511\,keV data as black histogram, together with the model posterior of sky (blue) and background (red), and the PPC as summarised into these times bins as green shading.
Naturally, the total count rate is dominated by the background, and consequently so is the PPC.
In all panels, the best-fit Bayesian block time nodes, $\hat{\mathscr{B}}$, are indicated by dashed orange lines.
In the middle panel, the absolute residuals in count space ($d_t-\tilde{d}_t$) are shown, together with the PPC as scattering around $0$, however with changing variance according to Poisson statistics.
In order to normalise the absolute residuals and to provide a common frame of comparison, we calculate the z-score for each time bin ($(d_t-\tilde{d}_t)/\sqrt{\mrm{var}(\tilde{d}_t)}$).
Clearly, the data points scatter around $0$, with seldom outliers beyond the 99th percentile of the PPC.
In this fit, only $12$ out of $603$ values are found outside this range, which is consistent with expectations.
In Fig.\,\ref{fig:PPC_phi_siegert2016} (Appendix), we show the PPC in other COSI data space dimensions, also finding that our model describes the data adequately.
We note that in various scattering angle bins, the number statistics is very small, which naturally leads to asymmetric residuals due to the nature of the Poisson statistics.

The results for the \citet{Skinner2014_511} model, i.e. a thin disk, are of similar nature, providing a total 511\,keV line flux of $(2.3 \pm 0.9) \times 10^{-3}\,\mrm{ph\,cm^{-2}\,s^{-1}}$.
There is no significant likelihood difference found between the two models.
This is expected since the exposure for any disk-like morphology is restricted to longitudes $-150^{\circ} \lesssim l \lesssim -90^{\circ}$ and $30^{\circ} \lesssim l \lesssim 60^{\circ}$, i.e. small sub-regions for which the disk-luminosity is expected to drop significantly compared to the bulge (cf. Fig.\,\ref{fig:exposure}).
We summarise our fitted models, corresponding flux values, and likelihoods in Tab.\,\ref{tab:model_results}.
In addition, the question of how extended the 511\,keV disk actually is, is still under debate, and whether a more-structured morphology, e.g. including spiral arms, is present.
See Sec.\,\ref{sec:discussion} for further discussion. 

\begin{table}
	\centering
	\begin{tabular}{lcccr}
		\hline
		Model & Bulge & Disk & Total & $\Delta\ln\mathscr{L} $ \\
		\hline
		$\mrm{G_{12}}$ & $1.9 \pm 0.4$ & $-$ & $1.9 \pm 0.4$ & $-954.9$ \\
		$\mrm{G_{12}}$ + Thin Disk & $1.5 \pm 0.5$ & $< 3.1$ & $2.6 \pm 0.7$ & $-953.3$ \\
		$\mrm{G_{12}}$ + Thick Disk  & $1.4 \pm 0.6$ & $<4.3$ & $2.9 \pm 0.8$ & $-953.6$ \\
		Siegert et al.\,(2016a) & $-$ & $-$ & $2.7 \pm 0.7$ & $-958.3$ \\
		Skinner et al.\,(2014) & $-$ & $-$ & $2.3 \pm 0.9$ & $-960.0$ \\
		\hline
	\end{tabular}
	\caption{Summary of model fitting results. Fluxes for bulge, disk, total components are given in units of $10^{-3}\,\mrm{ph\,cm^{-2}\,s^{-1}}$, with $1\sigma$ uncertainties, including the uncertainties in the extension of the spatial distribution. Upper limits are given as $99.85\,\%$ percentile of the posterior ($3\sigma$). The log-likelihood is shown as relative value with respect to a constant. While each map performs about equally-well, it should be noted that the smaller bulge components by \citet{Skinner2014_511} and \citet{Siegert2016_511} are slightly disfavoured by the COSI data.}
	\label{tab:model_results}
\end{table}

\section{Discussion}\label{sec:discussion}

\subsection{Positron annihilation puzzle}\label{sec:positron_puzzle}

The measured extent of the central 511\,keV emission ($\mrm{FWHM} \approx 28_{-12}^{+19} \,\mrm{deg}$) is found to be at least 2--3 times larger compared to previous measurements by INTEGRAL/SPI \citep[e.g. $8^{\circ}$ FWHM by][]{Knoedlseder2005_511}, however consistent with WIND/TGRS measurements by \citet{Harris1998_TGRS511}, finding $24_{-9}^{+8}$\,deg.
Our fitted extent is in agreement with COSI data analysis from \citet{Kierans2019_511COSI} ($\approx 33^{\circ}$), who focussed on spectral analyses.
We note, however, that such large-scale emission regions naturally (due to the nature of the fit) capture more flux than smaller regions.
In addition, this large 2D-Gaussian might capture not only  the `narrow bulge' emission component, but rather a superposition of the true emission morphology, including the disk and/or a halo, which might not have been seen in other instruments.
For example, \citet{Skinner2012_511} on the one hand noted that a halo component would be favoured by instruments like OSSE, TGRS, or SMM with large field of views, compared to a SPI-only description of the data.
On the other hand, as already shown by \citet{Albernhe1981_511}, \citet{Leventhal1986_511}, \citet{Lingenfelter1989_511}, or \citet{Purcell1997_511}, this field of view issue (larger field of views lead to larger fluxes) has been addressed correctly in both analyses.
This means a correct forward-implementation of the effective area as a function of zenith and azimuth for each instrument should yield the same results, because the field of view is already included, and cannot come into play a second time.
Rather, the analysis methods themselves have to be carefully investigated, as it is typically assumed, for example for SPI, that everything `outside' the Galactic bulge and disk \citep[e.g., at high latitudes, ][]{Bouchet2015_26Al} will provide no contribution to the expected counts.
For this reason, halo components would not be visible.
In addition, halo-like emission, or any emission with a shallow gradient or isotropy, is almost impossible for a coded-mask instrument to observe, because the coding would result in an equal response for all times.
If not accounted for, this ultimately would be disregarded as being due to background.
Similar statements apply for collimators as well.
A possible step towards observing large-scale diffuse emission with collimating or coded-mask telescopes would be occultation observations, for example being shadowed by Earth.
While the sensitivity of current \gr telescopes is probably too low to detect halo-like or isotropic emission at 511\,keV in general, Compton telescopes, such as COSI, provide a direct response to single photons so that low-gradient emission could be identified.
This would provide a major step in estimating the true extents of soft \gr emission in the MeV regime in general.

The 511\,keV flux measurement of $(1.90 \pm 0.45) \times 10^{-3}\,\mrm{ph\,cm^{-2}\,s^{-1}}$ for the best-fit 2D-Gaussian component to describe the bulge is about twice as large as in previous measurements, however consistent within uncertainties.
Adding a disk-like component reduces this value to about $1.5 \times 10^{-3}\,\mrm{ph\,cm^{-2}\,s^{-1}}$ (see Tab.\,\ref{tab:model_results}), which is more consistent with earlier measurements.
We find no evidence, however, for an additional disk component, probably due to its low surface brightness nature, and provide upper limits which are only about twice the measured values from recent SPI studies.

Our Richardson-Lucy deconvolution algorithm finds a 511\,keV flux in the central regions of the Galaxy of about $2 \times 10^{-3}\,\mrm{ph\,cm^{-2}\,s^{-1}}$, consistent with our model fits, and affirming a robust analysis.
We apply additional filters to our data set (Appendix\,\ref{sec:appendix_reliability}), from which we find that observations that include only night times or when the balloon altitude was above 30\,km result in the most noise-free maps (cf. Fig\,\ref{fig:RL_filter_maps}).
In addition, we find that the second third of the 603 observing hours provides the cleanest image, in which also the emission peak appears closer to the Galactic centre. 

We find a general consistency between COSI measurements and earlier studies regarding the absolute 511\,keV flux estimates, however with tendencies towards higher fluxes and more extended emission.
This could either be due to systematic mismatches between simulations for the imaging response and true effective area, unaccounted systematics in the background modelling procedure, or, alternatively, because of the better imaging capabilities of Compton telescope apertures in general, being able to capture also flux from emission regions with shallow gradients.

\subsection{The future of Compton telescopes}\label{sec:compton_telescopes}

A reliable, robust, and versatile background modelling for soft \gr telescopes is in general difficult to achieve.
Similar to earlier CGRO/COMPTEL procedures \citep[e.g.][]{Knoedlseder1996_COMPTELimaging,vonDijk1996_PhD_COMPTELBG,Bloemen1999_26AlOrion_revised}, and together with the experiences from the INTEGRAL/SPI spectrometer \cite{Diehl2018_BGRDB,Siegert2019_SPIBG}, we developed a method to infer a flexible background model for COSI, inferred from the measurements themselves.
As opposed to a rather well-defined space environment with fixed observation patterns, a free-floating balloon-borne telescope poses additional difficulties in estimating the time-dependent background contributions.
We showed that it is possible to infer imaging information in a full-forward modelling manner, by allowing the strongly variable background to be determined simultaneously with the celestial 511\,keV \gr signal.

We found that, naturally, the largest impact on the background count rate at 511\,keV is the balloon altitude as more air mass increases the interaction rate of cosmic-rays with the atmosphere, leading to more secondary particles and $\gamma$-rays.
Additionally, southern latitudes lead to an increased 511\,keV count rate due to the strong latitudinal dependence of the geomagnetic cutoff rigidity \citep[see, e.g.][]{Ling1975_MeVBG,Ling1977_511keV_atmosphericBG,Kierans2018_PhD}.
The short-term variability can be predicted from independent count rates of COSI's CsI veto-shield, for example, or by the photo events (single-site events) which provide an exceptionally good predictor for 511\,keV Compton event photons.
Scargle's Bayesian blocks algorithm \citep{Scargle1998_BayesianBlocks,Scargle2012_BayesianBlocks} provides a useful tool to identify additional changes of the background rate that are missed by any variability tracer.
Finally, our complete background model describes a semi-empirical and modular approach to tackle the unknown MeV background, being based on the instrument-specific data space as well as expertise from balloon enterprises.

In this work, we thus confirmed earlier studies regarding the Galactic 511\,keV emission and provided a scheme to approach the individual difficulties that such a complex instrument in a complex environment inherits.
The ultimate measurement of the 511\,keV emission would nevertheless be best from space.
This can be accomplished with the COSI-SMEX space mission
which is currently in a Phase A study\footnote{\url{https://www.nasa.gov/press-release/nasa-selects-proposals-to-study-volatile-stars-galaxies-cosmic-collisions}}.
In nearly-equatorial low-Earth orbit, a COSI satellite would have significant advantages compared to COSI on its balloon platform:
COSI-SMEX's reduced strip pitch will lead to better angular resolution and better background identification due to fewer incorrectly reconstructed events.
In addition, more events with close-by interactions can be used in the analysis, and, together with more detectors (16 instead of 9 in this work), less stringent event cuts due to better shielding will lead to a significantly larger effective area.
The lower and more stable background conditions in space, along with no atmospheric absorption and a larger field of view, will ultimately result in a better sensitivity than any previous MeV \gr telescope.

Considering also the longer mission duration (2 years, plus possible extensions), COSI-SMEX would readily be able to answer the still unsolved questions about the true 511\,keV morphology, allow the study of individual regions in 511\,keV, and the connections to its sources.
Due to its high-purity Ge detectors, it would still resolve \gr lines for high-resolution spectroscopy \citep{Tomsick2019_COSI}.

\acknowledgments

Compton binned-mode response developments were sponsored under NASA APRA grant NXX17AC84G. This research used resources of the National Energy Research Scientific Computing Center (NERSC), a U.S. Department of Energy Office of Science User Facility operated under Contract No. DE-AC02-05CH11231, for COSI response simulations. The COSI instrument developments and balloon flights are supported through NASA APRA grants NNX14AC81G \& 80NSSC19K1389. This work is also supported in part by CNES. Carolyn Kierans is supported by a NASA Postdoctoral Program Fellowship. Thomas Siegert is supported by the German Research Society (DFG-Forschungsstipendium SI 2502/1-1).

%

\vspace{5mm}
\facilities{COSI}


\software{MEGAlib \citep{Zoglauer2006_MEGAlib},
				numpy \citep{Oliphant2006_numpy},
				matplotlib \citep{Hunter2007_matplotlib},
				astropy \citep{astropy2013_astropy},
				scipy \citep{Virtanen2019_scipy},
				stan/pystan \citep{Carpenter2017_stan},
				arviz \citep{Kumar2019_arviz}
          }



\clearpage

\appendix

\section{Reliability of the image reconstruction}\label{sec:appendix_reliability}

In order to investigate the reliability and stability of our modified Richardson-Lucy deconvolution algorithm, we apply different filters to the full data set to estimate systematics and check for consistency.

From our 603 hour long data set, we select eight subsets to investigate the robustness of the deconvolution algorithm and how this is related to the data quality and environmental conditions:
Because lower altitudes increase the background count rate, we select times when the balloon was above \verb|30km|.
We study the influence and possible contributions of the Sun by splitting the data set either in only \verb|night| or \verb|day| times, defined by the sunset at each Earth latitude and time.
For a temporal distinction, we use the \verb|first|, \verb|second|, and \verb|third| 201 hours of observation as individual data sets.
Finally, to study if the Moon albedo is strong enough to show an imprint in the current COSI measurements, we select times when the Moon was in (\verb|moon|) and outside (\verb|nomoon|) the field of view.

\begin{figure}[ht!]
	\centering
	\includegraphics[trim=0.0in 0.0in 0.0in 0.0in, clip=True, width=0.8\textwidth]{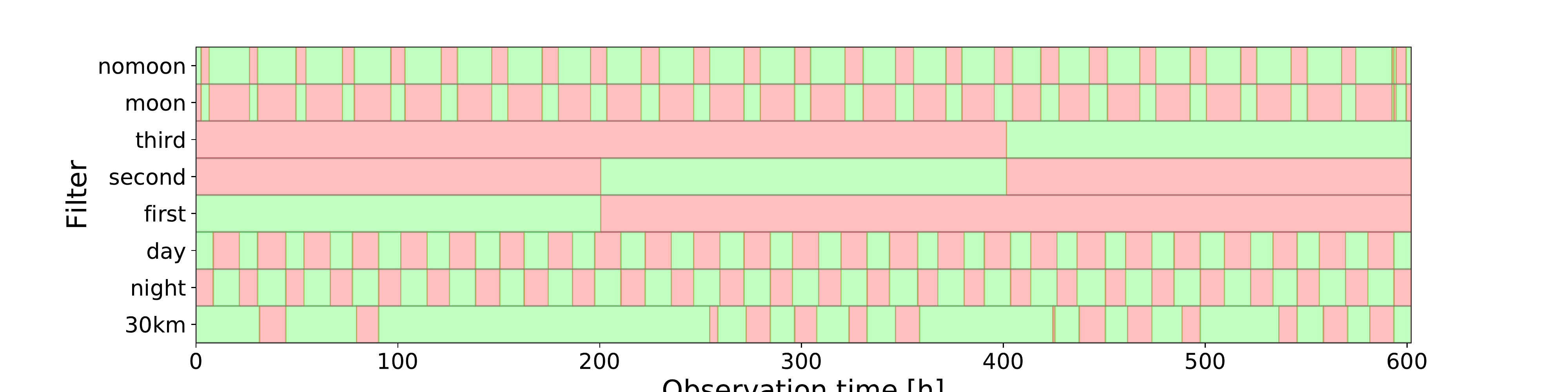}\\
	\includegraphics[trim=0.0in 0.0in 0.0in 0.4in, clip=True, width=0.8\textwidth]{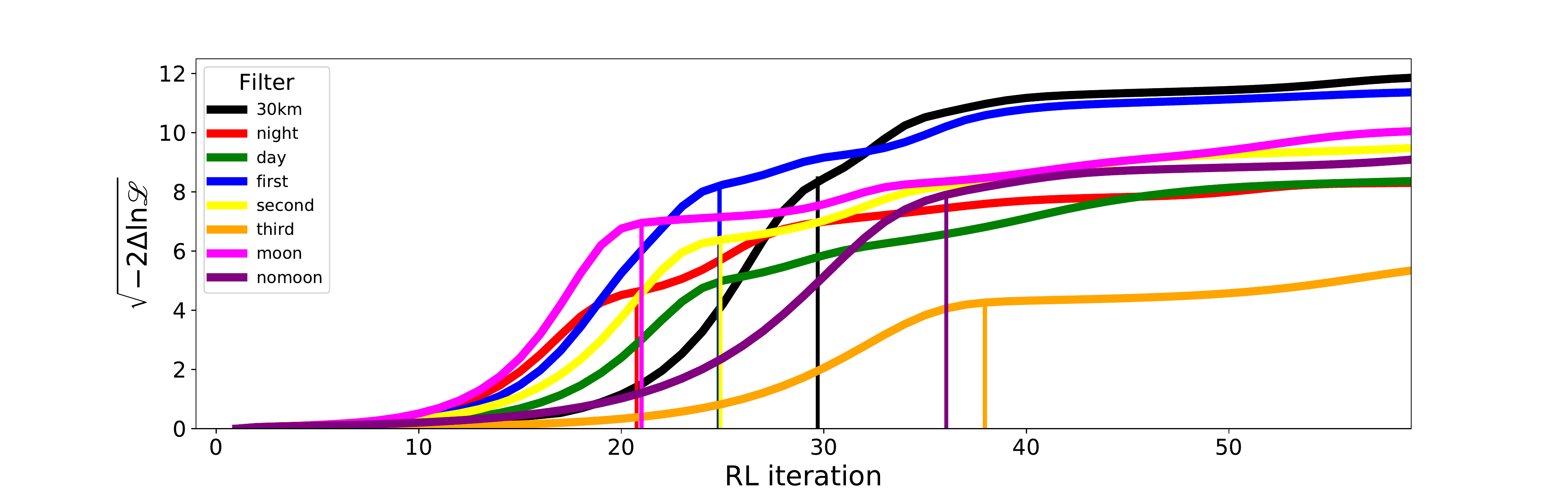}
	\caption{Overview of additionally-filtered data sets with green (accepted) and red (rejected) times (\textit{top}), and significance of the current Richardson-Lucy iteration versus a background-only fit, for all eight filters with chosen iteration indicated as vertical line (\textit{bottom}).}
	\label{fig:RL_filters}
\end{figure}

An overview of the selected times is given in Fig.\,\ref{fig:RL_filters}, top panel.
For each data subset, we use the same modified Richardson-Lucy algorithm as presented in Sec.\,\ref{sec:arldobags}, and apply the same choice for which the iteration to select as representative.
In Fig.\,\ref{fig:RL_filters}, bottom, the test statistics as a function of iteration is shown, together with the chosen iteration.
The range of acceptable deconvolutions spans iterations $\approx 20$ to $\approx 40$, similar to the full data set.
The significance of the chosen maps versus a background-only fit ranges between $4$ and $9\sigma$.

\begin{figure}[ht!]
	\centering
	\includegraphics[trim=1.9in 2.0in 1.1in 1.1in, clip=True, width=1.0\textwidth]{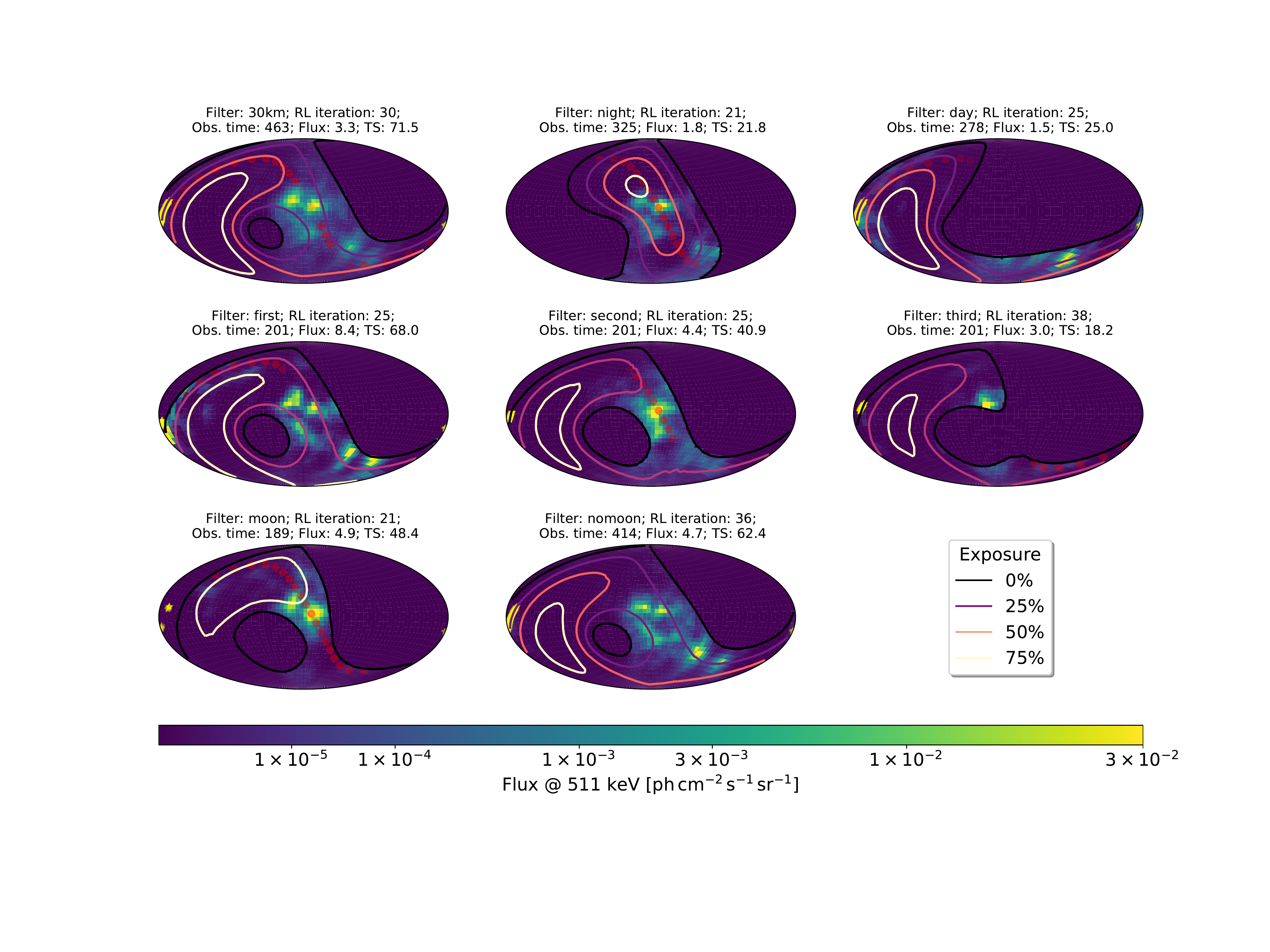} 
	\caption{Chosen Richardson-Lucy iterations for the eight additionally-filtered data sets. Colour scheme and exposure map contours similar to Fig.\,\ref{fig:RL_maps}. In each panel, the filter, the chosen iteration, the observation time in hours, the integrated 511\,keV flux in units of $10^{-3}\,\mrm{ph\,cm^{-2}\,s^{-1}}$, and the test statistics, $\mrm{TS} = -2 \Delta \ln\mathscr{L}$, of the map versus a background-only fit is provided. The positions of the Moon and Sun during the respective data set are indicated in red and yellow.}
	\label{fig:RL_filter_maps}
\end{figure}

In Fig.\,\ref{fig:RL_filter_maps}, we show the resulting maps for each filter, together with the respective exposure maps.
Clearly, when the bulge region is not masked out due to specific selections, the bulge always appears bright in 511\,keV emission.
This is reassuring that the deconvolution algorithm consistently finds the signal, and does not pick individual times to assign counts in different sky regions.
We note, however, that also the noise peaks at high negative latitudes appear for some selections.
The significance of any individual feature is between $1$ and $3\sigma$, as tested by masking out the feature and then performing a maximum likelihood test to determine significance the additional component.
The selections \verb|second| and \verb|moon| provide the cleanest images, with significances between $6.4$ and $7.0\sigma$, and 511\,keV fluxes between $4.4$ and $4.9 \times 10^{-3}\,\mrm{ph\,cm^{-2}\,s^{-1}}$.
The fluxes are considerably higher than for the total data set, however naturally come with a larger uncertainty since the exposure time is one third or less.
Night time (\verb|night|), selections on the altitude (\verb|30km|), and the \verb|first| 201 observation hours result in images very similar to the full data set (see Fig.\,\ref{fig:RL_maps}).

As a result, we consistently recover emission from the Galactic bulge region, and determine the significance of individual emission hotspots outside this region to be between $1$ and $3\sigma$ (cf. exposure maps in each panel of Fig.\,\ref{fig:RL_filter_maps}).
We therefore cannot claim any additional detection beyond the the central region of the Milky Way with COSI measurements from the balloon campaign in 2016.

\section{Simulating data sets}\label{sec:appendix_simulations}

\begin{figure}[ht!]
	\centering
	\includegraphics[trim=0.5in 0.5in 0.5in 0.0in, clip=True,width=0.5\columnwidth]{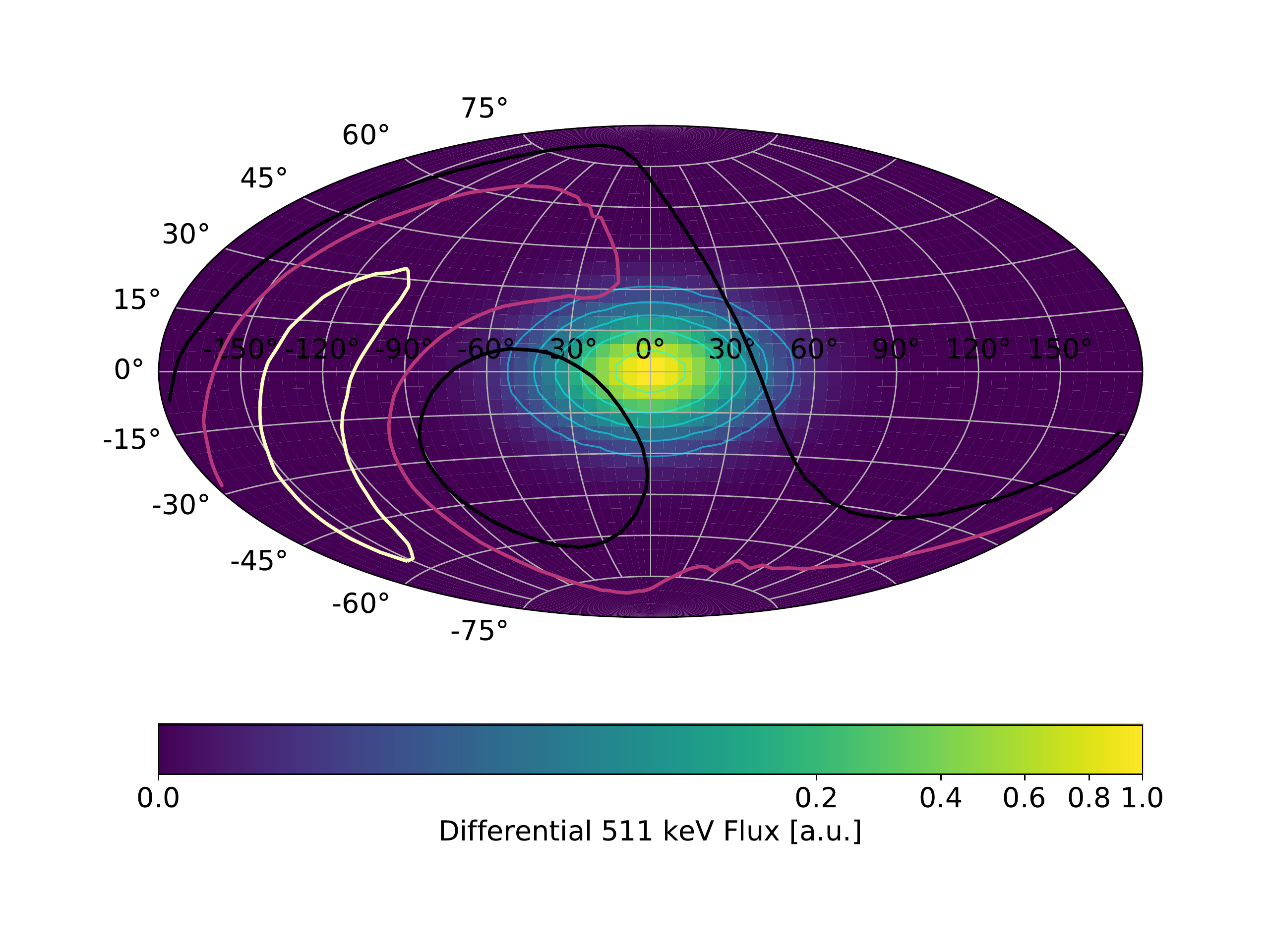}~
	\includegraphics[trim=0.0in 0.0in 0.4in 0.5in, clip=True,width=0.5\columnwidth]{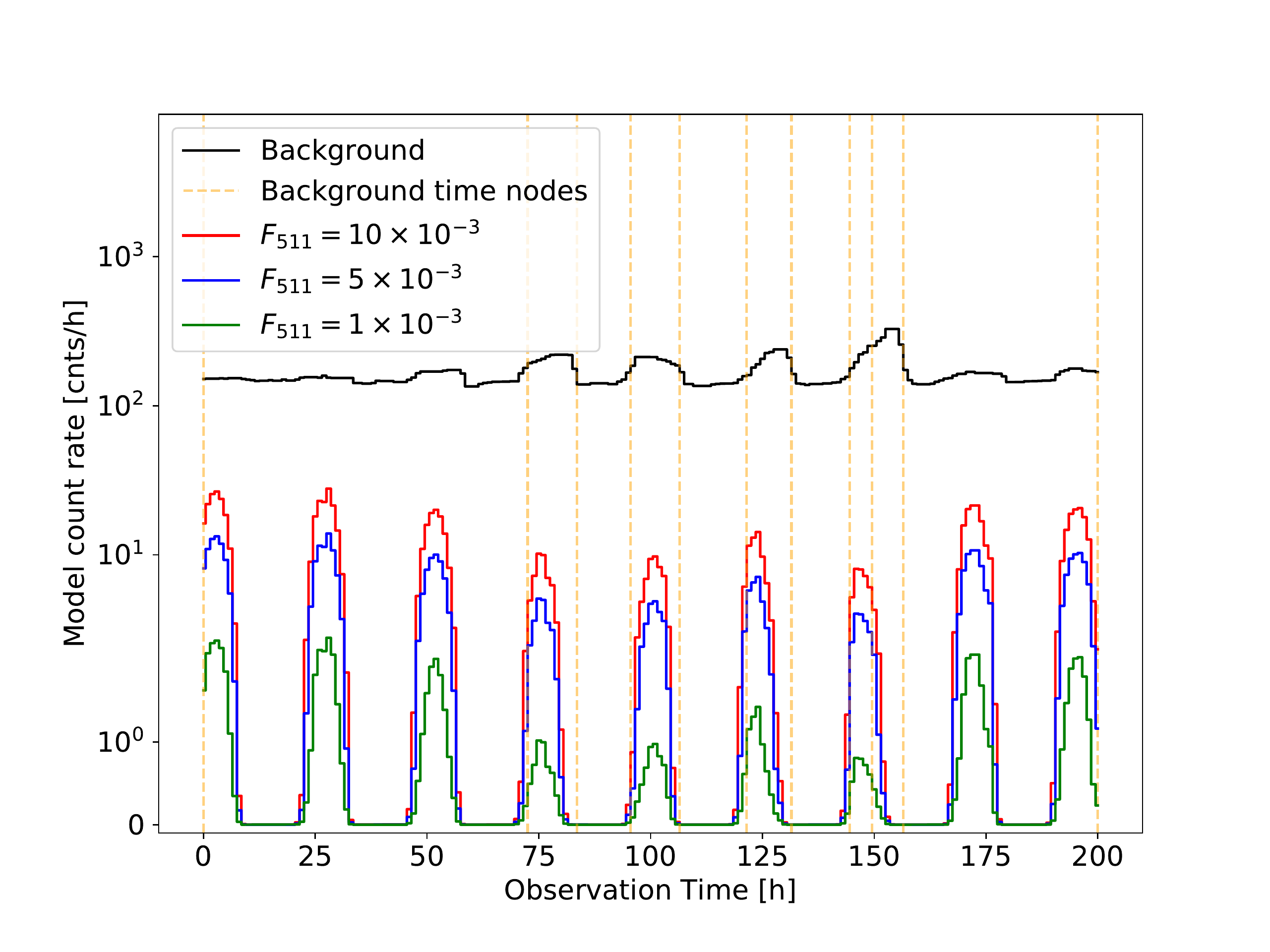}
	\caption{Simulated sky model (\textit{left}) and count rates for background and different total fluxes (\textit{right}). The exposure map of the simulated data set is indicated (similar to Fig.\ref{fig:RL_filter_maps}), showing that emission would be expected outside the exposed regions (black).}
	\label{fig:simulated_sky}
\end{figure}

For statistical and visual comparisons, we simulate a known celestial signal on top of a known background model.
We use a subset of $201$ observation hours (cf. \textit{second} subset from Appendix\,\ref{sec:appendix_reliability}) including the typical characteristics of the 2016 COSI balloon flight to better assess the general quality of image reconstructions with our modified Richardson-Lucy algorithm and maximum likelihood fits.
The background is modelled using the defined background response from Sec.\,\ref{sec:bg_response} and using a median filter with a width of 5 hours of the total count rate to define an absolute number.
The sky is modelled according to a 2D Gaussian function, located at the Galactic centre with longitudinal and latitudinal widths of $14^{\circ}$ and $8^{\circ}$ ($1\sigma$-values), respectively.
We simulated three different map-integrated fluxes of $10$, $5$, and $1 \times 10^{-3}\,\mrm{ph\,cm^{-2}\,s^{-1}}$ to obtain characteristic reconstructions and likelihood profiles for varying significances.
As the sky model emission extends beyond unexposed regions, this approach describes both: i) how structured the resulting map in the image reconstruction algorithm will be; and ii) how accurately a template fitting approach can determine the emission extents.
The true sky model, together with the exposure map and the expected count rates for background and the three different fluxes are shown in Fig.\,\ref{fig:simulated_sky}.

\begin{figure}[ht!]
	\centering
	\includegraphics[trim=0.7in 1.4in 0.7in 0.9in, clip=True,width=0.33\columnwidth]{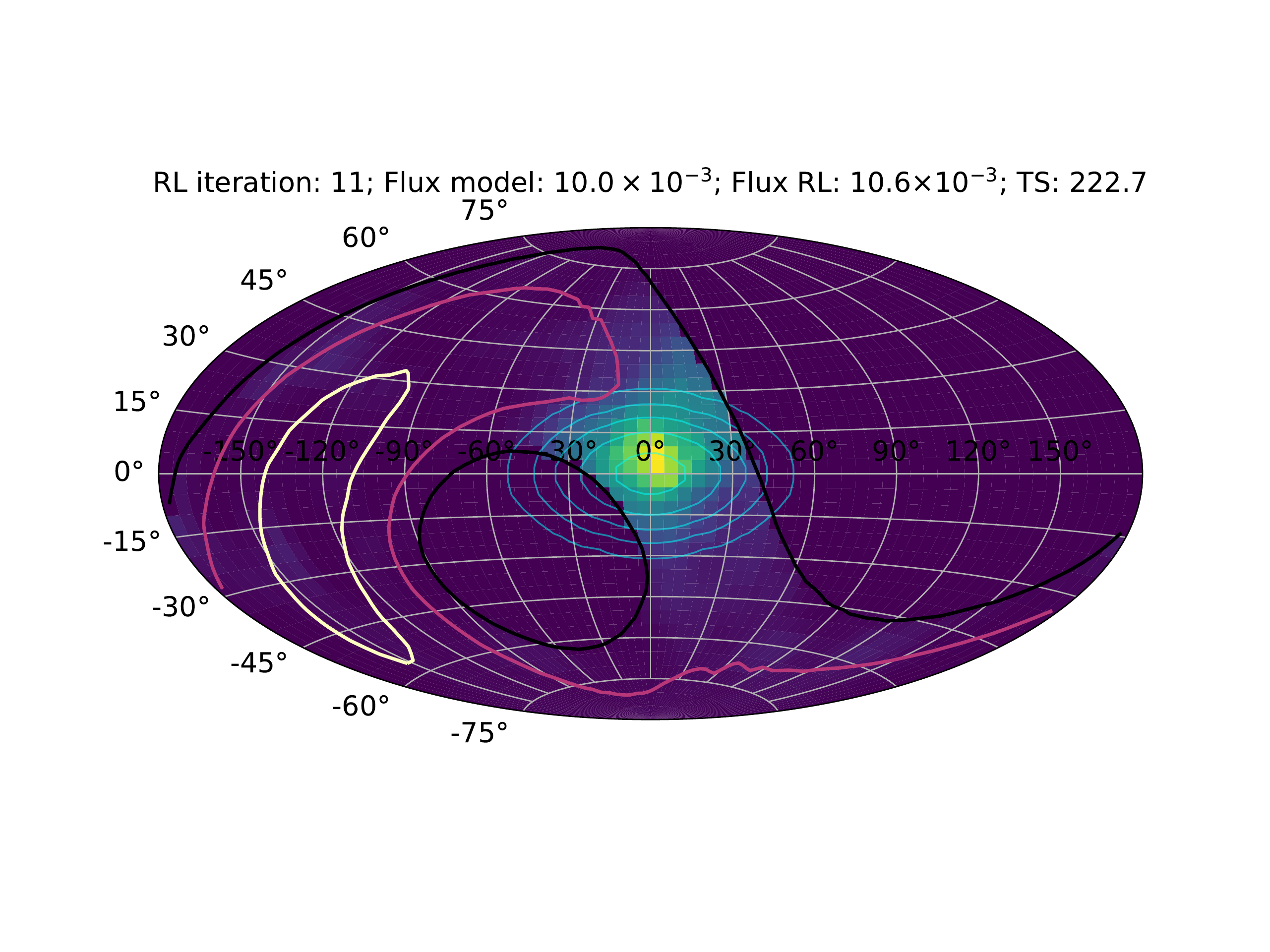}~
	\includegraphics[trim=0.7in 1.4in 0.7in 0.9in, clip=True,width=0.33\columnwidth]{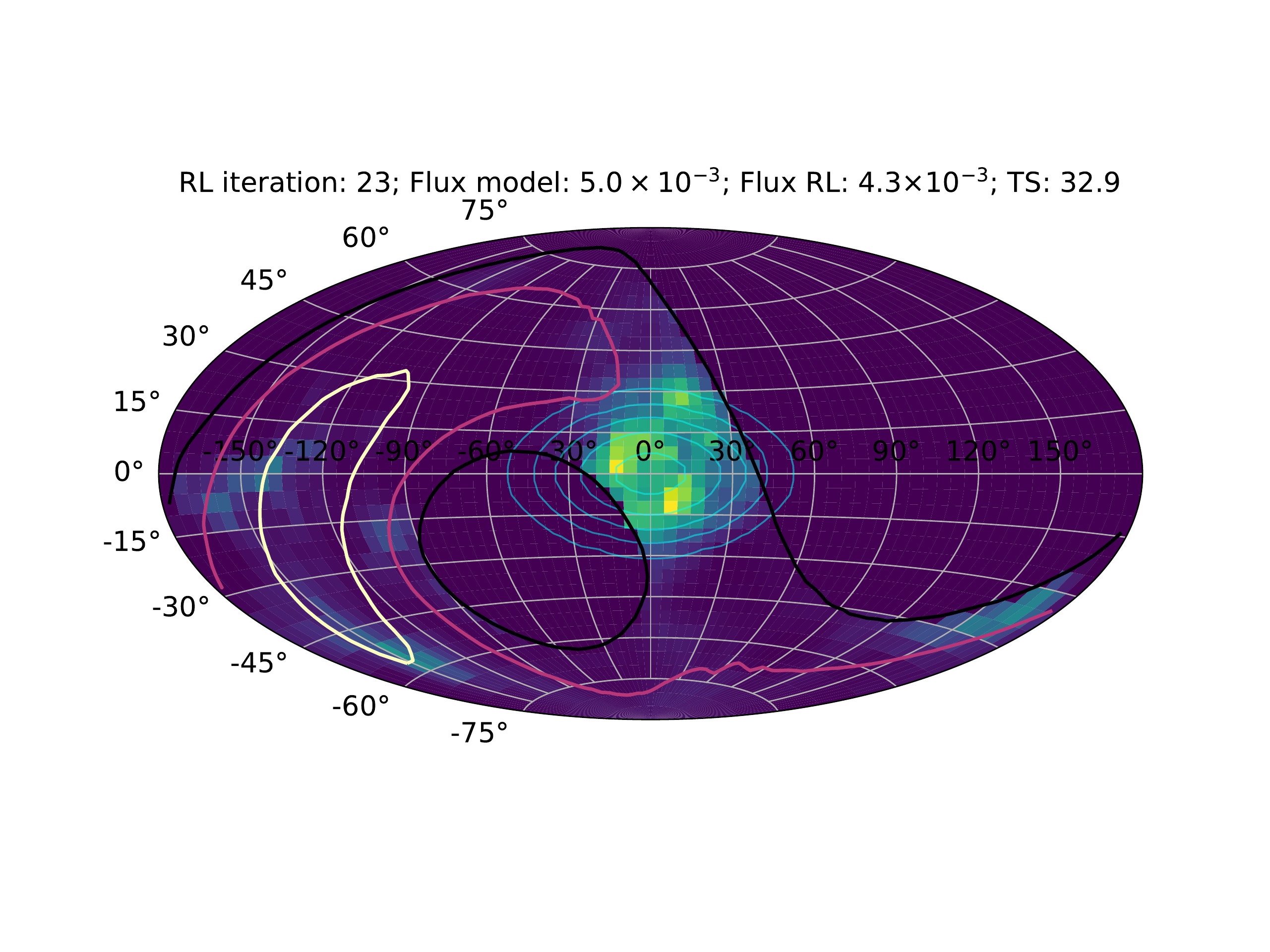}~
	\includegraphics[trim=0.7in 1.4in 0.7in 0.9in, clip=True,width=0.33\columnwidth]{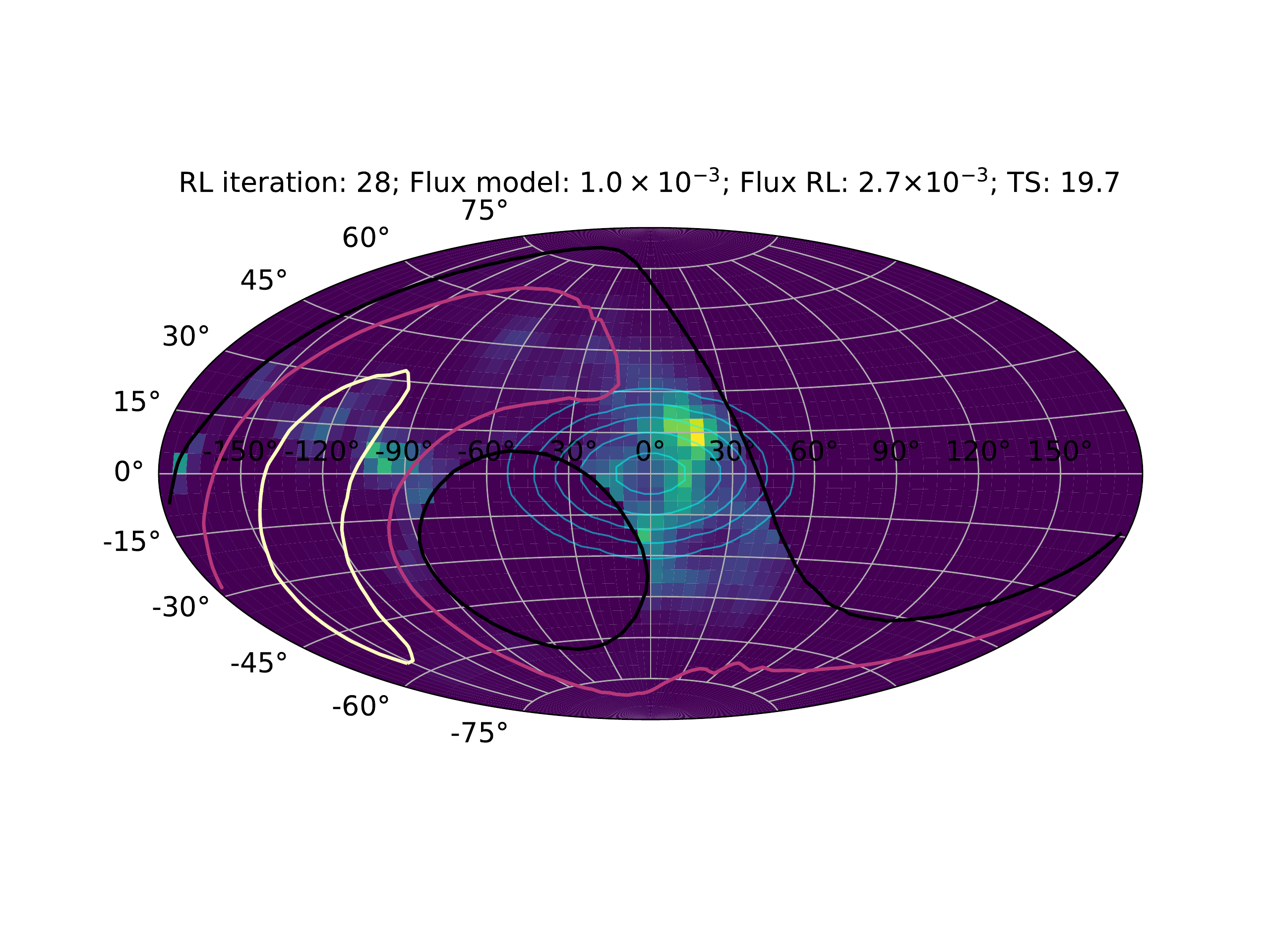}\\
	\includegraphics[trim=5.2in 3.5in 5.2in 3.5in, clip=True,width=0.33\columnwidth]{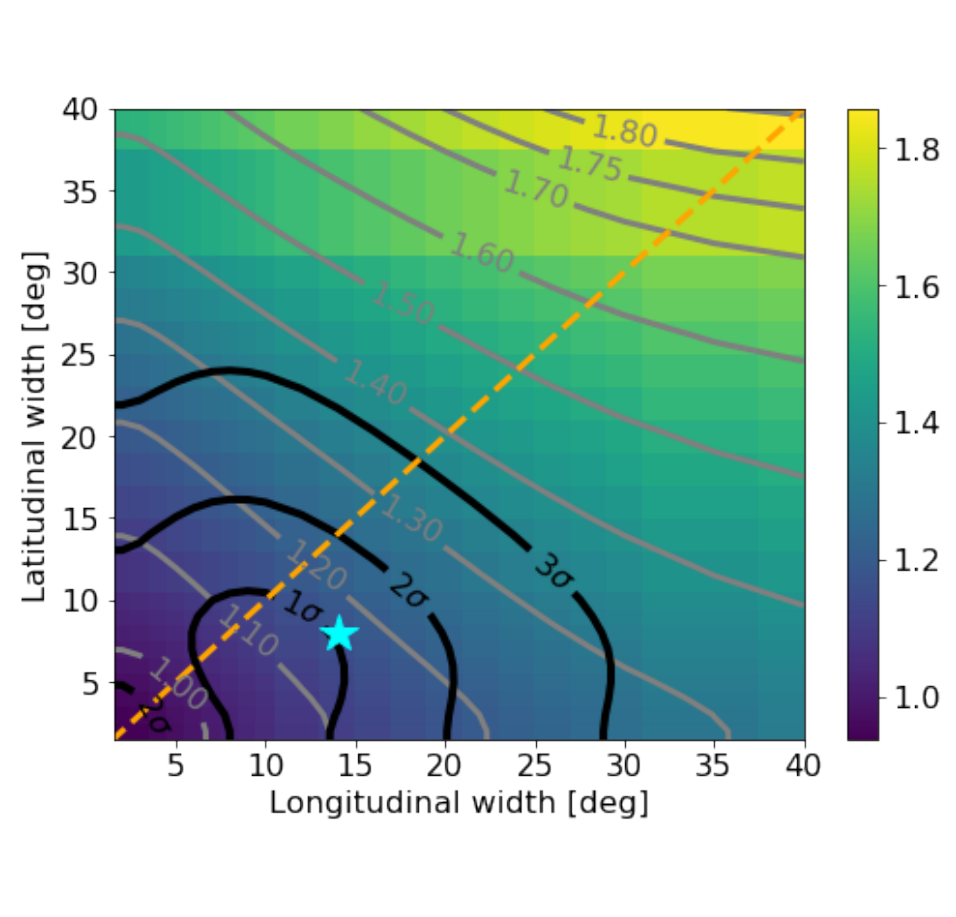}~
	\includegraphics[trim=5.2in 3.5in 5.2in 3.5in, clip=True,width=0.33\columnwidth]{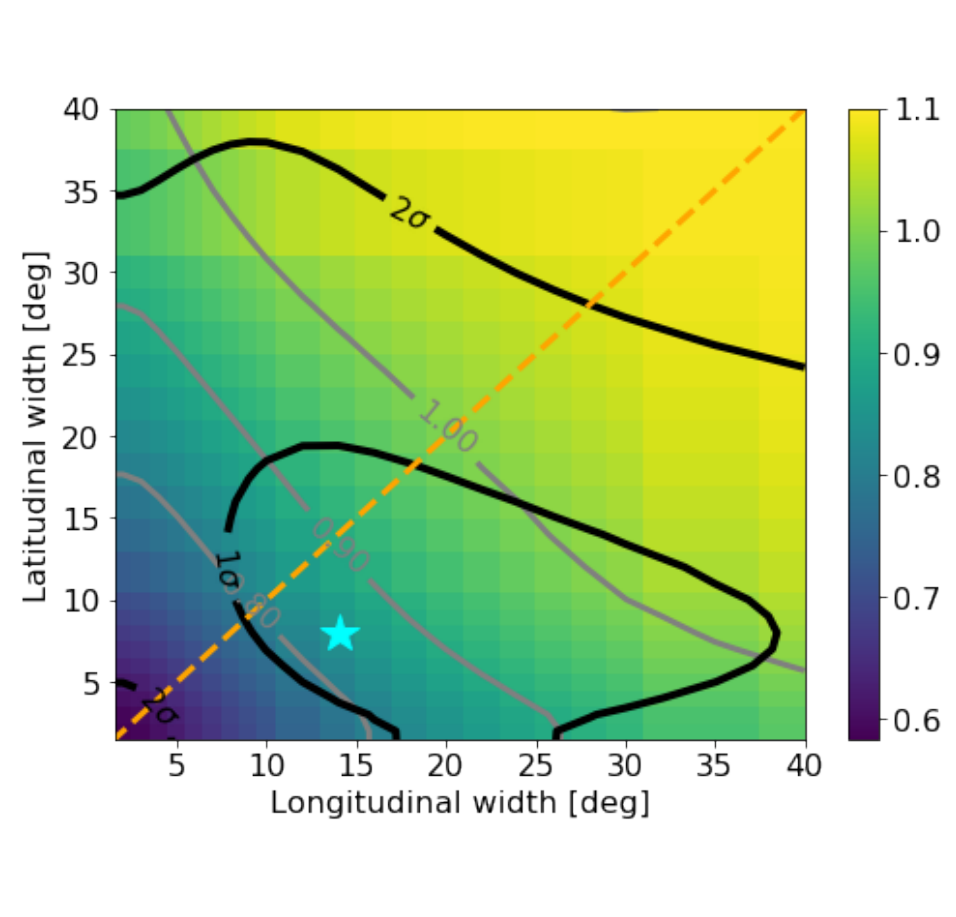}~
	\includegraphics[trim=5.2in 3.5in 5.2in 3.5in, clip=True,width=0.33\columnwidth]{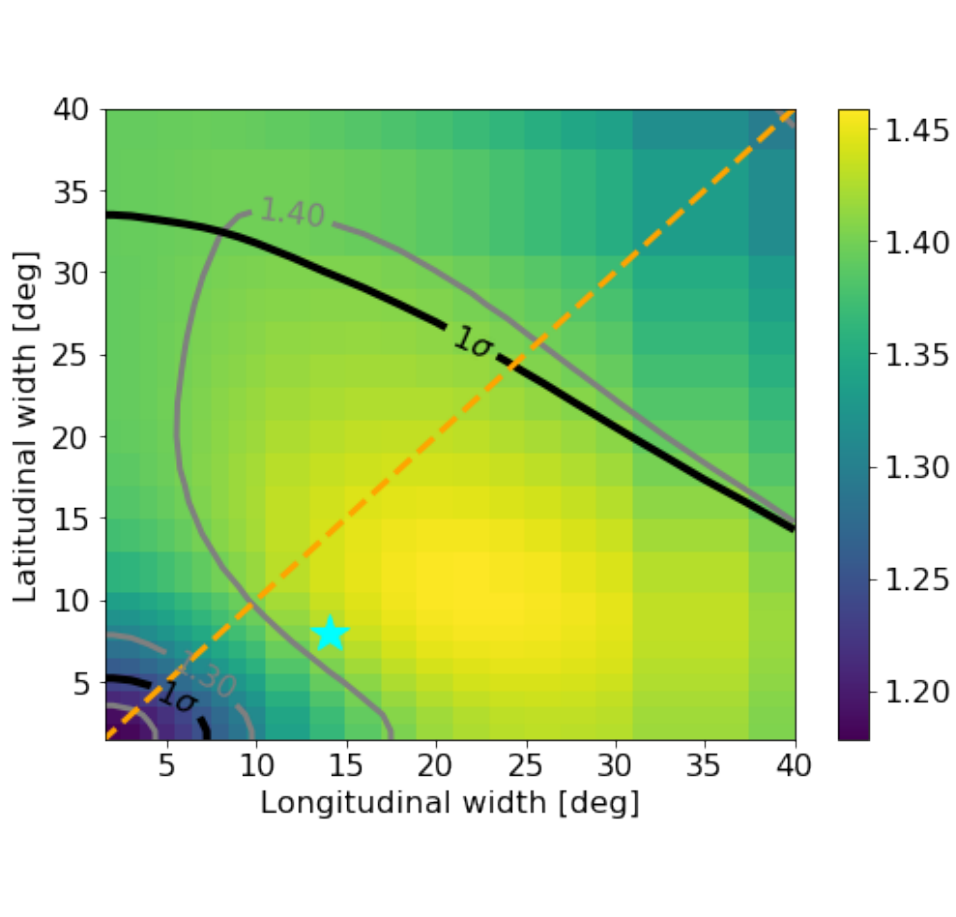}
	\caption{Reconstruction (\textit{top}) and maximum likelihood (\textit{bottom}) results for three different simulated flux levels. From left to right, the model fluxes are $10$, $5$, and $1 \times 10^{-3}\,\mrm{ph\,cm^{-2}\,s^{-1}}$, respectively. The true model parameters are indicated in cyan in all panels. The flux normalisations (coloured) are shown in units of the total flux. The optimal fit is therefore $1.0$.}
	\label{fig:simulation_results}
\end{figure}

In particular, we draw Poisson samples of the combined models, background plus convolved sky, and run the same image reconstruction as described in Sec.\,\ref{sec:arldobags}, including a background fit in each iteration.
For determining the image extent, we perform the same profile likelihood as in Sec.\,\ref{sec:2Dgaussians}.
We note that using exactly one third of the full data set does not result in a factor of 9 less sensitive measurements as the frequent altitude drops further decrease the instrument's sensitivity.
A summary of reconstructed images and likelihood profiles is shown in Fig.\,\ref{fig:simulation_results}.

Clearly, the strongest case (\textit{left} panels in Fig.\,\ref{fig:simulation_results}) is reliably recovered using our methods and nearly no image artifacts emerge.
The resulting image appears more concentrated when reconstructed with the Richardson-Lucy algorithm.
Nevertheless, the correct emission extents are recovered within $1\sigma$ as expected.
The \textit{middle} panels are similar to the real data case: here, the emission appears more structured and stronger artifacts can appear.
The uncertainties in emission extent flux are increased, mainly because the information from the underexposed regions is not enough and the flux is too low.
The last case (\textit{right} panels) represents a marginal detection of the signal.
Still, the emission is found in the regions close the Galactic centre, however more artifacts emerge, which results in a skewed flux distribution as well as an overestimate of the total flux.
This is mainly driven by the dominance of the instrumental background over the sky signal.

\newpage

\section{Additional figures}\label{sec:appendix_figure}

\begin{figure}[ht!]
	\centering
	\includegraphics[trim=0.7in 0.6in 1.3in 0.9in, clip=True, width=0.6\columnwidth]{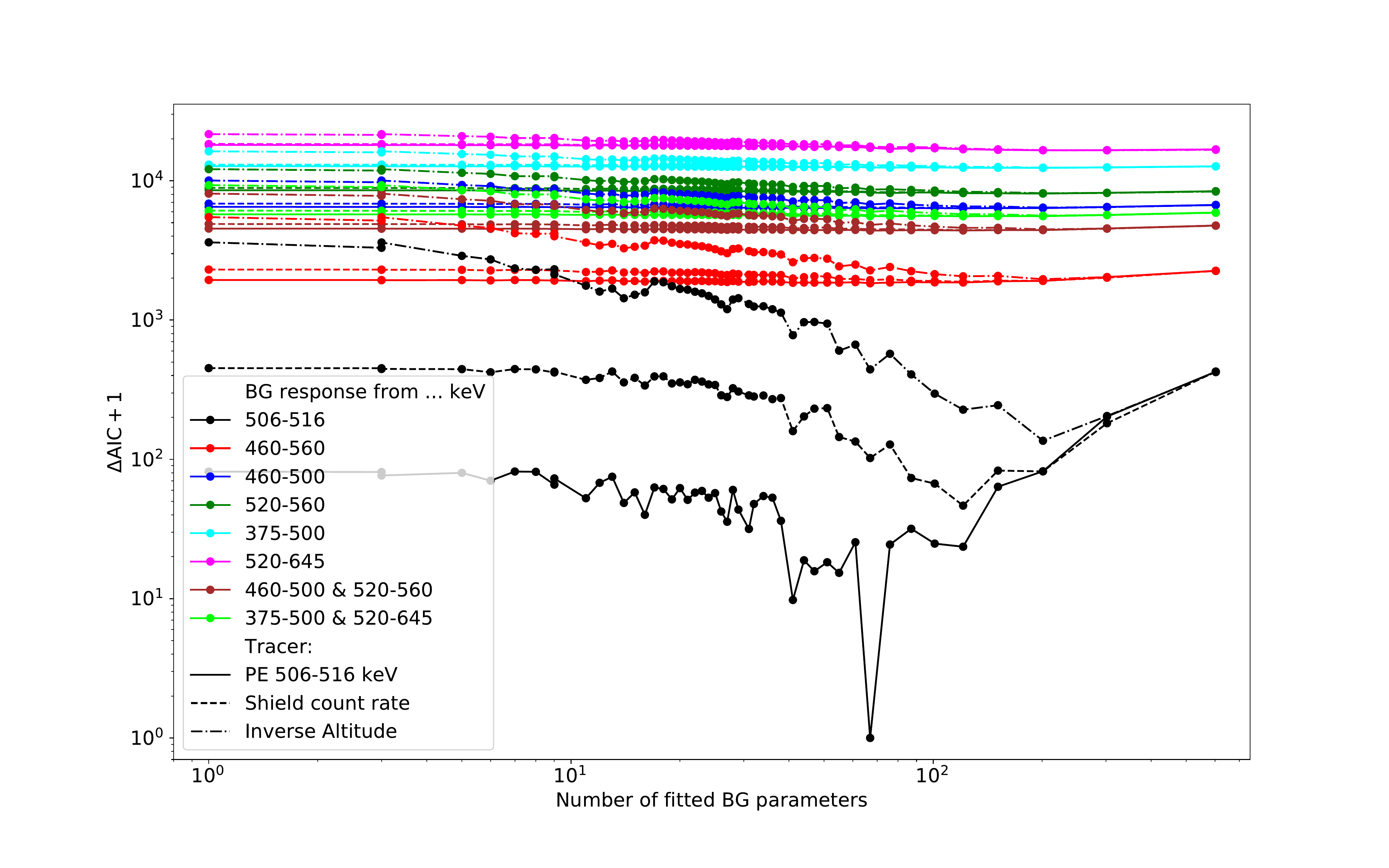}
	\caption{Performance of all background model combinations.}
	\label{fig:allBGresponses_AIC_performace}
\end{figure}

\begin{figure}[ht!]
	\centering
	\includegraphics[trim=0.8in 0.9in 1.0in 1.2in, clip=True, width=0.4\columnwidth]{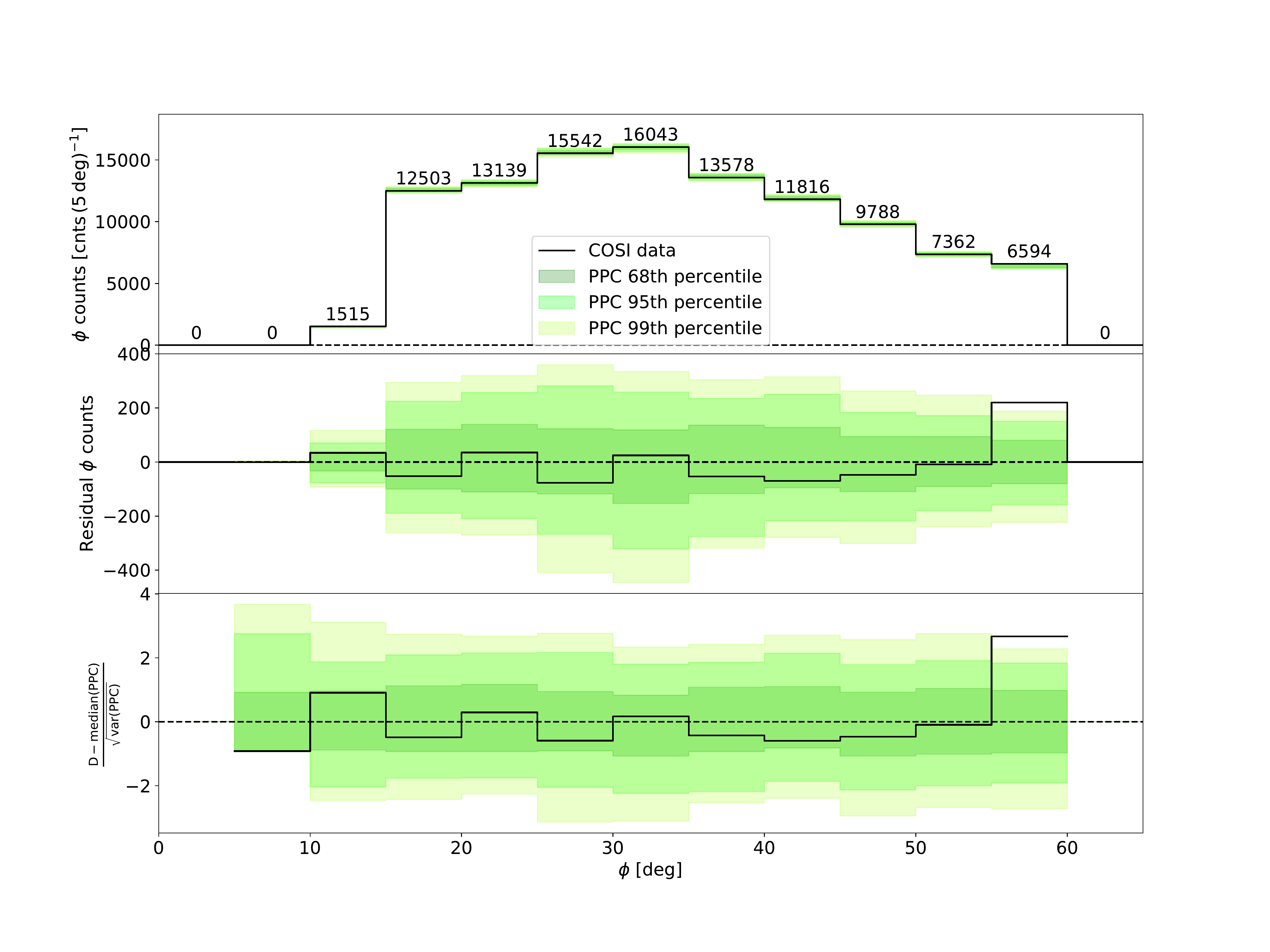}~
	\includegraphics[trim=0.8in 0.9in 1.0in 1.2in, clip=True, width=0.4\columnwidth]{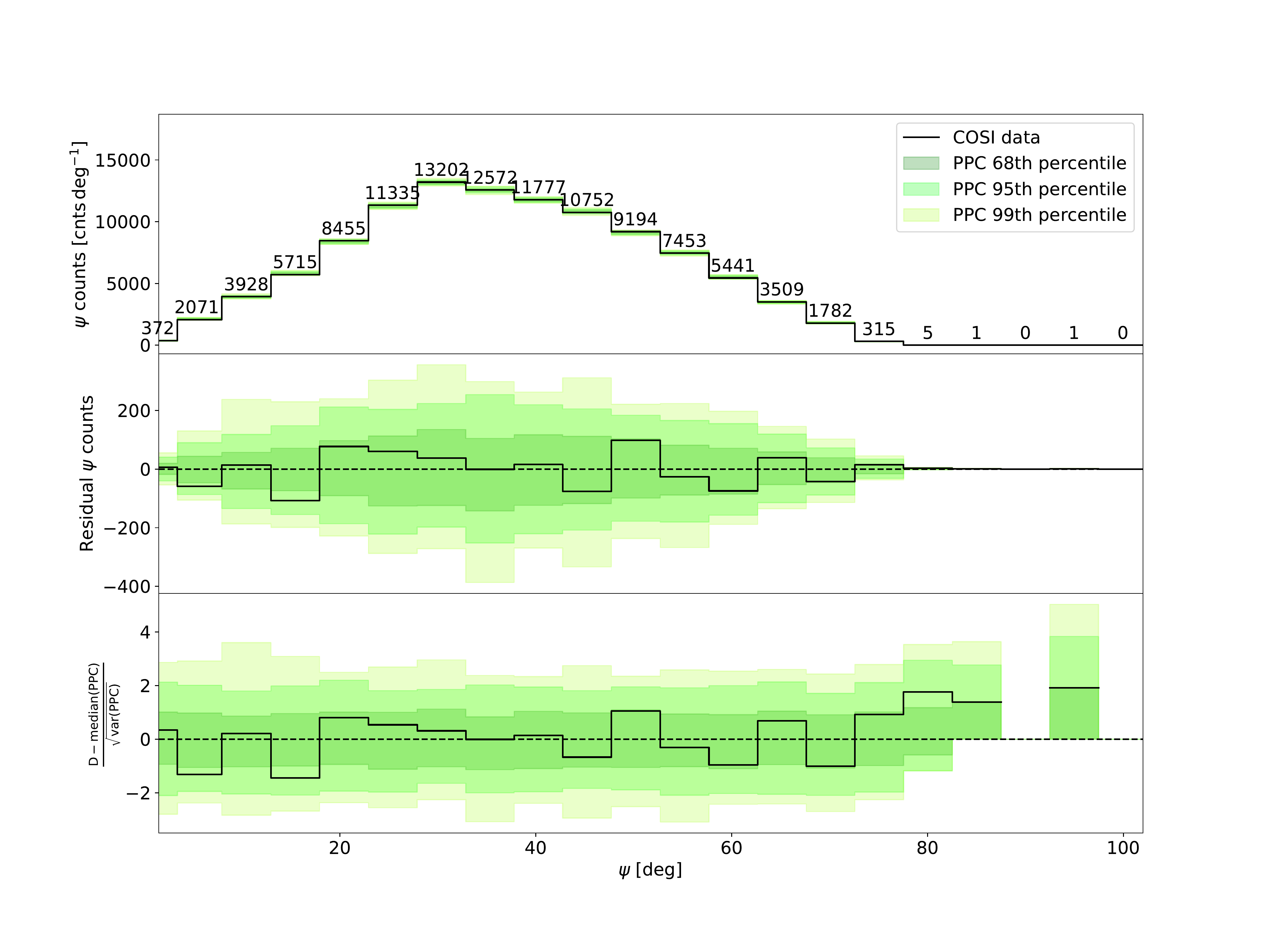}\\
	\includegraphics[trim=0.8in 0.9in 1.0in 1.2in, clip=True, width=0.4\columnwidth]{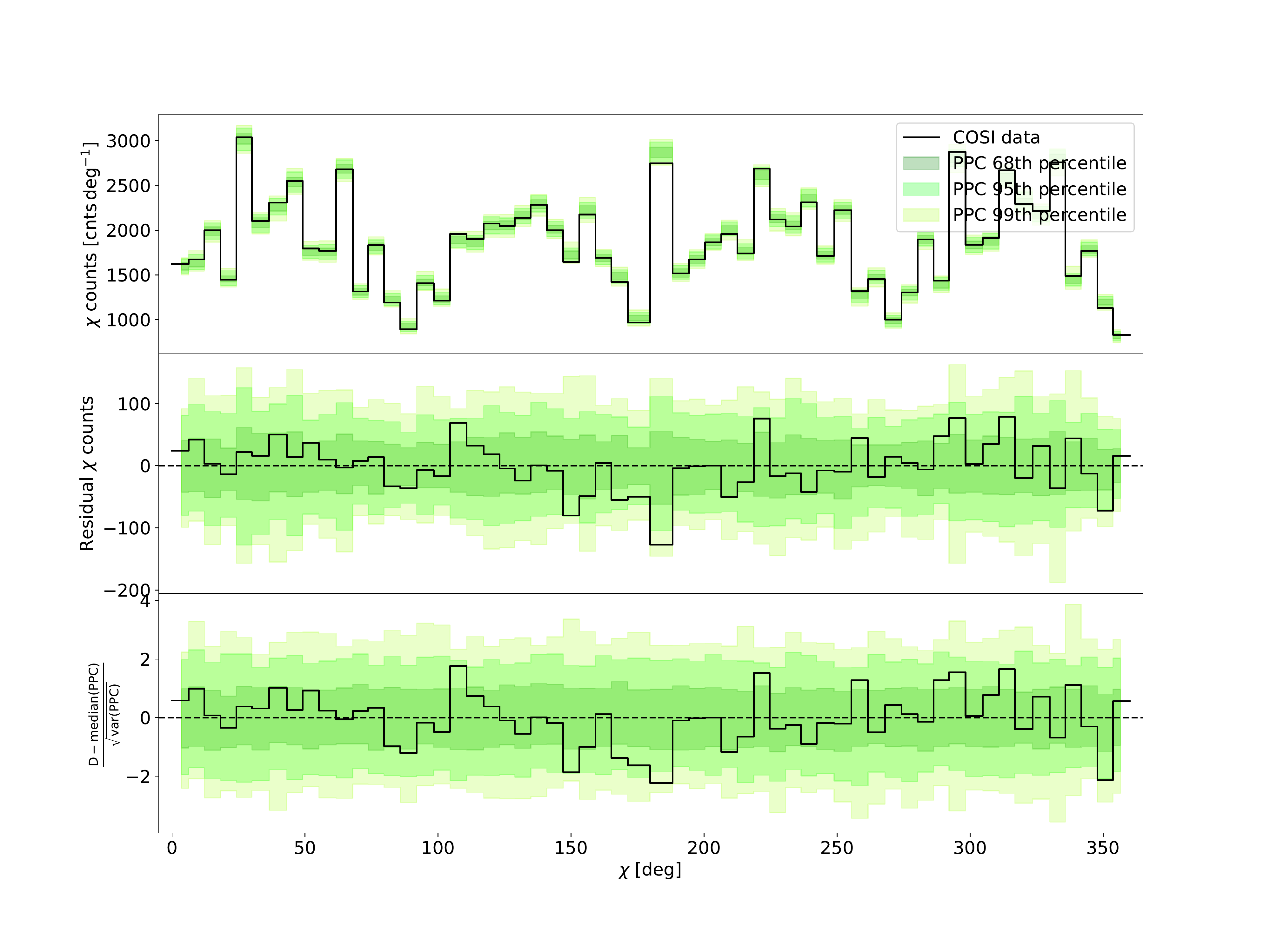}
	\caption{Same as Fig.\,\ref{fig:PPC_time_siegert2016} but for the Compton scattering angle $\phi$ (\textit{top left}), polar scattering angle $\psi$ (\textit{top right}), and azimuthal scattering angle $\chi$ (\textit{bottom}). The number above the summed data bins indicate the photons summed over time and the remaining COSI data space angles. Note the asymmetric residuals between at small $\phi$ or large $\psi$ due to the Poisson character of the counting experiment, leading to a heavily skewed distribution for low numbers.}
	\label{fig:PPC_phi_siegert2016}
\end{figure}

\newpage

\bibliography{alles}{}

\begin{thebibliography}{}
\expandafter\ifx\csname natexlab\endcsname\relax\def\natexlab#1{#1}\fi

\bibitem[{Akaike(1974)}]{Akaike1974_AIC}
Akaike, H. 1974, IEEE Transactions on Automatic Control, 19, 716

\bibitem[{Albernhe {et~al.}(1981)Albernhe, Le~Borgne, Vedrenne, Boclet,
  Durouchoux, \& da~Costa}]{Albernhe1981_511}
Albernhe, F., Le~Borgne, J.~F., Vedrenne, G., {et~al.} 1981, 94, 214

\bibitem[{Alexis {et~al.}(2014)Alexis, Jean, Martin, \&
  Ferri{\`e}re}]{Alexis2014_511ISM}
Alexis, A., Jean, P., Martin, P., \& Ferri{\`e}re, K. 2014, Astronomy {\&}
  Astrophysics, 564, A108

\bibitem[{Allain \& Roques(2006)}]{Allain2006_gammaimaging}
Allain, M., \& Roques, J.~P. 2006, Astronomy {\&} Astrophysics, 447, 1175

\bibitem[{Bandstra {et~al.}(2011)Bandstra, Bellm, Boggs, Perez-Becker,
  Zoglauer, Chang, Chiu, Liang, Chang, Liu, Hung, Huang, Chiang, Run, Lin,
  Amman, Luke, Jean, von Ballmoos, \& Wunderer}]{Bandstra2011_NCT}
Bandstra, M.~S., Bellm, E.~C., Boggs, S.~E., {et~al.} 2011, The Astrophysical
  Journal, 738, 8

\bibitem[{Bisnovatyi-Kogan \& Pozanenko(2017)}]{Bisnovatyi-Kogan2017_511}
Bisnovatyi-Kogan, G.~S., \& Pozanenko, A.~S. 2017, Astrophysics, 60, 223

\bibitem[{Bloemen {et~al.}(1999)Bloemen, Morris, Knoedlseder, Bennett, Diehl,
  Hermsen, Lichti, van~der Meulen, Oberlack, Ryan, Sch{\"o}nfelder, Strong,
  de~Vries, \& Winkler}]{Bloemen1999_26AlOrion_revised}
Bloemen, H., Morris, D., Knoedlseder, J., {et~al.} 1999, The Astrophysical
  Journal, 521, L137

\bibitem[{Boggs \& Jean(2000)}]{Boggs2000_EventReconstruction}
Boggs, S.~E., \& Jean, P. 2000, Astronomy and Astrophysics Supplement, 145, 311

\bibitem[{Boggs {et~al.}(2002)Boggs, Jean, Slassi-Sennou, Coburn, Lin, Madden,
  McBride, Pelling, Primbsch, \& von Ballmoos}]{Boggs2002_SPIBG}
Boggs, S.~E., Jean, P., Slassi-Sennou, S., {et~al.} 2002, Nuclear Instruments
  and Methods in Physics Research Section A: Accelerators, Spectrometers,
  Detectors and Associated Equipment, 491, 390

\bibitem[{Boggs {et~al.}(2015)Boggs, Harrison, Miyasaka, Grefenstette,
  Zoglauer, Fryer, Reynolds, Alexander, An, Barret, Christensen, Craig,
  Forster, Giommi, Hailey, Hornstrup, Kitaguchi, Koglin, Madsen, Mao, Mori,
  Perri, Pivovaroff, Puccetti, Rana, Stern, Westergaard, \&
  Zhang}]{Boggs2015_SN1987A}
Boggs, S.~E., Harrison, F.~A., Miyasaka, H., {et~al.} 2015, Science, 348, 670

\bibitem[{Bouchet {et~al.}(2015)Bouchet, Jourdain, \&
  Roques}]{Bouchet2015_26Al}
Bouchet, L., Jourdain, E., \& Roques, J.-P. 2015, The Astrophysical Journal,
  801, 142

\bibitem[{Bouchet {et~al.}(2010)Bouchet, Roques, \& Jourdain}]{Bouchet2010_511}
Bouchet, L., Roques, J.~P., \& Jourdain, E. 2010, The Astrophysical Journal,
  720, 1772

\bibitem[{Bouchet {et~al.}(1991)Bouchet, Mandrou, Roques, Vedrenne, Cordier,
  Goldwurm, Lebrun, Paul, Sunyaev, Churazov, Gilfanov, Pavlinsky, Grebenev,
  Babalyan, Dekhanov, \& Khavenson}]{Bouchet1991_mq511}
Bouchet, L., Mandrou, P., Roques, J.~P., {et~al.} 1991, 383, L45

\bibitem[{Burnham \&
  Anderson(2004{\natexlab{a}})}]{Burnham2004_ModelSelectionBook}
Burnham, K.~P., \& Anderson, D.~R., eds. 2004{\natexlab{a}}, {Model Selection
  and Multimodel Inference} (New York, NY: Springer New York),
  doi:10.1007/b97636

\bibitem[{Burnham \& Anderson(2004{\natexlab{b}})}]{Burnham2004_AICBIC}
Burnham, K.~P., \& Anderson, D.~R. 2004{\natexlab{b}}, Sociological Methods and
  Research, 33, 261

\bibitem[{Carpenter {et~al.}(2017)Carpenter, Gelman, Hoffman, Lee, Goodrich,
  Betancourt, Brubaker, Guo, Li, \& Riddell}]{Carpenter2017_stan}
Carpenter, B., Gelman, A., Hoffman, M.~D., {et~al.} 2017, Journal of
  Statistical Software, 76, 1

\bibitem[{Churazov {et~al.}(2011)Churazov, Sazonov, Tsygankov, Sunyaev, \&
  Varshalovich}]{Churazov2011_511}
Churazov, E., Sazonov, S., Tsygankov, S., Sunyaev, R., \& Varshalovich, D.
  2011, Monthly Notices of the Royal Astronomical Society, 411, 1727

\bibitem[{Churazov {et~al.}(2005)Churazov, Sunyaev, Sazonov, Revnivtsev, \&
  Varshalovich}]{Churazov2005_511}
Churazov, E., Sunyaev, R., Sazonov, S., Revnivtsev, M., \& Varshalovich, D.
  2005, 357, 1377

\bibitem[{Churazov {et~al.}(2014)Churazov, Sunyaev, Isern, Knoedlseder, Jean,
  Lebrun, Chugai, Grebenev, Bravo, Sazonov, \& Renaud}]{Churazov2014_SN2014J}
Churazov, E., Sunyaev, R., Isern, J., {et~al.} 2014, Nature, 512, 406

\bibitem[{Churazov {et~al.}(2015)Churazov, Sunyaev, Isern, Bikmaev, Bravo,
  Chugai, Grebenev, Jean, Knoedlseder, Lebrun, \&
  Kuulkers}]{Churazov2015_2014JCo}
---. 2015, The Astrophysical Journal, 812, 62

\bibitem[{Collaboration {et~al.}(2013)Collaboration, Robitaille, Tollerud,
  Greenfield, Droettboom, Bray, Aldcroft, Davis, Ginsburg, Price-Whelan,
  Kerzendorf, Conley, Crighton, Barbary, Muna, Ferguson, Grollier, Parikh,
  Nair, Unther, Deil, Woillez, Conseil, Kramer, Turner, Singer, Fox, Weaver,
  Zabalza, Edwards, Azalee~Bostroem, Burke, Casey, Crawford, Dencheva, Ely,
  Jenness, Labrie, Lim, Pierfederici, Pontzen, Ptak, Refsdal, Servillat, \&
  Streicher}]{astropy2013_astropy}
Collaboration, A., Robitaille, T.~P., Tollerud, E.~J., {et~al.} 2013, Astronomy
  {\&} Astrophysics, 558, A33

\bibitem[{Cumani {et~al.}(2019)Cumani, Hernanz, Kiener, Tatischeff, \&
  Zoglauer}]{Cumani2019_MeVBG}
Cumani, P., Hernanz, M., Kiener, J., Tatischeff, V., \& Zoglauer, A. 2019,
  Experimental Astronomy, 47, 273

\bibitem[{Diehl {et~al.}(1992)Diehl, Bennett, Collmar, Connors, den Herder,
  Hermsen, Lichti, Lockwood, Macri, McConnell, Morris, Ryan, Sch{\"o}nfelder,
  Steinle, Strong, Swanenburg, de~Vries, \& Winkler}]{Diehl1992_CDS}
Diehl, R., Bennett, K., Collmar, W., {et~al.} 1992, In NASA. Goddard Space
  Flight Center, 3137

\bibitem[{Diehl {et~al.}(2006)Diehl, Halloin, Kretschmer, Lichti,
  Sch{\"o}nfelder, Strong, von Kienlin, Wang, Jean, Kn{\"o}dlseder, Roques,
  Weidenspointner, Schanne, Hartmann, Winkler, \& Wunderer}]{Diehl2006_26Al}
Diehl, R., Halloin, H., Kretschmer, K., {et~al.} 2006, Nature, 439, 45

\bibitem[{Diehl {et~al.}(2014)Diehl, Siegert, Hillebrandt, Grebenev, Greiner,
  Krause, Kromer, Maeda, R{\"o}pke, \& Taubenberger}]{Diehl2014_SN2014J_Ni}
Diehl, R., Siegert, T., Hillebrandt, W., {et~al.} 2014, Science, 345, 1162

\bibitem[{Diehl {et~al.}(2015)Diehl, Siegert, Hillebrandt, Krause, Greiner,
  Maeda, R{\"o}pke, Sim, Wang, \& Zhang}]{Diehl2015_SN2014J_Co}
---. 2015, Astronomy {\&} Astrophysics, 574, A72

\bibitem[{Diehl {et~al.}(2018)Diehl, Siegert, Greiner, Krause, Kretschmer,
  Lang, Pleintinger, Strong, Weinberger, \& Zhang}]{Diehl2018_BGRDB}
Diehl, R., Siegert, T., Greiner, J., {et~al.} 2018, 611, A12

\bibitem[{Gabry {et~al.}(2019)Gabry, Simpson, Vehtari, Betancourt, \&
  Gelman}]{Gabry2019_BayesianWorkflow}
Gabry, J., Simpson, D., Vehtari, A., Betancourt, M., \& Gelman, A. 2019,
  Journal of the Royal Statistical Society: Series A (Statistics in Society),
  182, 389

\bibitem[{Gehrels(1985)}]{Gehrels1985_balloonBG}
Gehrels, N. 1985, Nuclear Instruments and Methods in Physics Research Section
  A, 239, 324

\bibitem[{Gelman {et~al.}(1996)Gelman, Meng, \& Stern}]{Gelman1996_PPC}
Gelman, A., Meng, X.-L., \& Stern, H. 1996, Statistica Sinica, 6, 733

\bibitem[{Grebenev {et~al.}(2012)Grebenev, Lutovinov, Tsygankov, \&
  Winkler}]{Grebenev2012_SN1987A}
Grebenev, S.~A., Lutovinov, A.~A., Tsygankov, S.~S., \& Winkler, C. 2012,
  arXiv.org, 490, 373

\bibitem[{Grefenstette {et~al.}(2014)Grefenstette, Harrison, Boggs, Reynolds,
  Fryer, Madsen, Wik, Zoglauer, Ellinger, Alexander, An, Barret, Christensen,
  Craig, Forster, Giommi, Hailey, Hornstrup, Kaspi, Kitaguchi, Koglin, Mao,
  Miyasaka, Mori, Perri, Pivovaroff, Puccetti, Rana, Stern, Westergaard, \&
  Zhang}]{Grefenstette2014_CasA}
Grefenstette, B.~W., Harrison, F.~A., Boggs, S.~E., {et~al.} 2014, Nature, 506,
  339

\bibitem[{Grefenstette {et~al.}(2017)Grefenstette, Fryer, Harrison, Boggs,
  DeLaney, Laming, Reynolds, Alexander, Barret, Christensen, Craig, Forster,
  Giommi, Hailey, Hornstrup, Kitaguchi, Koglin, Lopez, Mao, Madsen, Miyasaka,
  Mori, Perri, Pivovaroff, Puccetti, Rana, Stern, Westergaard, Wik, Zhang, \&
  Zoglauer}]{Grefenstette2017_CasA}
Grefenstette, B.~W., Fryer, C.~L., Harrison, F.~A., {et~al.} 2017, The
  Astrophysical Journal, 834, 19

\bibitem[{Guessoum {et~al.}(2006)Guessoum, Jean, \&
  Prantzos}]{Guessoum2006_MQ511}
Guessoum, N., Jean, P., \& Prantzos, N. 2006, Astronomy {\&} Astrophysics, 457,
  753

\bibitem[{Guttman(1967)}]{Guttman1967_PPC}
Guttman, I. 1967, Journal of the Royal Statistical Society: Series B
  (Methodological), 29, 83

\bibitem[{Halloin(2009)}]{spiorthomodel}
Halloin, H. 2009, {|spiorthomodel| Explanatory Guide and Users Manual}, version
  2.0 edn., Max Planck Institut f{\"u}r extraterrestrische Physik,
  Giessenbachstra{\ss}e 1, 85748 Garching, Germany

\bibitem[{Harris {et~al.}(1998)Harris, Teegarden, Cline, Gehrels, Palmer,
  Ramaty, \& Seifert}]{Harris1998_TGRS511}
Harris, M.~J., Teegarden, B.~J., Cline, T.~L., {et~al.} 1998, The Astrophysical
  Journal, 501, L55

\bibitem[{Higdon {et~al.}(2009)Higdon, Lingenfelter, \&
  Rothschild}]{Higdon2009_511}
Higdon, J.~C., Lingenfelter, R.~E., \& Rothschild, R.~E. 2009, The
  Astrophysical Journal, 698, 350

\bibitem[{Hoffman \& Gelman(2011)}]{Hoffman2011_NUTS}
Hoffman, M.~D., \& Gelman, A. 2011, arXiv.org, 1111.4246v1

\bibitem[{Hoffman \& Gelman(2014)}]{Hoffman2014_NUTS}
---. 2014, Journal of Machine Learning Research, 15, 1593

\bibitem[{Hunter(2007)}]{Hunter2007_matplotlib}
Hunter, J.~D. 2007, Computing in Science {\&} Engineering, 9, 90

\bibitem[{Isern {et~al.}(2016)Isern, Jean, Bravo, Knoedlseder, Lebrun,
  Churazov, Sunyaev, Domingo, Badenes, Hartmann, Hoeflich, Renaud, Soldi,
  Elias-Rosa, Hernanz, Dom{\'\i}nguez, Garc{\'\i}a-Senz, Lichti, Vedrenne, \&
  von Ballmoos}]{Isern2016_SN2014J}
Isern, J., Jean, P., Bravo, E., {et~al.} 2016, Astronomy {\&} Astrophysics,
  588, A67

\bibitem[{Iyudin {et~al.}(1997)Iyudin, Diehl, Lichti, Sch{\"o}nfelder, Strong,
  Bloemen, Hermsen, Ryan, Bennett, \& Winkler}]{Iyudin1997_CasA}
Iyudin, A.~F., Diehl, R., Lichti, G.~G., {et~al.} 1997, in The Transparent
  Universe, ed. C.~Winkler, T.~J.~L. Courvoisier, \& P.~Durouchoux, 37

\bibitem[{Jean {et~al.}(2009)Jean, Gillard, Marcowith, \&
  Ferri{\`e}re}]{Jean2009_511ISM}
Jean, P., Gillard, W., Marcowith, A., \& Ferri{\`e}re, K. 2009, Astronomy {\&}
  Astrophysics, 508, 1099

\bibitem[{Jean {et~al.}(2006)Jean, Kn{\"o}dlseder, Gillard, Guessoum,
  Ferri{\`e}re, Marcowith, Lonjou, \& Roques}]{Jean2006_511}
Jean, P., Kn{\"o}dlseder, J., Gillard, W., {et~al.} 2006, Astronomy {\&}
  Astrophysics, 445, 579

\bibitem[{Jean {et~al.}(2003)Jean, Vedrenne, Roques, Sch{\"o}nfelder,
  Teegarden, von Kienlin, Kn{\"o}dlseder, Wunderer, Skinner, Weidenspointner,
  Atti{\'e}, Boggs, Caraveo, Cordier, Diehl, Gros, Leleux, Lichti, Kalemci,
  Kiener, Lonjou, Mandrou, Paul, Schanne, \& von Ballmoos}]{Jean2003_SPIBG}
Jean, P., Vedrenne, G., Roques, J.~P., {et~al.} 2003, 411, L107

\bibitem[{Johnson {et~al.}(1993)Johnson, Kinzer, Kurfess, Strickman, Purcell,
  Grabelsky, Ulmer, Hillis, Jung, \& Cameron}]{Johnson1993_OSSE}
Johnson, W.~N., Kinzer, R.~L., Kurfess, J.~D., {et~al.} 1993, Astrophysical
  Journal Supplement Series (ISSN 0067-0049), 86, 693

\bibitem[{Johnson \& Haymes(1973)}]{Johnson1973_511}
Johnson, W. N.~I., \& Haymes, R.~C. 1973, Astrophysical Journal, 184, 103

\bibitem[{Kaufman(1987)}]{Kaufman1987_RL}
Kaufman, L. 1987, IEEE Transactions on Medical Imaging, 6, 37

\bibitem[{Kierans(2018)}]{Kierans2018_PhD}
Kierans, C. 2018, UC Berkeley Electronic Theses and Dissertations,
  doi:https://escholarship.org/uc/item/1244t3h7

\bibitem[{Kierans {et~al.}(2016)Kierans, Boggs, Chiu, Lowell, Sleator, Tomsick,
  Zoglauer, Amman, Chang, Tseng, Yang, Lin, Jean, \& von
  Ballmoos}]{Kierans2016_COSI}
Kierans, C., Boggs, S., Chiu, J.~L., {et~al.} 2016, in Proceedings of the 11th
  INTEGRAL Conference Gamma-Ray Astrophysics in Multi-Wavelength Perspective.
  10-14 October 2016 Amsterdam, 75

\bibitem[{Kierans {et~al.}(2019)Kierans, Boggs, Zoglauer, Lowell, Sleator,
  Beechert, Brandt, Jean, Lazar, Roberts, Siegert, Tomsick, \& von
  Ballmoos}]{Kierans2019_511COSI}
Kierans, C.~A., Boggs, S.~E., Zoglauer, A., {et~al.} 2019, arXiv.org,
  arXiv:1912.00110

\bibitem[{Knoedlseder {et~al.}(1996)Knoedlseder, von Ballmoos, Diehl, Oberlack,
  Schoenfelder, Bloemen, Hermsen, Ryan, \&
  Bennett}]{Knoedlseder1996_COMPTELimaging}
Knoedlseder, J., von Ballmoos, P., Diehl, R., {et~al.} 1996, in SPIE's 1996
  International Symposium on Optical Science, Engineering, and Instrumentation,
  ed. B.~D. Ramsey \& T.~A. Parnell (SPIE), 386--397

\bibitem[{Knoedlseder {et~al.}(1999)Knoedlseder, Dixon, Bennett, Bloemen,
  Diehl, Hermsen, Oberlack, Ryan, Sch{\"o}nfelder, \& von
  Ballmoos}]{Knoedlseder1999_26AlCOMPTEL}
Knoedlseder, J., Dixon, D., Bennett, K., {et~al.} 1999, Astronomy {\&}
  Astrophysics, 345, 813

\bibitem[{Knoedlseder {et~al.}(2005)Knoedlseder, Jean, Lonjou, Weidenspointner,
  Guessoum, Gillard, Skinner, von Ballmoos, Vedrenne, Roques, Schanne,
  Teegarden, Sch{\"o}nfelder, \& Winkler}]{Knoedlseder2005_511}
Knoedlseder, J., Jean, P., Lonjou, V., {et~al.} 2005, Astronomy {\&}
  Astrophysics, 441, 513

\bibitem[{Koehler(1993)}]{Koehler1993_priorbeliefs}
Koehler, J.~J. 1993, Organizational Behavior and Human Decision Processes, 56,
  28

\bibitem[{Kretschmer {et~al.}(2013)Kretschmer, Diehl, Krause, Burkert,
  Fierlinger, Gerhard, Greiner, \& Wang}]{Kretschmer2013_26Al}
Kretschmer, K., Diehl, R., Krause, M., {et~al.} 2013, Astronomy {\&}
  Astrophysics, 559, A99

\bibitem[{Kumar {et~al.}(2019)Kumar, Carroll, Hartikainen, \&
  Martin}]{Kumar2019_arviz}
Kumar, R., Carroll, C., Hartikainen, A., \& Martin, O. 2019, Journal of Open
  Source Software, 4, 1143

\bibitem[{Leventhal {et~al.}(1986)Leventhal, MacCallum, Huters, \&
  Stang}]{Leventhal1986_511}
Leventhal, M., MacCallum, C.~J., Huters, A.~F., \& Stang, P.~D. 1986, 302, 459

\bibitem[{Leventhal {et~al.}(1978)Leventhal, MacCallum, \&
  Stang}]{Leventhal1978_511}
Leventhal, M., MacCallum, C.~J., \& Stang, P.~D. 1978, Astrophysical Journal,
  225, L11

\bibitem[{Ling(1975)}]{Ling1975_MeVBG}
Ling, J.~C. 1975, Journal of Geophysical Research, 80, 3241

\bibitem[{Ling {et~al.}(1977)Ling, Mahoney, Willett, \&
  Jacobson}]{Ling1977_511keV_atmosphericBG}
Ling, J.~C., Mahoney, W.~A., Willett, J.~B., \& Jacobson, A.~S. 1977, Journal
  of Geophysical Research, 82, 1463

\bibitem[{Lingenfelter \& Ramaty(1989)}]{Lingenfelter1989_511}
Lingenfelter, R.~E., \& Ramaty, R. 1989, 343, 686

\bibitem[{Lucy(1974)}]{Lucy1974_RichardsonLucy}
Lucy, L.~B. 1974, 79, 745

\bibitem[{Lucy(1992)}]{Lucy1992_RL}
---. 1992, Astronomical Journal (ISSN 0004-6256), 104, 1260

\bibitem[{Milne \& Leising(1997)}]{Milne1997_511}
Milne, P.~A., \& Leising, M.~D. 1997, in Proceedings of the Fourth Compton
  Symposium, ed. C.~D. Dermer, M.~S. Strickman, \& J.~D. Kurfess, 1017--1021

\bibitem[{Morris {et~al.}(2006)Morris, Bennett, Bloemen, Hermsen, Lichti,
  McConnell, Ryan, \& Sch{\"o}nfelder}]{Morris1995_SN1991TCOMPTEL}
Morris, D.~J., Bennett, K., Bloemen, H., {et~al.} 2006, Annals of the New York
  Academy of Sciences, 759, 397

\bibitem[{Nickerson(1998)}]{Nickerson1998_confirmationbias}
Nickerson, R.~S. 1998, Review of General Psychology, 2, 175

\bibitem[{Oberlack {et~al.}(1996)Oberlack, Bennett, Bloemen, Diehl, Dupraz,
  Hermsen, Knoedlseder, Morris, Schoenfelder, Strong, \&
  Winkler}]{Oberlack1996_26Al}
Oberlack, U., Bennett, K., Bloemen, H., {et~al.} 1996, 120, C311

\bibitem[{Oliphant(2006)}]{Oliphant2006_numpy}
Oliphant, T.~E. 2006, {A guide to NumPy}, Vol.~1 (Trelgol Publishing USA)

\bibitem[{Panther(2018)}]{Panther2018_pos_transport}
Panther, F. 2018, Galaxies, 6, 39

\bibitem[{Pleintinger {et~al.}(2019)Pleintinger, Siegert, Diehl, Fujimoto,
  Greiner, Krause, \& Krumholz}]{Pleintinger2019_26Al}
Pleintinger, M. M.~M., Siegert, T., Diehl, R., {et~al.} 2019, Astronomy {\&}
  Astrophysics, 632, A73

\bibitem[{Pohl(2004)}]{Pohl2004_biases}
Pohl, R. 2004, {Cognitive Illusions}, A Handbook on Fallacies and Biases in
  Thinking, Judgement and Memory (Psychology Press)

\bibitem[{Prantzos(2006)}]{Prantzos2006_511}
Prantzos, N. 2006, Astronomy {\&} Astrophysics, 449, 869

\bibitem[{Prantzos {et~al.}(2011)Prantzos, Boehm, Bykov, Diehl, Ferri{\`e}re,
  Guessoum, Jean, Knoedlseder, Marcowith, Moskalenko, Strong, \&
  Weidenspointner}]{Prantzos2011_511}
Prantzos, N., Boehm, C., Bykov, A.~M., {et~al.} 2011, Reviews of Modern
  Physics, 83, 1001

\bibitem[{Purcell {et~al.}(1993)Purcell, Grabelsky, Ulmer, Johnson, Kinzer,
  Kurfess, Strickman, \& Jung}]{Purcell1993_511}
Purcell, W.~R., Grabelsky, D.~A., Ulmer, M.~P., {et~al.} 1993, 413, L85

\bibitem[{Purcell {et~al.}(1997)Purcell, Cheng, Dixon, Kinzer, Kurfess,
  Leventhal, Saunders, Skibo, Smith, \& Tueller}]{Purcell1997_511}
Purcell, W.~R., Cheng, L.~X., Dixon, D.~D., {et~al.} 1997, 491, 725

\bibitem[{Richardson(1972)}]{Richardson1972_RichardsonLucy}
Richardson, W.~H. 1972, Journal of the Optical Society of America (1917-1983),
  62, 55

\bibitem[{Rubin(1981)}]{Rubin1981_PPC}
Rubin, D.~B. 1981, Journal of Educational Statistics, 6, 377

\bibitem[{Rubin(1984)}]{Rubin1984_PPC}
---. 1984, The Annals of Statistics, 12, 1151

\bibitem[{Sato(2016)}]{Sato2016_EXPACS}
Sato, T. 2016, PLOS ONE, 11, e0160390

\bibitem[{Scargle(1998)}]{Scargle1998_BayesianBlocks}
Scargle, J.~D. 1998, The Astrophysical Journal, 504, 405

\bibitem[{Scargle {et~al.}(2012)Scargle, Norris, Jackson, \&
  Chiang}]{Scargle2012_BayesianBlocks}
Scargle, J.~D., Norris, J.~P., Jackson, B., \& Chiang, J. 2012, Astrophysics
  Source Code Library, ascl:1209.001

\bibitem[{Shepp \& Vardi(1982)}]{Shepp1982_RL}
Shepp, L.~A., \& Vardi, Y. 1982, IEEE Transactions on Medical Imaging, 1, 113

\bibitem[{Siegert {et~al.}(2019{\natexlab{a}})Siegert, Crocker, Diehl, Krause,
  Panther, Pleintinger, \& Weinberger}]{Siegert2019_lv511}
Siegert, T., Crocker, R.~M., Diehl, R., {et~al.} 2019{\natexlab{a}}, Astronomy
  {\&} Astrophysics, 627, A126

\bibitem[{Siegert {et~al.}(2016{\natexlab{a}})Siegert, Diehl, Khachatryan,
  Krause, Guglielmetti, Greiner, Strong, \& Zhang}]{Siegert2016_511}
Siegert, T., Diehl, R., Khachatryan, G., {et~al.} 2016{\natexlab{a}}, Astronomy
  {\&} Astrophysics, 586, A84

\bibitem[{Siegert {et~al.}(2015)Siegert, Diehl, Krause, \&
  Greiner}]{Siegert2015_CasA}
Siegert, T., Diehl, R., Krause, M. G.~H., \& Greiner, J. 2015, Astronomy {\&}
  Astrophysics, 579, A124

\bibitem[{Siegert {et~al.}(2019{\natexlab{b}})Siegert, Diehl, Weinberger,
  Pleintinger, Greiner, \& Zhang}]{Siegert2019_SPIBG}
Siegert, T., Diehl, R., Weinberger, C., {et~al.} 2019{\natexlab{b}}, Astronomy
  {\&} Astrophysics, 626, A73

\bibitem[{Siegert {et~al.}(2016{\natexlab{b}})Siegert, Diehl, Greiner, Krause,
  Beloborodov, Bel, Guglielmetti, Rodriguez, Strong, \&
  Zhang}]{Siegert2016_V404}
Siegert, T., Diehl, R., Greiner, J., {et~al.} 2016{\natexlab{b}}, Nature, 531,
  341

\bibitem[{Skinner {et~al.}(2014)Skinner, Diehl, Zhang, Bouchet, \&
  Jean}]{Skinner2014_511}
Skinner, G., Diehl, R., Zhang, X., Bouchet, L., \& Jean, P. 2014, in
  Proceedings of the 10th INTEGRAL Workshop: ''A Synergistic View of the
  High-Energy Sky'' (INTEGRAL 2014). 15-19 September 2014. Annapolis, MD, USA.
  Published online at http://pos.sissa.it/cgi-bin/reader/conf.cgi?confid=228,
  id.054, 054

\bibitem[{Skinner {et~al.}(2012)Skinner, Jean, Knoedlseder, von Ballmoos,
  Leising, Milne, \& Weidenspointner}]{Skinner2012_511}
Skinner, G., Jean, P., Knoedlseder, J., {et~al.} 2012, in Proceedings of ''An
  INTEGRAL view of the high-energy sky (the first 10 years)'' - 9th INTEGRAL
  Workshop and celebration of the 10th anniversary of the launch (INTEGRAL
  2012). 15-19 October 2012. Bibliotheque Nationale de France, Paris, France.
  Published online at http://pos.sissa.it/cgi-bin/reader/conf.cgi?confid=176,
  id.112, 112

\bibitem[{Sleator(2019)}]{Sleator2019_PhD}
Sleator, C. 2019, PhD thesis, UC Berkeley, Published online at
  https://escholarship.org/uc/item/0zn566rj

\bibitem[{Sleator {et~al.}(2019)Sleator, Zoglauer, Lowell, Kierans, Pellegrini,
  Beechert, Boggs, Brandt, Lazar, Roberts, Siegert, \&
  Tomsick}]{Sleator2019_COSI_DEE}
Sleator, C.~C., Zoglauer, A., Lowell, A.~W., {et~al.} 2019, Nuclear Inst. and
  Methods in Physics Research, 946, 162643

\bibitem[{Sunyaev {et~al.}(1992)Sunyaev, Churazov, Gilfanov, Dyachkov,
  Khavenson, Grebenev, Kremnev, Sukhanov, Goldwurm, Ballet, Cordier, Paul,
  Denis, Vedrenne, Niel, \& Jourdain}]{Sunyaev1992_xrb511}
Sunyaev, R., Churazov, E., Gilfanov, M., {et~al.} 1992, 389, L75

\bibitem[{Tomsick {et~al.}(2019)Tomsick, Zoglauer, Sleator, Lazar, Beechert,
  Boggs, Roberts, Siegert, Lowell, Wulf, Grove, Phlips, Brandt, Smale, Kierans,
  Burns, Hartmann, Leising, Ajello, Fryer, Amman, Chang, Jean, \& von
  Ballmoos}]{Tomsick2019_COSI}
Tomsick, J.~A., Zoglauer, A., Sleator, C., {et~al.} 2019, arXiv.org,
  arXiv:1908.04334

\bibitem[{Tsygankov {et~al.}(2016)Tsygankov, Krivonos, Lutovinov, Revnivtsev,
  Churazov, Sunyaev, \& Grebenev}]{Tsygankov2016_44Ti}
Tsygankov, S.~S., Krivonos, R.~A., Lutovinov, A.~A., {et~al.} 2016, Monthly
  Notices of the Royal Astronomical Society, 458, 3411

\bibitem[{van Dijk(1996)}]{vonDijk1996_PhD_COMPTELBG}
van Dijk, R. 1996, PhD thesis, -

\bibitem[{Vedrenne {et~al.}(2003)Vedrenne, Roques, Sch{\"o}nfelder, Mandrou,
  Lichti, von Kienlin, Cordier, Schanne, Kn{\"o}dlseder, Skinner, Jean,
  Sanchez, Caraveo, Teegarden, von Ballmoos, Bouchet, Paul, Matteson, Boggs,
  Wunderer, Leleux, Weidenspointner, Durouchoux, Diehl, Strong, Cass{\'e},
  Clair, \& Andr{\'e}}]{Vedrenne2003_SPI}
Vedrenne, G., Roques, J.~P., Sch{\"o}nfelder, V., {et~al.} 2003, 411, L63

\bibitem[{Vink {et~al.}(2001)Vink, Laming, Kaastra, Bleeker, Bloemen, \&
  Oberlack}]{Vink2001_CasA}
Vink, J., Laming, J.~M., Kaastra, J.~S., {et~al.} 2001, 560, L79

\bibitem[{Virtanen {et~al.}(2019)Virtanen, Gommers, Oliphant, Haberland, Reddy,
  Cournapeau, Burovski, Peterson, Weckesser, Bright, van~der Walt, Brett,
  Wilson, Jarrod~Millman, Mayorov, Nelson, Jones, Kern, Larson, Carey, Polat,
  Feng, Moore, VanderPlas, Laxalde, Perktold, Cimrman, Henriksen, Quintero,
  Harris, Archibald, Ribeiro, Pedregosa, van Mulbregt, \&
  Contributors}]{Virtanen2019_scipy}
Virtanen, P., Gommers, R., Oliphant, T.~E., {et~al.} 2019, arXiv.org,
  arXiv:1907.10121

\bibitem[{von Ballmoos {et~al.}(1989)von Ballmoos, Diehl, \&
  Schoenfelder}]{vonBallmoos1989_ComptonTelescope}
von Ballmoos, P., Diehl, R., \& Schoenfelder, V. 1989, Astronomy and
  Astrophysics (ISSN 0004-6361), 221, 396

\bibitem[{Weidenspointner {et~al.}(2008)Weidenspointner, Skinner, Jean,
  Kn{\"o}dlseder, von Ballmoos, Diehl, Strong, Cordier, Schanne, \&
  Winkler}]{Weidenspointner2008_511b}
Weidenspointner, G., Skinner, G.~K., Jean, P., {et~al.} 2008, 52, 454

\bibitem[{Winkler {et~al.}(2003)Winkler, Courvoisier, Di~Cocco, Gehrels,
  Gim{\'e}nez, Grebenev, Hermsen, Mas-Hesse, Lebrun, Lund, Palumbo, Paul,
  Roques, Schnopper, Sch{\"o}nfelder, Sunyaev, Teegarden, Ubertini, Vedrenne,
  \& Dean}]{Winkler2003_INTEGRAL}
Winkler, C., Courvoisier, T. J.~L., Di~Cocco, G., {et~al.} 2003, Astronomy {\&}
  Astrophysics, 411, L1

\bibitem[{Zoglauer {et~al.}(2006)Zoglauer, Andritschke, \&
  Schopper}]{Zoglauer2006_MEGAlib}
Zoglauer, A., Andritschke, R., \& Schopper, F. 2006, New Astronomy Reviews, 50,
  629

\bibitem[{Zoglauer {et~al.}(2007)Zoglauer, Boggs, Andritschke, \&
  Kanbach}]{Zoglauer2007_EventReconstruction}
Zoglauer, A., Boggs, S.~E., Andritschke, R., \& Kanbach, G. 2007, Mathematics
  of Data/Image Pattern Recognition, 6700, 67000I

\bibitem[{Zoglauer \& Kanbach(2003)}]{Zoglauer2003_DopplerCompton}
Zoglauer, A., \& Kanbach, G. 2003, X-Ray and Gamma-Ray Telescopes and
  Instruments for Astronomy. Edited by Joachim E. Truemper, 4851, 1302

\bibitem[{Zoglauer(2006)}]{Zoglauer2006_PhD}
Zoglauer, A.~C. 2006, PhD Thesis

\end{thebibliography}
\bibliographystyle{aasjournal}

\end{document}